\documentclass[fleqn,usenatbib]{mnras}

\usepackage[T1]{fontenc}
\usepackage{multirow}
\DeclareRobustCommand{\VAN}[3]{#2}
\let\VANthebibliography\thebibliography
\def\thebibliography{\DeclareRobustCommand{\VAN}[3]{##3}\VANthebibliography}

\usepackage{graphicx}	
\usepackage{amsmath}	
\usepackage{amssymb}	

\usepackage{newtxtext,newtxmath}

\newcommand{\rulesep}{\unskip\ \vrule\ }

\title[Machine learning in 21-cm cosmology]{Emulation of the Cosmic Dawn 21-cm Power Spectrum and Classification of Excess Radio Models Using an Artificial Neural Network}

\author[S. Sikder et al.]{
Sudipta Sikder,$^{1}$\thanks{E-mail: sudiptas@mail.tau.ac.il}
Rennan Barkana,$^{1,2}$
Itamar Reis$^{1}$
and Anastasia Fialkov$^{3}$
\\
$^{1}$School of Physics and Astronomy, Tel-Aviv University, Tel-Aviv, 69978, Israel \\
$^{2}$ Institute for Advanced Study, 1 Einstein Drive, Princeton, New Jersey 08540, USA
\\
$^{3}$Institute of Astronomy, University of Cambridge, Madingley Road, Cambridge, CB3 0HA, UK
}

\date{Accepted XXX. Received YYY; in original form ZZZ}

\pubyear{2021}

\begin{document}
\label{firstpage}
\pagerange{\pageref{firstpage}--\pageref{lastpage}}
\maketitle

\begin{abstract}

The cosmic 21-cm line of hydrogen is expected to be measured in detail by the next generation of radio telescopes. The enormous dataset from future 21-cm surveys will revolutionize our understanding of early cosmic times. We present a machine learning approach based on an Artificial Neural Network that uses emulation in order to uncover the astrophysics in the epoch of reionization and cosmic dawn. Using a seven-parameter astrophysical model that covers a very wide range of possible 21-cm signals, over the redshift range 6 to 30 and wavenumber range $0.05$ to $1 \ \rm{Mpc}^{-1}$ we emulate the 21-cm power spectrum with a typical accuracy of $10 - 20\%$. As a realistic example, we train an emulator using the power spectrum with an optimistic noise model of the Square Kilometre Array (SKA). Fitting to mock SKA data results in a typical measurement accuracy of $2.8\%$ in the optical depth to the cosmic microwave background, $34\%$ in the star-formation efficiency of galactic halos, and a factor of 9.6 in the X-ray efficiency of galactic halos. Also, with our modeling we reconstruct the true 21-cm power spectrum from the mock SKA data with a typical accuracy of $15 - 30\%$.
In addition to standard astrophysical models, we consider two exotic possibilities of strong excess radio backgrounds at high redshifts. We use a neural network to identify the type of radio background present in the 21-cm power spectrum, with an accuracy of $87\%$ for mock SKA data.
\end{abstract}

\begin{keywords}
methods: numerical -- methods: statistical --  dark ages, reionization, first stars -- cosmology: theory
\end{keywords}



\section{Introduction}

The redshifted 21-cm signal from neutral hydrogen is the most promising probe of the Epoch of Reionization (EoR) and cosmic dawn. This 21-cm emission or absorption originates from the hyperfine splitting of the hydrogen atom. As this signal depends on both cosmological and astrophysical parameters, it should be possible to decipher abundant information about the early universe from the signal once it is observed. The Low Frequency Array \citep[LOFAR,][]{lofar_gehlot}, the Precision Array to Probe the Epoch of Reionization \citep[PAPER,][]{paper_kolopanis}, the Murchison Wide-field Array \citep[MWA,][]{mwa_trott},  the Owens Valley Radio Observatory Long Wavelength Array \citep[OVRO-LWA,][]{lwa_eastwood}, the Large-aperture Experiment to detect the Dark Age \citep[LEDA,][]{leda_price, leda_garsden} and the Hydrogen Epoch of Reionization Array \citep[HERA,][]{hera_DeBoer} are experiments that have analyzed data in an attempt to detect the power spectrum from the epoch of reionization. Although the existing upper limits are weak, they already provide interesting constraints on some of the exotic scenarios (e.g. with extra radio background as considered here) \citep{LOFARnew, HERAnew}. HERA along with the New Extension in Nan\c{c}ay Upgrading LOFAR  \citep[NenuFAR,][]{nenufar_zarka} and the Square Kilometre Array \citep[SKA,][]{Koopmans} will aim to measure the power spectrum over a wide range of redshifts including cosmic dawn. Thus, we expect a great deal of data from observations in the upcoming decade.

The question arises as to what are the possible ways to infer the astrophysical parameters from the observed 21-cm power spectrum data. Since the characteristic astrophysical parameters at high redshifts are currently almost entirely unconstrained, the 21-cm signal must be calculated for a large number of parameter sets that cover a wide range of possibilities. Given the complexity of the 21-cm signal \citep[see][]{barkana18book, Mesingerbook} and its highly non-linear dependence on the astrophysical parameters, Artificial Neural Networks (ANNs) are a useful method for emulation and fitting. \citet{Shimabukuro} used an ANN to estimate the astrophysical parameters from 21-cm observations. They trained the ANN using 70 data sets where each set consists of the 21-cm power spectrum obtained using \textsc{21cmfast} \citep{mesinger11} as input, with three EoR parameters used in the simulation as output. They applied the trained ANN to 54 data sets to evaluate how the algorithm performs. \citet{Kern} used a machine learning algorithm to emulate the 21-cm power spectrum and perform Bayesian analysis for parameter constraints over eleven parameters which included six parameters of the EoR and X-ray heating and five additional cosmological parameters. \citet{Schmit} built an emulator using a neural network to emulate the 21-cm power spectrum where they generated the training and test data sets using the \textsc{21cmfast} simulation and compared their results with 21CMMC. \citet{cohen19} introduced the first all sky averaged (global) 21-cm signal emulator using an ANN. Recently, \citet{21cmvae, globalemu} proposed two different approaches for emulating the global 21-cm signal. The astrophysical parameters and reionization history can also be recovered from 21-cm images directly using convolutional neural networks (CNN) \citep{Gillet2019, Plante2019}. In this paper, we use an emulation method to constrain the 21-cm power spectrum for the seven-parameter astrophysical model. Using ANN, we construct an emulator that is trained on a large dataset of models that cover a very wide range of the astrophysical parameter space. Given the seven-parameter astrophysical model, the emulator is able to predict the 21-cm power spectrum over a wide redshift range ($z = 6$ to $30$). We construct our algorithm in a way that approximately accounts for emulation error (i.e., the uncertainty due to the finite size of the training set), and also test the accuracy (and improve the error estimates) using 5-fold cross validation. We also explore a more realistic case of the observational measurements expected for the SKA, as well as extended models that also include an excess early radio background. 
We note that any supervised machine learning algorithm is specific to the particular semi-numerical model that is used to generate the training data. Therefore, the results obtained from such algorithms may not be directly applicable to simulations conducted using other semi-numerical methods (such as \textsc{21cmfast}) or to the real Universe; in further work we plan to test with data generated with other methods.

This paper is organised as follows: We present in section \ref{sec: signal} a description of the theory and methods used to generate the datasets (2.1 -- 2.4) and build the ANN and maximum likelihood estimator (2.5 -- 2.6). Section \ref{sec: results} presents our results, for standard astrophysical models (3.1 -- 3.4) and ones with an early radio background (3.5 -- 3.6). Finally, we summarize our results and discuss our conclusions in section \ref{sec: conclusions}.


\section{Theory and methods}\label{sec: signal}
\subsection{21-cm signals}
\subsubsection{Astrophysical parameters} 
We use seven key parameters to parameterize the high redshift astrophysics: the star formation efficiency ($f_{\star}$), the minimum circular velocity of star-forming halos ($V_C$), the X ray radiation efficiency ($f_X$), the power law slope ($\alpha$) and the low energy cutoff ($E_{\rm{min}}$) of the X ray spectral energy distribution (SED), the optical depth ($\tau$) of the cosmic microwave background (CMB) and the mean free path ($R_{\rm{mfp}}$) of ionizing photons.
Here we briefly discuss these astrophysical parameters.

\ \ $\bullet$ \textbf{The star formation efficiency}, $f_{\star}$, quantifies the fractional amount of gas in star-forming dark matter halos that is converted into stars \citep{Tegmark1997}. The value of $f_{\star}$ depends on the details of star formation that are unknown at high redshift, so we treat it as a free parameter. We assume a constant star formation efficiency in halos heavier than the atomic cooling mass and a logarithmic cutoff in the efficiency in lower mass halos \citep{fialkov2013}. We cover a wide range of $f_{\star}$ values, from 0.0001 to 0.5.

\ \ $\bullet$ \textbf{The circular velocity}, $V_C$, is another parameter that encodes the information about star formation. Star formation takes place in dark matter halos that are massive enough to radiatively cool the in-falling gas \citep{Tegmark1997}. This is the main element in setting the minimum mass of star-forming halos, $M_{\rm{min}}$. We equivalently use the minimum circular velocity as one of our free parameters. Since the cooling and the internal feedback depend on the depth of the potential and the potential is directly related to $V_C$, it is more physical to use a fixed $V_C$ versus redshift rather than a fixed $M_{\rm{min}}$. Since complex feedback \citep[e.g.,][]{Schauer} of various types can suppress star formation in low-mass halos, we treat $V_C$ as a free parameter. In practice the actual threshold is not spatially homogeneous in our simulation since individual pixels are affected by feedback processes including Lyman-Werner feedback on small halos, photoheating feedback during the EoR and the streaming velocity between dark matter and baryons. The relation between the circular velocity ($V_C$) and the minimum mass of the dark matter halo ($M_{\rm{min}}$) is given (in the Einstein de-Sitter limit which is valid at high redshift) by
\begin{equation}
V_C = 16.9\left(\frac{M_{\rm{min}}}{10^8M_{\odot}}\right)^{1/3}\left(\frac{1+z}{10}\right)^{1/2}\left(\frac{\Omega_m}{0.0316}\right)^{1/6}\ \rm{km\ s^{-1}} \ .
\end{equation}

\ \ $\bullet$ \textbf{The X ray radiation efficiency}, $f_X$, is defined by the standard expression of the ratio of the X-ray luminosity to the star formation rate ($\rm{L_X}-$SFR relation) [see \citet{fialkov14a},  \citet{Cohen2017} for more details]
\begin{equation}
\frac{\rm{L_X}}{\rm{SFR}} = 3\times10^{40}f_X \ \ \rm{erg\ s^{-1}\ M_\odot^{-1}\ yr}\ \ .
\end{equation}
In the above expression $\rm{L_X}$ is the bolometric luminosity and $f_X$ is the X-ray efficiency of the source. The normalization is such that $f_X=1$ corresponds to the typical observed value for low-metallicity galaxies. Given the almost total absence of observational constraints at the relevant redshifts, we vary $f_X$  from 0.0001 to 1000.

\ \ $\bullet$ \textbf{The power law slope $\alpha$ and the low energy cutoff $E_{\rm{min}}$} determine the shape of the spectral energy distribution (SED). We parameterize the X-ray SED by the power law slope $\alpha$ (where $d$log($E_X$)/$d$log($\nu$) = $-\alpha$) and the low energy cutoff ($E_{\rm{min}}$). These two parameters have significant degeneracy, so we vary $\alpha$ in the narrow range $1-1.5$ and $E_{\rm{min}}$ in the broad range of $0.1-3.0$ keV. The SEDs of the early X-ray sources strongly affect the 21-cm signal from both the EoR and cosmic dawn \citep[][]{fialkov14a, fialkov14b}. Soft X-ray sources (emitting mostly below 1 keV) produce strong fluctuations on relatively small scales (up to a few tens of Mpc) whereas hard X-ray sources produce milder fluctuations on larger scales. X-Ray binaries (XRB) \citep{Mirabel2011,Fragos2013} are major sources that are expected to have a hard X-ray spectral energy distribution.

\ \ $\bullet$ \textbf{The optical depth of the CMB}, $\tau$, is one of two parameters that describe the epoch of reionization. For given values of the other astrophysical parameters, the CMB optical depth has a one to one relation with the ionizing efficiency $\zeta$ which is defined by
\begin{equation}
\zeta = f_{\star}f_{\rm{esc}}N_{\rm{ion}}\frac{1}{1 + \bar{n}_{\rm{rec}}}\ ,
\end{equation}
where $f_{\star}$ is the star formation efficiency, $f_{\rm{esc}}$ is the fraction of ionizing photons that escape from their host galaxy, $N_{\rm{ion}}$ is the number of ionizing photons produced per stellar baryon in star-forming halos, and $\bar{n}_{\rm{rec}}$ is the mean number of recombinations per ionized hydrogen atom. We choose to include the CMB optical depth ($\tau$) in our seven-parameter astrophysical model instead of the ionizing efficiency ($\zeta$) because $\tau$ is directly constrained by CMB observations \citep{planckcollaboration18}.

\ \ $\bullet$ \textbf{The mean free path of ionizing photons}, $R_{\rm{mfp}}$, is the other EoR parameter \citep{Alvarez2012}. $R_{\rm{mfp}}$ sets the maximum distance travelled by ionizing photons. Due to the process of structure formation, dense regions of neutral hydrogen (Lyman-limit systems) effectively absorb all the ionizing radiation and thus limit the sphere of influence of each ionizing source. The mean free path parameter approximately accounts for the effect of these dense neutral hydrogen pockets during reionization. In our simulations, we vary $R_{\rm{mfp}}$ from 10 to 70 comoving Mpc \citep{Wyithe_loeb2004,Songaila2010}.

\subsubsection{Power spectrum}
It is possible in principle to map the distribution of neutral hydrogen three dimensionally in the early universe by observing the brightness temperature contrast of the 21-cm line. In order to infer the information about the astrophysical processes in the epoch of reionization and cosmic dawn, there are a variety of approaches one can follow to characterize the 21-cm signal. Other than the global signal, the most straightforward approach is to use the statistical description of the 21-cm fluctuations, i.e., the 21-cm power spectrum.

The 21-cm power spectrum encodes a great deal of information about the underlying physical processes related to reionization and cosmic dawn. We define the power spectrum $P(k)$ of fluctuation of the 21-cm brightness temperature (relative to the radio background, which is the CMB in standard models) by 
\begin{equation}
\langle \tilde{\delta}_{T_b}(\mathbf{k})\tilde{\delta}^*_{T_b}(\mathbf{k^{\prime}}) \rangle = (2\pi)^3\delta_D(\mathbf{k}-\mathbf{k^{\prime}})P(k)\ ,
\end{equation}
where $\mathbf{k}$ is the comoving wave vector, $\delta _D$ is the Dirac delta function, and the angular brackets denote the ensemble average. $\tilde{\delta}_{T_b} (\mathbf{k})$ is the Fourier transform of $\delta_{T_b} (\mathbf{x})$ which is defined by $\delta_{T_b} (\mathbf{x}) = (\delta T_b(\mathbf{x}) - \bar{\delta T_b})/\bar{\delta T_b}$. Finally we express the power spectrum in terms of the variance, in mK$^2$ units:
\begin{equation}
\Delta^2 = \langle \delta T_b \rangle ^2 \frac{k^3P(k)}{2\pi^2}\ ,
\end{equation}
where the expression $k^3P(k)/2\pi^2$ is dimensionless. The 21-cm signal is significantly non-Gaussian because of both large-scale and small-scale processes during reionization and cosmic dawn. Thus, the power spectrum does not reveal all the statistical information that is available. Nevertheless, a wealth of astrophysical information can be extracted from the 21-cm power spectrum and it can be measured relatively easily from observations.

\subsection{The Excess radio background}\label{sec: excess_radio}
The first observational signature of the HI 21-cm line from cosmic dawn was tentatively detected by the EDGES collaboration \citep{bowman18}. The shape and magnitude of this signal are not consistent with the standard astrophysical expectation. The reported 21-cm signal is centered at $\nu = 78.2$ MHz with an absorption trough of $\delta T_b = -500^{+200}_{-500}$ mK \citep{bowman18}. The amplitude of absorption is more than a factor of two larger than that predicted from standard astrophysics based on the $\Lambda$CDM cosmology and hierarchical structure formation. The SARAS 3 experiment recently reported the upper limit of the global signal that is inconsistent with the EDGES signal \citep{SARAS3} at 95\%, so it will be some time before we can be confident that the global 21-cm signal has been reliably measured. 

If EDGES is confirmed, one possible explanation of this observed signal is that there is an additional cooling mechanism that makes the neutral hydrogen gas colder than expected; a novel dark matter interaction with the cosmic gas \citep{barkana18} is a viable option, but it likely requires a somewhat elaborate dark matter model \citep{Berlin, barkana18a, munoz18, Liu19}.
Another possibility, which we consider in detail in this paper, is an excess radio background above the CMB \citep{bowman18, feng18, ewall18, fialkov19, mirocha19, ewall20, Reis2020}. This excess radio background increases the contrast between the spin temperature and the background radiation temperature. In this case the basic equation for the observed 21-cm brightness temperature from redshift $z$ relative to the background is
\begin{equation}
\delta T_b = \frac{T_{\rm{S}} - T_{\rm{rad}}}{1 + z}(1 - e^{-\tau_{\nu}})\ ,
\end{equation}
where $T_{\rm{rad}} = T_{\rm{CMB}} + T_{\rm{radio}}$, with $T_{\rm{radio}}$ being the brightness temperature of the excess radio background and $T_{\rm{CMB}} =  2.725(1 + z)$ K. We consider two distinct types of extra radio models, which we have considered in previous publications. The external radio model assumes a homogeneous background that is not directly related to astrophysical sources, i.e., may be generated by exotic processes (such as dark matter decay) in the early universe. In this model, we assume that the brightness temperature of the excess radio background at the 21-cm rest frame frequency at redshift $z$ is given by \citep{fialkov19}
\begin{equation}
T_{\rm{radio}} = A_r \times 2.725(1 + z)\left[ \frac{1420}{78(1 + z)}\right]^{\beta} \ \hbox{K}\ \ ,
\end{equation}
where the spectral index $\beta = -2.6$ (set to match the slope of the observed extragalactic radio background observed by ARCADE2 \citep{fixsen11, seiffert11} and confirmed by LWA1 \citep{dowell18}) and $A_r$ is the amplitude of the radio background. Here 1420~MHz/$(1+z)$ is the observed frequency corresponding to 
redshift $z$, and $A_r$ measures the amplitude (relative to the CMB) at the central frequency of the EDGES feature (78~MHz). Thus, the external radio model has eight free parameters: $f_{\star}$, $V_C$, $f_X$, $\alpha$, $E_{\rm{min}}$, $\tau$, $R_{\rm{mfp}}$ and $A_r$.

In contrast to this external radio background, astrophysical sources such as supermassive black holes or supernovae could in principle produce such an extra radio background due to synchrotron radiation. In such a case, the radio emission would originate from within high redshift radio galaxies and would thus result in a spatially varying radio background, as computed accurately on large scales within our semi-numerical simulations \citep{Reis2020}. The galaxy radio luminosity can be written as 
\begin{equation}
L_{\rm{radio}}(\nu, z) = f_{R} \times 10^{22} \left(\frac{\nu}{150~\hbox{MHz}} \right)^{-\alpha_{\rm{radio}}} \left(\frac{\rm{SFR}}{\rm{M_\odot yr^{-1}}}\right) \ \ \ \ \rm{W\,Hz}^{-1}\ ,
\end{equation}
where $\alpha_{\rm{radio}}$ is the spectral index in the radio band, $\rm{SFR}$ is the star formation rate and $f_{R}$ is the normalization of the radio emissivity.
Based on observations of low-redshift galaxies, we set $\alpha_{\rm{radio}} = 0.7$
and note that $f_R=1$ roughly corresponds to the expected value \citep{gurkan18,mirocha19}. Since extrapolating low-redshift observations to cosmic dawn may be wildly inaccurate, in our analysis we allow $f_{R}$ to vary over a wide range. Thus, the galactic radio model is also based on eight parameters: $f_{\star}$, $V_C$, $f_X$, $\alpha$, $E_{\rm{min}}$, $\tau$, $R_{\rm{mfp}}$, and $f_R$. 

Both types of radio background, if they exist, can affect the 21-cm power spectrum, leading to a strong amplification of the 21-cm signal during cosmic dawn and the EoR in models in which the radio background is significantly brighter than the CMB. However, there are some major differences between the two models. The external radio background is spatially uniform,
is present at early cosmic times (prior to the formation of the first stars),
and increases with redshift (i.e., it is very strong at cosmic dawn and
weakens during the EoR). On the other hand, the galactic radio background is non-uniform, and its intensity generally rises with time as it follows the formation of galaxies (as long as $f_{R}$ is assumed to be constant with redshift). 

\subsection{Mock SKA data}\label{sec:mock_ska_data}

To consider a more realistic case study, we create mock SKA data by including several expected observational effects in the 21-cm power spectrum, which we refer to as the case "mock SKA data". To incorporate the SKA noise case within the data, (i) we include the effect of the SKA angular resolution, (ii) we add a pure Gaussian noise smoothed over the SKA resolution as a realization of the SKA thermal noise. The strength of the SKA thermal noise (for a frequency depth corresponding to 3 comoving Mpc and assuming a 1000 hour integration) is approximated by 
\begin{equation}
\centering
\sigma_{\rm{thermal}} = a\left( \frac{1+z}{17}\right)^b \ ,
\end{equation}
where $a$ and $b$ depend on the resolution used. Here we use a smoothing radius, $R_{\rm{SKA}} = 20$ Mpc, for which $a = 4$ mK and $b = 5.1$ for $z>16$, while for lower redshifts, $b=2.7$ (following \citet{Banet}, see also \citet{Koopmans}) and (iii) we remove (i.e., set to zero) modes from part of the $k$-space (the "foreground wedge") that is dominated by foregrounds (following \citet{discrete}, see also \citet{datta10, Morales2012,  dillon14, pober14, liu2014a, liu2014b, pober15, jensen15}). Each of the three effects is included along with its expected redshift dependence. Regarding the foreground avoidance/filtering, we note that we assume that the high-resolution maps of the SKA will enable a first step of reasonably accurate foreground subtraction, so that the remaining wedge-like region that must be set to zero will be limited (corresponding to the "optimistic model" of \citet{pober14}). In order to gain some understanding of the separate SKA effects, we also consider a case that we label "SKA thermal noise". In this case, we add the effect of SKA resolution and thermal noise, i.e., the same as "mock SKA data" except without foreground avoidance. 

Given the lower sensitivity, for cases with mock SKA effects we use coarser binning, as we do not expect the results to depend on the detailed power spectrum shape that would come out in finer binning. Specifically, we used eight redshift bins and five $k$ bins. The five $k$-bins are spaced evenly in log scale between $k$ = 0.05 Mpc$^{-1}$ and $k$ = 1.0 Mpc$^{-1}$; we average the 21-cm power spectrum at each redshift over the range of $k$ values within each bin. To fix the redshift bins, we imagine placing our simulation box multiple times along the line of sight, so that our comoving box size fixes the redshift range of each bin. For example, we start with $z = 27.4$, which corresponds to $50$ MHz (the limit of the SKA), as the far side of the highest-redshift bin. Then the center of the box is 192 comoving Mpc (half of our 384 Mpc box length) closer to us. The redshift corresponding to the center is taken as the central redshift of the first bin. The next redshift bin is $384$ Mpc closer and so on. As the total comoving distance between $z = 27.4$ and $z = 6$ is around $3000$ Mpc, we obtain $8$ redshift bins that naturally correspond to a line of sight filled with simulation boxes. We then average the 21-cm power spectrum over the redshift range spanned by each box along the line of sight, by using the simulation outputs which we have at finer resolution in redshift. This averaging is part of the effect of observing a light cone; while there is also an associated anisotropy \citep{lightcone,Datta12}, in this paper we only consider the spherically-averaged 21-cm power spectrum. 

\subsection{Method to generate the dataset}\label{sec: simulation}

We use our own semi numerical simulation code \citep{visbal12, fialkov14b}, which we have named 21cmSPACE (21-cm Semi-numerical Predictions Across Cosmological Epochs), to predict the 21-cm signal for each possible model. The simulation generates realizations of the universe in a large cosmological volume ($384^3$ comoving Mpc$^3$) with a resolution of 3 comoving Mpc over a wide range of redshifts (6 to 50). The simulation follows the hierarchical structure formation and the evolution of the Ly$\alpha$, X-ray, Lyman-Werner (LW), and ionizing ultra-violet radiation. The extended Press-Schechter formalism is used to compute the star formation rate in each cell at each redshift \citep{barkana04}. The 21-cm brightness temperature cubes are output by the simulation and we use them to calculate the 21-cm power spectrum at each redshift. While this semi-numerical simulation was inspired by \textsc{21cmFAST} \citep{mesinger11}, it is an entirely independent implementation with various differences such as more accurate X-ray heating (including the effect of local reionization on the X-ray absorption) and Ly$\alpha$ fluctuations (including the effect of multiple scattering and Ly$\alpha$ heating). Inhomogeneous processes such as the streaming velocity, LW feedback, and photo-heating feedback are also included in the code. We created a mock 21-cm signal using the code for a large number of astrophysical models and calculated the 21-cm power spectrum for each parameter combination. Considering first standard astrophysical models (without an excess radio background), we generated the 21-cm power spectrum for 3195 models that cover a wide range of possible values of the seven astrophysical parameters. The ranges of the parameters were $f_{\star} = 0.0001 - 0.50$, $V_C = 4.2 - 100$ km s$^{-1}$, $f_X = 0.0001 - 1000$, $\alpha = 1.0 - 1.5$, $E_{\rm{min}} = 0.1 - 3.0 $ keV, $\tau = 0.01 - 0.12$, and $R_{\rm{mfp}} = 10.0 - 70.0$ Mpc. The sampling of different parameters is done using randomly-selected values over these wide ranges in the seven-parameter space.

As explained above, our analysis involved two more datasets (3195 models each) of 21-cm power spectra, with either full SKA noise or SKA thermal noise only, in order to analyze a more realistic situation. In order to investigate the two scenarios of the excess radio background (where the number of free parameters is increased by one), we use two new datasets of models: 10158 models with the galactic radio background and 5077 models with the external radio background. Throughout this work, we adopt the $\Lambda$CDM cosmology with cosmological parameters from \citet{planck2014}.

\subsection{\textbf{Artificial Neural Network}}\label{sec: ann}

ANN (often simply called neural networks or NN) are computing systems that mimic in some ways the biological neural networks that constitute the human brain. We briefly summarize their properties. An ANN consists of a collection of artificial neurons. Each artificial neuron has inputs and produces a single output which can be the input of multiple other neurons. In our analysis, we utilize a Multi-layer Perceptron (MLP) which is a deep neural network consisting of fully connected layers. To define the architecture of the neural network, several parameters must be specified, including the number of hidden layers, the number of nodes in each layer, the activation function, the solver, and the maximum number of iterations.

A Multi-layer Perceptron \citep[MLP,][]{Ramchoun2016MultilayerPA} is a supervised learning algorithm that learns to fit a mapping  $f : X^m \rightarrow Y^n$ using a training dataset, where $m$ is the input dimension and $n$ is the output dimension. When we apply unknown data as a set of input features $X = x_1, x_2, x_3, ..., x_m$, the neural network uses the mapping to infer the target output ($Y$). This Multi-layer Perceptron can be used for both classification and regression problems. The advantage of a Multi-layer Perceptron is that it can learn highly non-linear models. Every neural network has three different types of layers each consisting of a set of nodes or neurons. They are the input layer, hidden layer and output layer. The input layer consists of a set of neurons that represent the input features
$X = x_1, x_2, x_3, ..., x_m$. Each neuron in the input layer is connected to all the neurons in the first hidden layer with some weights and each node in the first hidden layer is connected to all the nodes in the next hidden layer and so on. The output layer receives the values from the last hidden layer and transforms them to the output target value. A specific weight ($w_{ij}$) and a bias ($b_j$) are applied to every input or feature. Both the weight and the bias are initially chosen randomly. For a particular neuron in the $i$'th hidden layer, if $m_i$ is the input and $n_j$ is the output of that neuron, then $n_j = f(\sum_i w_{ij} m_i + b_j)$, where $f$ is called the activation function. Using linear activation functions would make the entire network linear in the inputs, and thus equivalent to a one layered network. Thus, non-linear activation functions such as a rectified linear unit \citep[ReLu,][]{relu}, sigmoid function \citep{Han1995TheIO}, etc., are typically used in order to provide the ability to handle complex, non-linear data. The precise nature of the non-linearity depends on the activation function used. We use the sigmoid activation function, $f(x) = 1/(1 + e^{-x})$. When the absolute value of the input is large, this function saturates and returns a constant output. The output obtained from the activated neuron is then utilized as the input for the next hidden layer. The primary objective of training an ANN is to determine a set of weights and biases that enable the ANN to generate output vectors that are consistent with the desired output, for a given set of input vectors. A backpropagation algorithm \citep{back_propagation} is usually used to train an artificial neural network. The training procedure for a network involves several steps:
\begin{itemize}
\item Initialization: Randomly chosen initial weights and biases are applied to all the nodes or neurons in each layer.
\item Forward propagation: The output is computed using the neural network based on the initial choices of the weights and biases given the input from the training dataset. Since the calculation progresses from the input to the output layer (through the hidden layers), this is known as forward propagation.
\item Error estimation: An error function (often called a loss function) is used to compute the difference between the predicted and the true (known) output of the model, given the current weights. MLP uses different loss functions based on the problem type. For regression, a common choice is the mean square error. 
\item Backpropagation and updating of the weights : A backpropagation algorithm minimizes the error function and finds the optimal weight values, typically by using the gradient descent technique \citep{adam_optimizer}. The outermost weights get updated first and then the updates propagates towards the input layer, hence the term backpropagation. 
\item Repetition until convergence: In each iteration, the weights get updated by a small amount, so to train a neural network several iterations are required. The number of iterations until convergence depends on the learning rate and the optimization method used in the network.
\end{itemize}
Once the network has been trained using the training dataset, the trained network can make predictions for arbitrary input data that were not a part of the training set. We note that in the various mappings presented below, when we train emulators using simulated training sets, we do this separately for each of our cases of ideal data, mock SKA data, and SKA thermal noise only. In each case the power spectra within the training set are generated from 21-cm images that include the observational effects that correspond to the appropriate case.
By applying the observational effects directly to the images, we are working in the spirit of a forward pipeline simulation, which is designed to be as realistic as possible.

\subsubsection{Astrophysical parameter predictions}
For the purpose of predicting the astrophysical parameters, we adopted a four-layer MLP (input layer + two hidden layers + output layer) from the \texttt{Scikit-learn} library \citep{scikit_learn}. Specifically, there are 150 neurons in the first hidden layer and 50 neurons in the second hidden layer. The network was expected to be somewhat complex as we want a mapping between the seven astrophysical parameters of the model and, on the other side, the 21-cm power spectrum for 32 values of the wavenumber in the range 0.05 Mpc$^{-1}$ $< k < $1.0 Mpc$^{-1}$ and for redshifts ranging from 6 to 30. For the purpose of predicting the parameters, the 21-cm power spectra are the input and the astrophysical parameters are the output.

Data pre-processing is an important step before applying machine learning. We have two pre-processing steps in the code: 1) Standardization of the datasets, and 2) Dimensionality reduction of the datasets. If individual features have large differences in their ranges, those with large ranges will dominate over those with small ranges, which leads to biased results. Thus it is important to transform all the features to comparable scales, i.e., perform standardization of the data. Mathematically, to do the standardization, we subtract the mean from each value of each feature and then divide it by the standard deviation. After standardization, all the features are centered around zero and have identical variances. For optimal performance of the learning algorithm, it is best that the individual features are as close as possible to standard normally distributed data (a Gaussian distribution with zero mean and unit variance). For standardization of the features, we use the \textit{StandardScaler} class from the \textit{preprocessing} module in the \texttt{Scikit-learn} library \citep{scikit_learn}. For dimensionality reduction, we use \textit{Principal Component Analysis} \citep[PCA, ][]{pca} to project our data into a lower dimensional space. In the case of predicting the seven parameter model, the input dimension for a particular model is $25 \times 32$, which is quite high and makes the learning algorithm too slow. Thus we use PCA to speed up the learning algorithm. Though the reduction in dimensionality results in loss of information, the datasets with the reduced dimension are sufficiently good to encode the most important relationships between the data points. Using PCA with 200 components, we are able to reduce the input dimension by $75\%$, and capture $> 99\%$ of the data variance; for the mock SKA and SKA thermal noise cases, the number of components is lower so we use all of them without dimensionality reduction. We transform each dataset (both the parameters and the 21-cm power spectra) in log scale ($\log_{10}$) before applying it to the MLP regressor.  In our MLP regressor, we set the maximum number of iterations to 10000. We choose the logistic sigmoid function as the activation function for the hidden layers, an adaptive learning rate initialized with 0.001 and the stochastic gradient-based optimizer \citep{adam_optimizer} for the weight optimization. No early stopping criteria has been used here. We use 2557 models (which is $80\%$ of our full dataset) to train the neural network, and we then apply the trained ANN to a test dataset consisting of 639 models ($20\%$ of our full dataset). Throughout this paper,
for simplicity we choose test cases that have non-zero power spectra from intergalactic hydrogen, i.e., that have not fully reionized by redshift 6.

The mapping presented in this subsection, of the power spectrum to the parameters, is included for completeness, and since it is useful whenever a quick but still accurate result of the parameter fitting is needed. However, in the rest of this paper we focus more on Markov Chain Monte Carlo (MCMC) fitting \citep{Hogg}, since that allows us to also generate estimates of errors as well as multi-parameter error contours; in that application, we make only limited use of the mapping presented in this subsection, as a method to obtain a good first guess of the seven astrophysical parameters given a 21-cm power spectrum.

\subsubsection{Emulation of the 21-cm power spectrum}

If the statistical description of the 21-cm signal (here the 21-cm power spectrum) is our main focus, then we hope to avoid the need to run a semi-numerical simulation for each parameter combination. We can instead construct an emulator that provides rapidly-computed output statistics that capture the important information in the signal given a set of astrophysical parameters.

We train a neural network to predict the 21-cm power spectrum based on the seven parameter astrophysical model specified above. This trained network/emulator is used primarily for inference using MCMC (hereafter, the word 'emulator' refers to the network that predicts the power spectrum given the seven parameter astrophysical model). As in the case of the ANN to predict the parameters, here also we standardize the features as part of data pre-processing. To reduce the dimension of the power spectrum data, we apply PCA transformation to the data; after experimentation we found that here 20 PCA components suffice to capture $> 98\%$ of the data variance. As before, we again use a log scale for both the dataset of the parameters and the 21-cm power spectra. Next we need to find the appropriate neural network architecture to construct the emulator. For this, we choose some specified hyperparameters for our multi layer perceptron estimator and search among all possible combinations to find the best one to use in our MLP regressor. To emulate the 21-cm power spectrum, we employ a three layer MLP from the \texttt{Scikit-learn} library \citep{scikit_learn} with 134 neurons in each layer. Thus, the full network architecture consists of five layers in total: input layer, three hidden layers and output layer. We use the logistic sigmoid function \citep{Han1995TheIO} 
as the activation function for the hidden layer and limited-memory  Broyden–Fletcher–Goldfarb–Shanno (BFGS) optimizer \citep{lbfgs_optimizer} for the weight optimization. As before, we fix the maximum number of iterations to 10000. After emulating the 21-cm power spectrum, we need to inverse PCA transform the predicted data to get back the power spectrum in original dimensions.

\subsection{Posterior distribution of the astrophysical model}\label{sec: bayesian_analysis_0}

Given the 21-cm power spectrum, we can predict the seven astrophysical parameters that describe the epoch of reionization and the cosmic dawn using the NN trained on the dataset consisting of the seven astrophysical parameters and the 21-cm power spectrum $P(k)$ over the wide range of redshifts of interest. This parameter estimation is very computationally fast. However, estimating uncertainties on the predicted parameters using a traditional neural network, as employed in this study, is not straightforward. One solution is to use a Bayesian neural network that incorporates uncertainties while making predictions on output parameters \citep{bnn_21cm}. Leaving this for future investigation, here we take a hybrid approach. In order to estimate the astrophysical parameters as well as their uncertainties (including complete information on covariances), our approach is as follows. First we use the NN to predict the parameters given the 21-cm power spectrum. Then we employ our power spectrum emulator to an MCMC sampler, using the predicted parameter values as the initial guesses; the emulator allows us to avoid a direct dependence on the semi-numerical simulation, thus making the MCMC process computationally fast. The final output of the sampler is the posterior probability distribution for the parameters. We report the median of each of the marginalized distributions of the parameters as the predicted value and calculate the uncertainty bound based on quantiles. This is discussed in further detail in the results section, below.

In order to estimate the uncertainties in predicting the parameters we follow a Bayesian analysis for finding the posterior probability distribution of the parameters. We use MCMC methods for sampling the probability distribution functions or probability density functions (pdfs). 

The posterior pdf for the parameters $\theta$ given the data $D$, $p(\theta|D)$, is, in general, the likelihood $p(D|\theta)$ (i.e., the pdf for the data $D$ given the parameters $\theta$) times the prior pdf $p(\theta)$ for the parameters, divided by the probability of the data $p(D)$:
\begin{equation}
p(\theta|D) = \frac{p(D|\theta)p(\theta)}{p(D)}\ ,
\end{equation}
where the denominator $p(D)$ can be thought of as a normalization factor that makes the posterior distribution function integrate to unity. If we assume that the noise is independent between data points, then the likelihood function is the product of the conditional probabilities
\begin{equation}
\mathcal{L} = \prod_{n=1}^N p(D_n|\theta)\ .
\end{equation}
Taking the logarithm, 
\begin{equation}
\ln \mathcal{L}  = -\frac{1}{2}\sum_{n=1}^N \left[ \frac{[D_n - D_{\rm{n, model}} (\theta)]^2}{s_n^2} + \ln(2\pi s_n^2)\right]\ ,
\end{equation}
where we set  

\begin{equation}
    s_n^2 = \sigma_n^2 + D_{\rm{n, model}}^2f^2 \ . \label{eq:sigma}
\end{equation}

The likelihood function here is assumed to be a Gaussian, where the variance is modelled as is common for the MCMC procedure, as a sum of a constant plus a multiple of the predicted data (i.e., as a combination of an absolute error and a relative error). Here $f$ is a free parameter that gives the MCMC procedure some flexibility, so we include it effectively as an additional model parameter. We apply the procedure to obtain the posterior distribution for all the parameters (seven astrophysical parameters and $f$) and then marginalize over the extra parameter ($f$) to obtain the properly marginalized posterior distribution for the seven astrophysical parameters \citep{Hogg}. Here the index $n$ denotes various $z$-bins and $k$-bins, where the data $D_n$ is the mock observation of the 21-cm power spectrum and $D_{\rm{n, model}}$ is the predicted 21-cm power spectrum from the emulator. In this work we adopt an effective constant error of:
\begin{equation}
\sigma_n = 0.25\ {\rm mK}^2\ . \label{eq:sn}
\end{equation}
This ensures that the algorithm does not try to achieve a low relative error when the fluctuation itself is low (below $\sim 0.5$~mK) and likely more susceptible to systematic errors in realistic data. What we have described here is a typical setup for MCMC. We emphasize that regardless of our detailed assumptions, in the end we have test models that allow us to independently test the reliability of the uncertainty estimates, as described further in the results section below. 

When we use one of the datasets with observational noise and other SKA effects, we modify equation \ref{eq:sigma} as follows:
\begin{equation}
    s_n^2 = \sigma_n^2 + \sigma_{\rm{var},n}^2 + D_{\rm{n, model}}^2f^2 \ ,
\end{equation}
where  $\sigma_{\rm{var},n}^2$ is the variance (for each bin of $z$ and $k$) of the SKA noise power spectrum, found separately for the mock SKA and SKA thermal noise cases. We found these expected variances by randomly generating observational (i.e., signal-free) data. We did not include co-variances among different bins since we found that the correlation coefficients are much smaller than unity (no more than a few percent), indeed so small that their values did not converge even with tens of thousands of samples.

We use the \texttt{emcee} sampler \citep{Foreman-Mackey} which is the affine-invariant ensemble sampler for MCMC \citep{Goodman}.  The MCMC sampler only computes the likelihood when the parameters are within the prior bounds. We set the prior bounds for the parameters according to Table~\ref{tab:table1} and we use flat priors for the parameter values (in log except for $\alpha$ and $E_{\rm{min}}$).

\begin{table}
\centering
\begin{tabular}{lcc} 
\hline
Parameters~                     & Lower bound & Upper bound  \\ 
\hline\hline
$f_{\star}$                        & 0.0001      & 0.50         \\
$V_C$ [km/s]                       & 4.2         & 100          \\
$f_X$                        & 0.0001      & 1000         \\
$\alpha$       & 0.9         & 1.6          \\
$E_{\rm{min}}$ [keV] & 0.09        & 3.1          \\
$\tau$         & 0.01       & 0.14       \\
$R_{\rm{mfp}}$ [Mpc]                 & 9           & 74 \\
\hline
\end{tabular}
\caption{The prior bounds for the astrophysical parameters.}
\label{tab:table1}
\end{table}


\section{Results}\label{sec: results}

\subsection{Performance analysis of the emulator}\label{sec: perform_emulators}

\begin{figure*}
    \centering
    \begin{minipage}{0.4\textwidth}
      \includegraphics[scale=0.42]{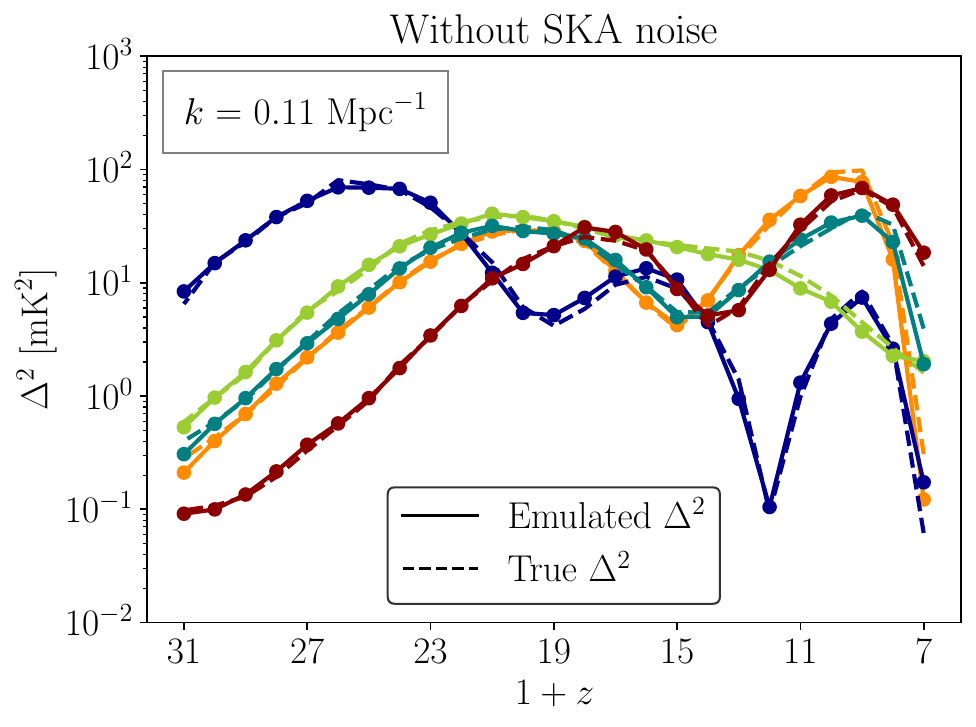}
    \end{minipage}
    \begin{minipage}{0.4\textwidth}
      \includegraphics[scale=0.42]{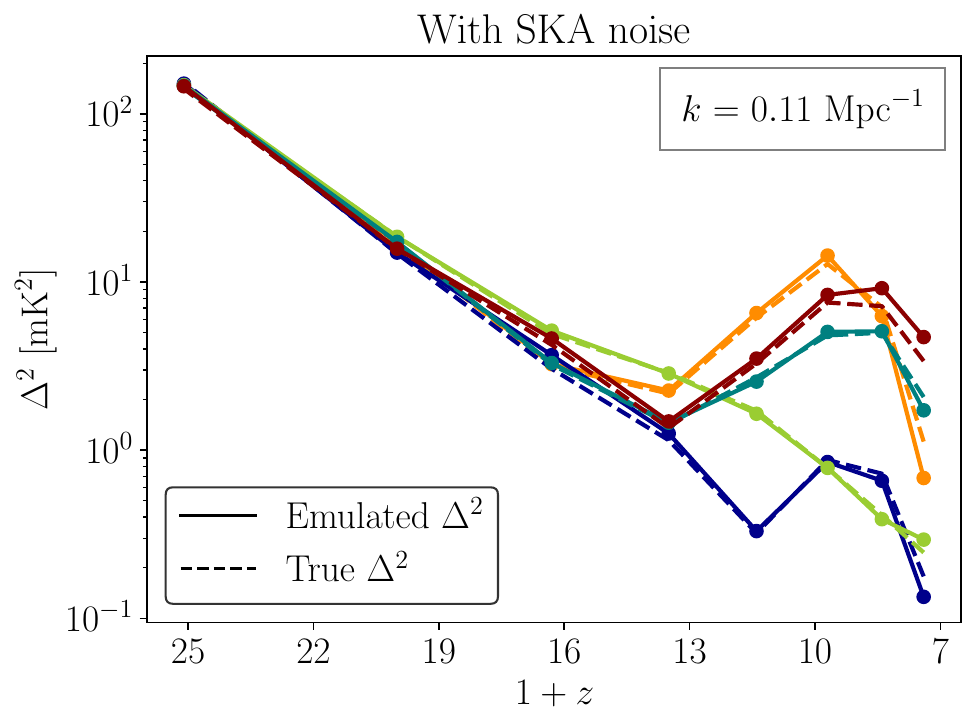}
    \end{minipage}
    \centering
    \begin{minipage}{0.4\textwidth}
       \includegraphics[scale=0.42]{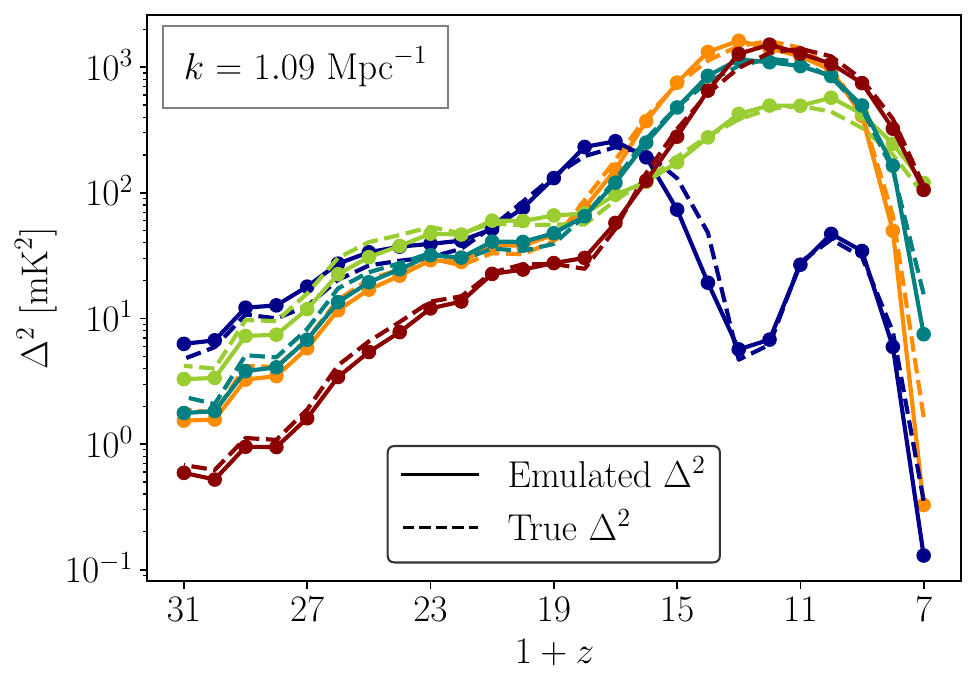}
    \end{minipage}
    \begin{minipage}{0.4\textwidth}
      \includegraphics[scale=0.42]{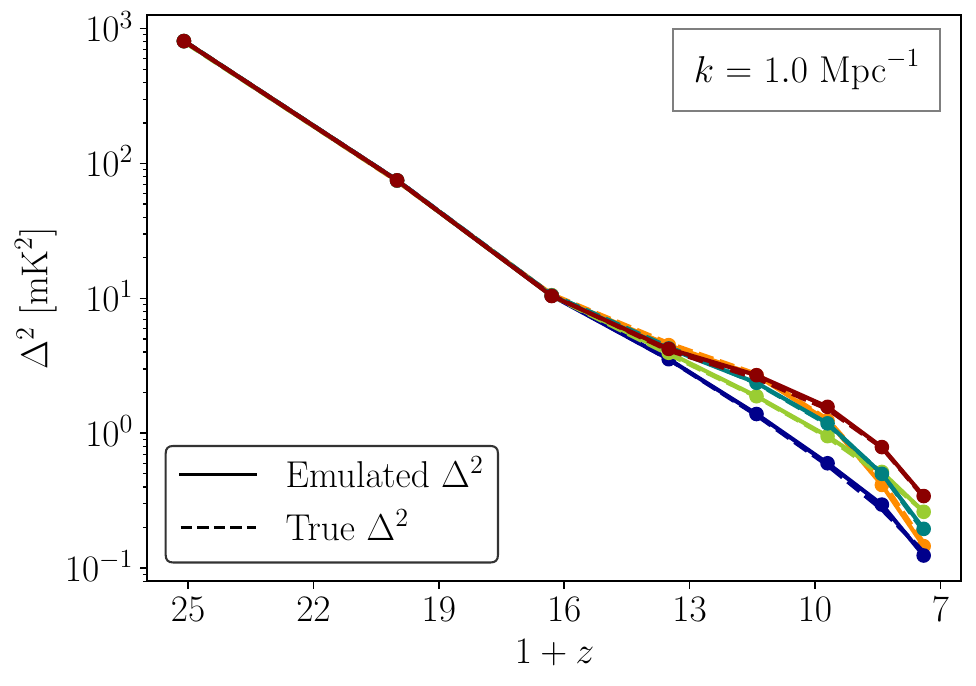}
    \end{minipage}
    \caption{A few random examples of the emulated power spectrum without SKA noise (left panel) and with SKA noise (right panel) at $k$ = 0.11 Mpc$^{-1}$ (upper panel) and $k$ $\approx 1.0$ Mpc$^{-1}$ (lower panel); note that the $k$-bin values and widths are different in the SKA case, as explained in the text. The dashed line is the true power spectrum from the simulation and the solid line is the emulated power spectrum (for combinations of astrophysical parameters that were not included in the training set). Different colors show different models.}\label{fig: example_ps}
\end{figure*}

We show the performance of the emulator of the 21-cm power spectrum in Fig.~\ref{fig: example_ps}. We compare the emulated power spectrum and the true power spectrum from the semi-numerical simulation for two particular $k$ values. The left panel shows a few random examples of the emulated power spectrum (solid lines) and the true power spectrum (dashed lines). The different colors denote different models. In this figure, we see that the accuracy of the emulator is generally good and tends to improve with the height of the power spectrum, although there is some random variation among different models. A more representative, statistical analysis of the accuracy is shown further below.

The right panel of Fig.~\ref{fig: example_ps} shows a few random examples of the comparison between the power spectrum emulated by the emulator trained using mock SKA dataset and the true power spectrum from the mock SKA dataset. The different colors denote different astrophysical models. Again, the emulation is seen to be reasonably accurate,  although in some cases the emulated 21-cm power spectrum significantly deviates from the actual one at low redshift and/or when the power spectrum is low. The variations intrinsic to the different models in the power spectra (left panels in Fig.~\ref{fig: example_ps}) are heavily suppressed once we include the expected observational effects of the SKA experiment into the power spectra (right panels in Fig.~\ref{fig: example_ps}). In particular, the thermal noise dominates at high redshift. However, as we find from the results below, when we fit the power spectrum with SKA noise there is still significant information in the data that allows the fitting procedure to reconstruct the input parameters. An advantage of machine learning is that the algorithm learns directly how to best deal with noisy data, and there is no need to try to explicitly model or fit the observational effects.

\subsection{Dependence of the emulation error on the redshift and wavenumber}

\begin{figure*}
    \centering
    \begin{minipage}{0.264\textwidth}
        \includegraphics[scale=0.31]{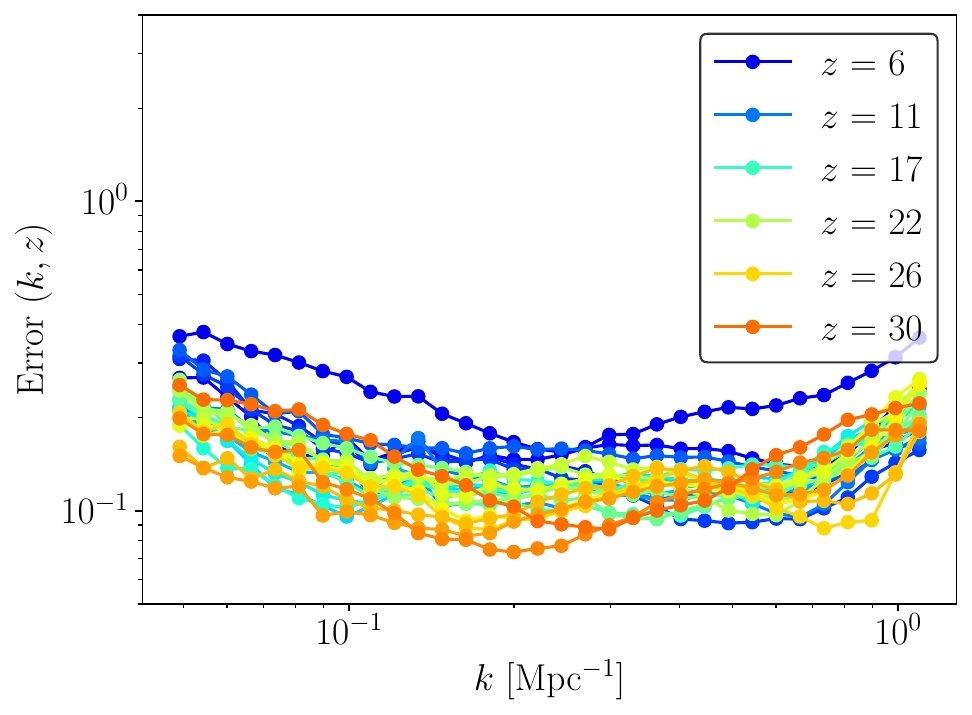}
    \end{minipage}\hfill    
    \begin{minipage}{0.23\textwidth}
       \includegraphics[scale=0.31]{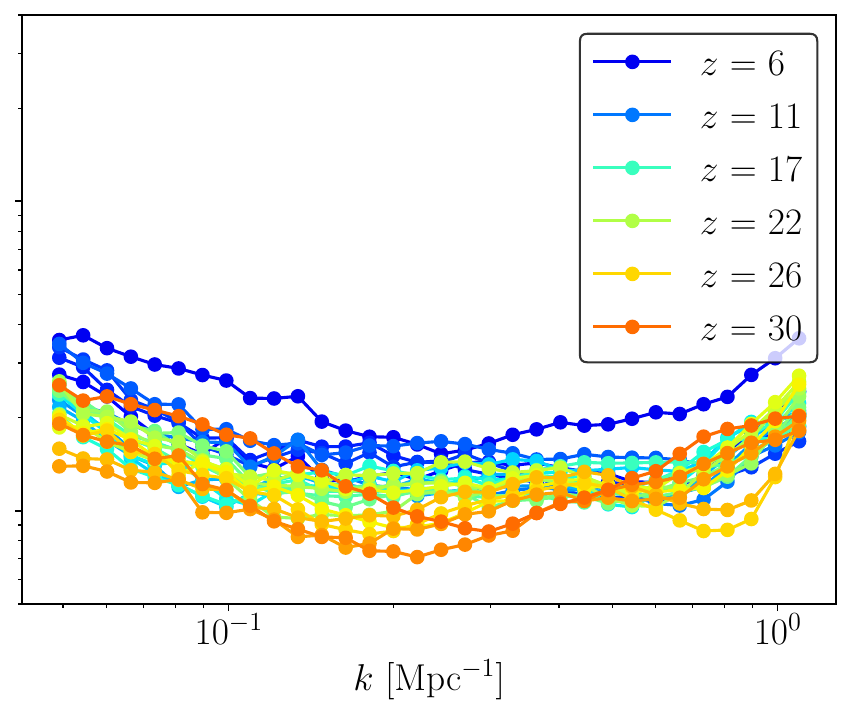}
    \end{minipage}\hfill
    \begin{minipage}{0.23\textwidth}
      \includegraphics[scale=0.31]{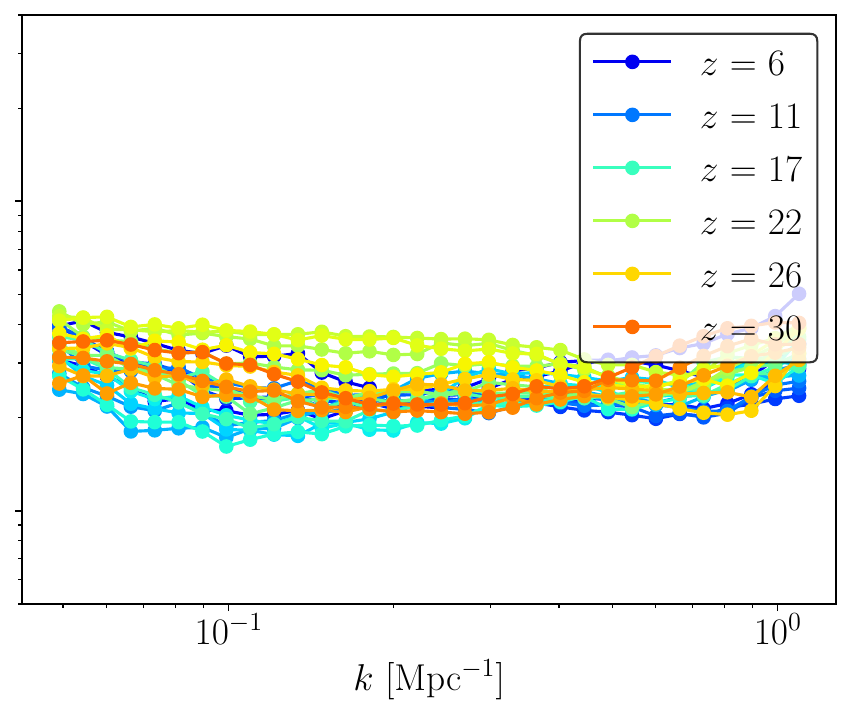}
    \end{minipage}\hfill
    \begin{minipage}{0.23\textwidth}
      \includegraphics[scale=0.31]{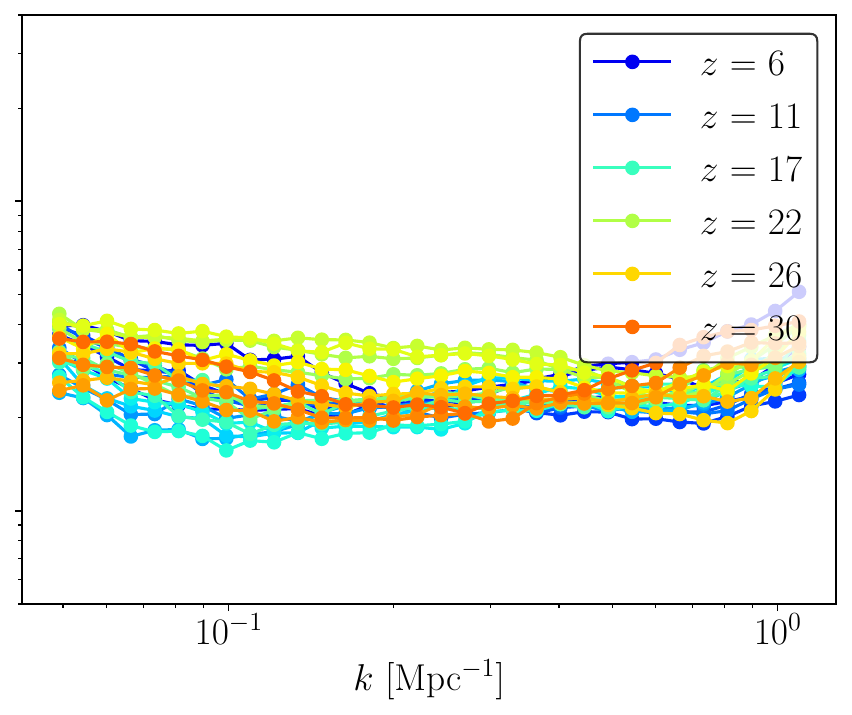}
    \end{minipage}
    \centering
    \begin{minipage}{0.265\textwidth}
       \includegraphics[scale=0.31]{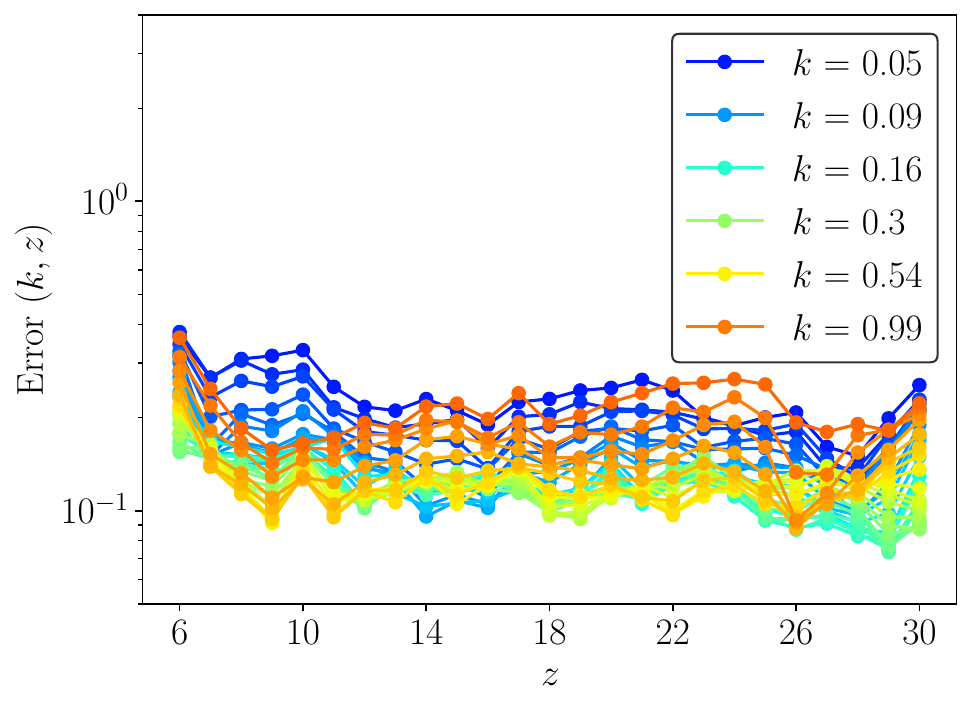}
    \end{minipage}\hfill 
    \begin{minipage}{0.23\textwidth}
       \includegraphics[scale=0.31]{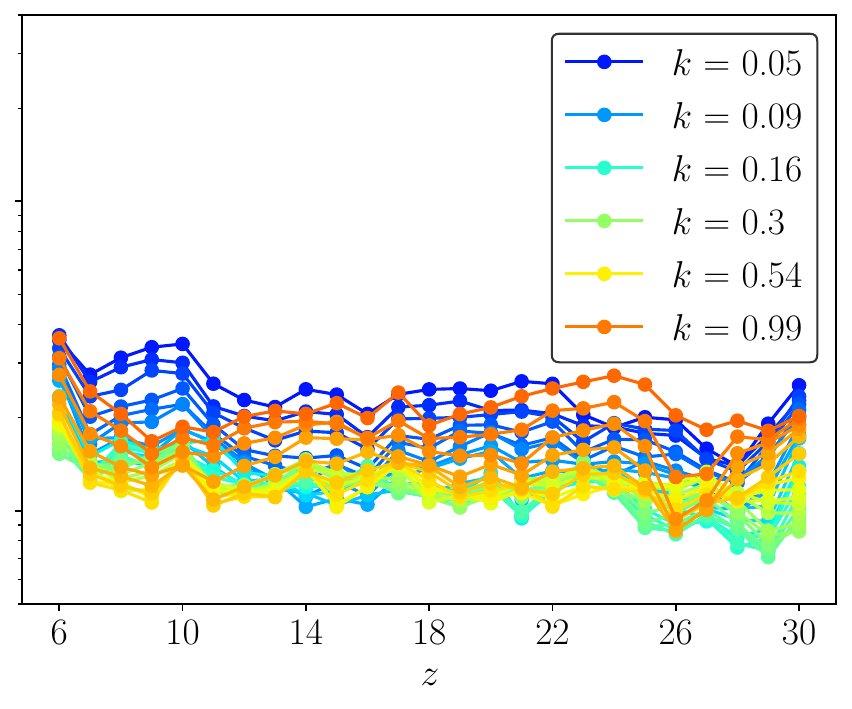}
    \end{minipage}\hfill
    \begin{minipage}{0.23\textwidth}
      \includegraphics[scale=0.31]{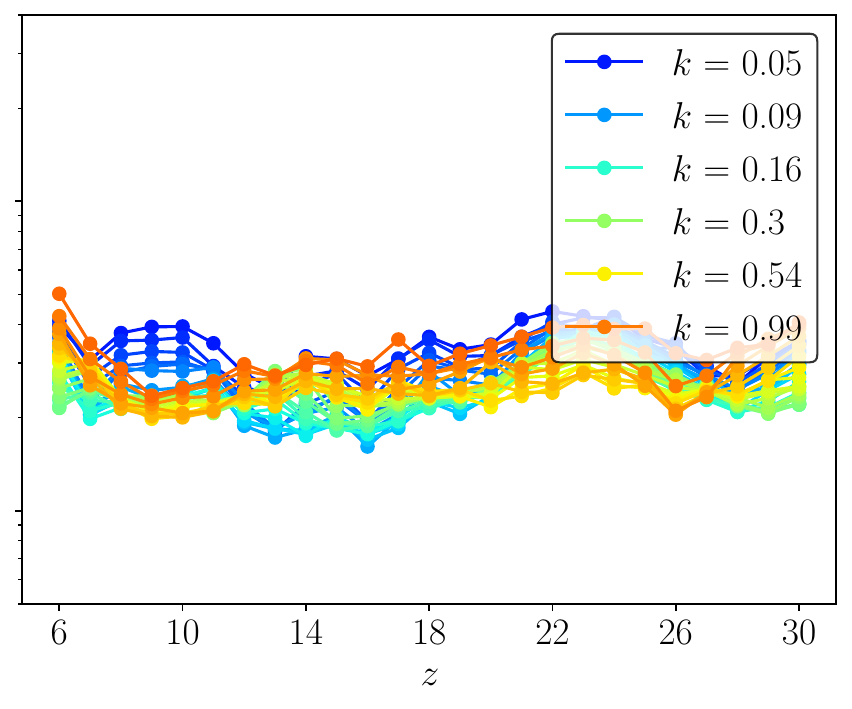}
    \end{minipage}\hfill
    \begin{minipage}{0.23\textwidth}
      \includegraphics[scale=0.31]{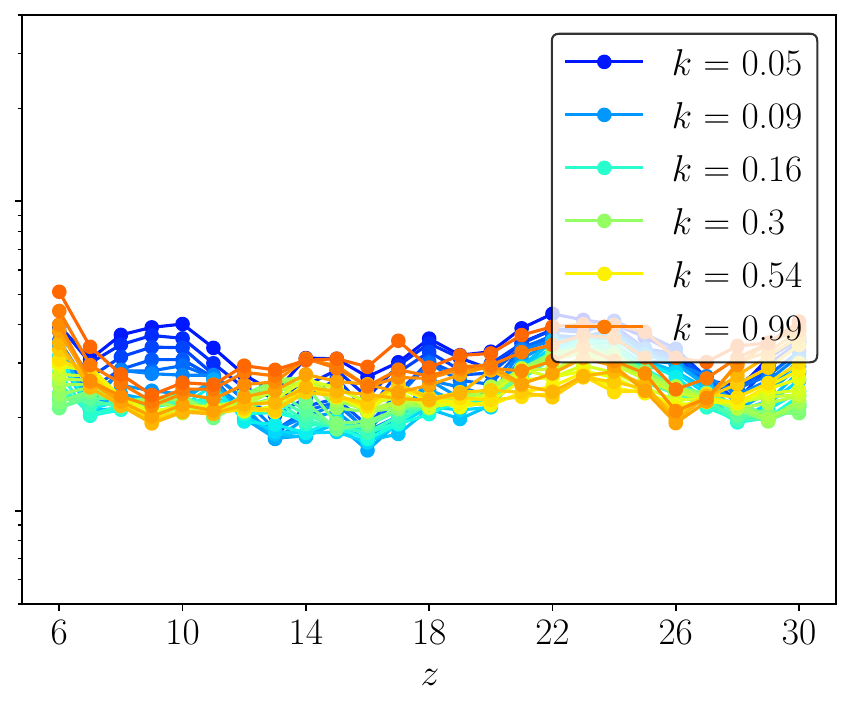}
    \end{minipage}
    \caption{Redshift and wavenumber dependence of the relative error in emulating the best-fit power spectrum. The upper panels shows the dependence on  wavenumber (for fixed redshift) and the lower panels depict the redshift dependence (for fixed wavenumber). For the left-most panels, we emulate the power spectrum using the true parameters from the test dataset. For the panels in the second column from the left, we emulate the power spectrum using the best-fit parameters derived from the network of ideal dataset. For the panels in the third column from the left, we use the best-fit parameters derived from the network of mock SKA data, but for the error we measure the prediction of the real power spectrum, i.e., we apply the emulator trained without SKA noise. For the panels in the right-most column, we use the best-fit parameters derived from the network of SKA thermal noise case and otherwise do the same as for the third column. Note that the plots in this figure show all 25 $z$ values and 32 $k$ values.}\label{fig: error_k_z}
\end{figure*}

For a more detailed assessment of the emulator, we calculate how the error varies with redshift and wavenumber. For this we use a test dataset of 639 models (which is $20\%$ of our full dataset) for each of the cases: ideal dataset, mock SKA data, and SKA thermal noise case. We first directly test the emulator by comparing the predicted power spectrum (feeding into the emulator the known true parameters) to the true simulated power spectrum (as in the previous subsection, but here divided separately into $k$ and $z$ bins). In addition, we test the complete framework by finding the best-fit astrophysical parameters to mock data using the MCMC sampler; feeding the best-fit parameters to the emulator of the power spectrum; and finding the error of this best-fit predicted power spectrum compared to the true simulated power spectrum. In cases with SKA noise (mock SKA dataset and SKA thermal noise case), we are not interested in finding the error in the predicted power spectrum with SKA noise (as the power spectrum is often dominated by noise, especially at high redshifts); instead, we make the more challenging comparison of the best-fit predicted power spectrum to the true power spectrum, both in their "clean" versions (i.e., without SKA noise). To be clear, this means taking the reconstructed best-fit astrophysical parameters (which were reconstructed from the mock SKA dataset, based on the NN trained using mock SKA power spectrum), and using it as input to the NN trained using power spectra from the ideal dataset (i.e., without SKA noise). Here we use the following definition to quantify the error as a function of redshift and wavenumber:
\begin{equation}
{\rm Error}(z, k) = \rm{Median\left|\frac{\Delta^2_{predicted\_clean} - \Delta^2_{true\_clean}}{\Delta^2_{true\_clean} + 0.25 \ \rm{mK^2}}\right|} \ \ ,
\label{eq: error_k_z}
\end{equation}
where we take the median over the test models; in this paper we often take the median in order to measure the typical error and reduce the sensitivity to outliers. This definition of the error measures the absolute value of the relative error, except that the denominator includes a constant in order not to demand a low relative error when the fluctuation itself is low (in agreement with eq.~\ref{eq:sn}). Note that here the errors are much larger than before because they are not normalized to the maximum value of the power spectrum but are measured separately at each bin, including when the power spectrum is low. 

In Fig.~\ref{fig: error_k_z}, we show how the error varies with wavenumber (top panels) and redshift (bottom panels), for both the ideal and SKA cases (mock SKA dataset and SKA thermal noise case). For the direct emulation case (left-most panels, where we emulate the power spectrum using the true parameters from the test dataset), the relative error decreases with wavenumber up to $k \sim 0.1 - 0.2$ Mpc$^{-1}$, then plateaus, and again increases above $k \sim 0.6$ Mpc$^{-1}$. The redshift dependence shows a less regular pattern, except that the errors tend to increase both at the low-redshift and high-redshift end. Overall, the typical emulation error of the power spectrum in each bin is $10-20\%$ over a broad range of $k$ and $z$, but it rises above $20\%$ at the lowest and highest $k$ values (for most redshifts), and at the lowest redshift for all $k$ values (i.e., at $z = 6$, near the end of reionization, when the power spectrum is highly variable and is sensitive to small changes in the parameters). For some perspective, we note that a $20\%$ error is typically adopted to represent the systematic theoretical modeling error in the 21-cm power spectrum \citep[e.g.,][]{21CMMC,Greig2017}. In the panels in the second column from the left, we use the best-fit parameters derived from the network trained using ideal dataset to emulate the power spectrum. From the comparison to the left-most panels, we see that the fitting of the astrophysical parameters (in this ideal case) is nearly perfect, in that the error that it adds is small compared to the error of the emulator itself. In the panels in the third column, the best-fit parameters are derived from the network trained using mock SKA dataset, but as noted above, the errors are calculated for the ability to predict the real power spectrum, i.e., by comparing the true power spectrum to the prediction of the emulator that was trained using the power spectrum without SKA noise. SKA noise reduces the accuracy of the reconstruction of the astrophysical parameters but not by too much, increasing the typical errors by a fairly uniform factor of $\sim 1.5$, to $15-30\%$ for most values of $k$ and $z$. For the panels in the last column, we use the best-fit parameters derived from the network trained using the power spectrum of SKA thermal noise case. The errors are nearly identical to the full SKA noise panels, showing that the foreground effects do not add substantial error beyond the angular resolution plus thermal noise, at least for the
optimistic foreground avoidance model that we have assumed.

\begin{figure*}
    \centering
    \begin{minipage}{0.38\textwidth}
    \includegraphics[scale=0.315]{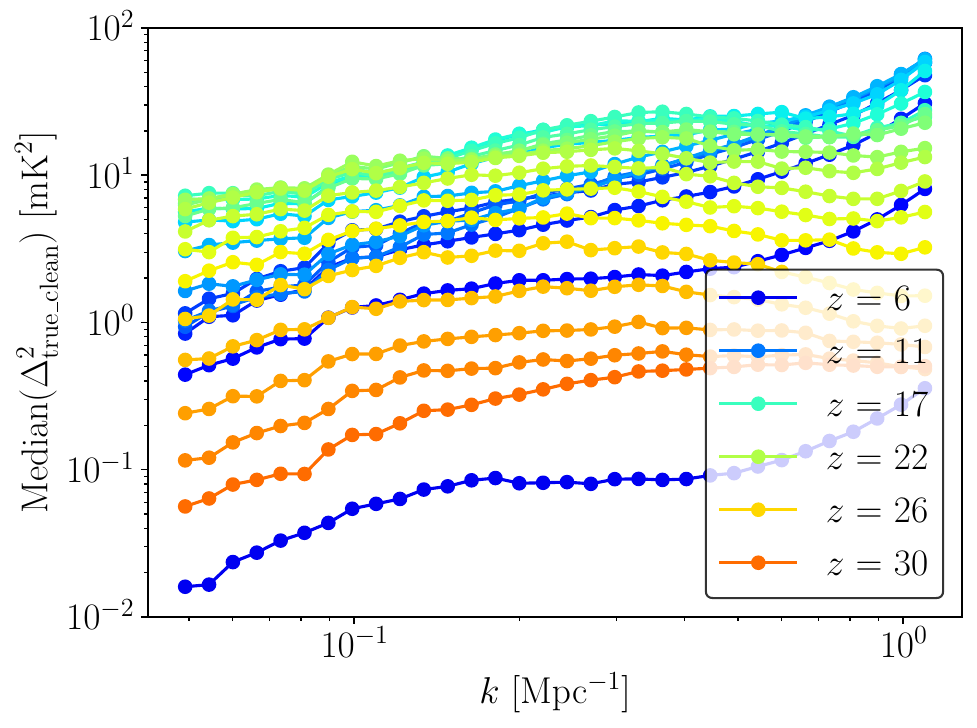}
    \hspace{0.25in}
    \rulesep
    \end{minipage}\hfill
    \begin{minipage}{0.30\textwidth}
    \includegraphics[scale=0.315]{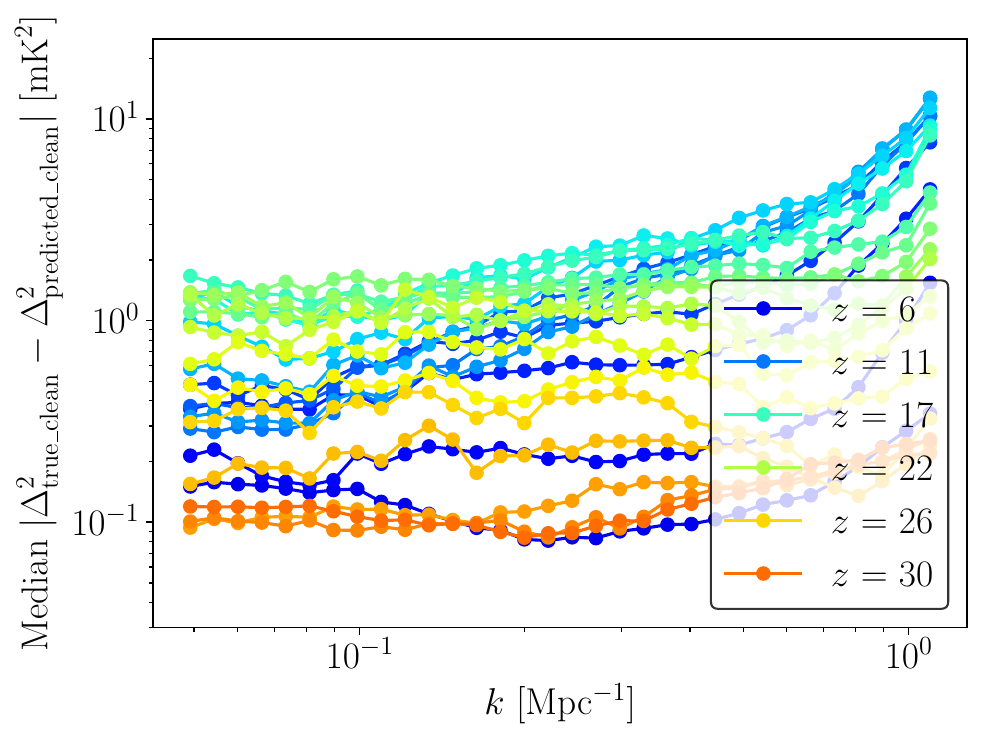}
    \end{minipage}\hfill
    \begin{minipage}{0.30\textwidth}
    \includegraphics[scale=0.315]{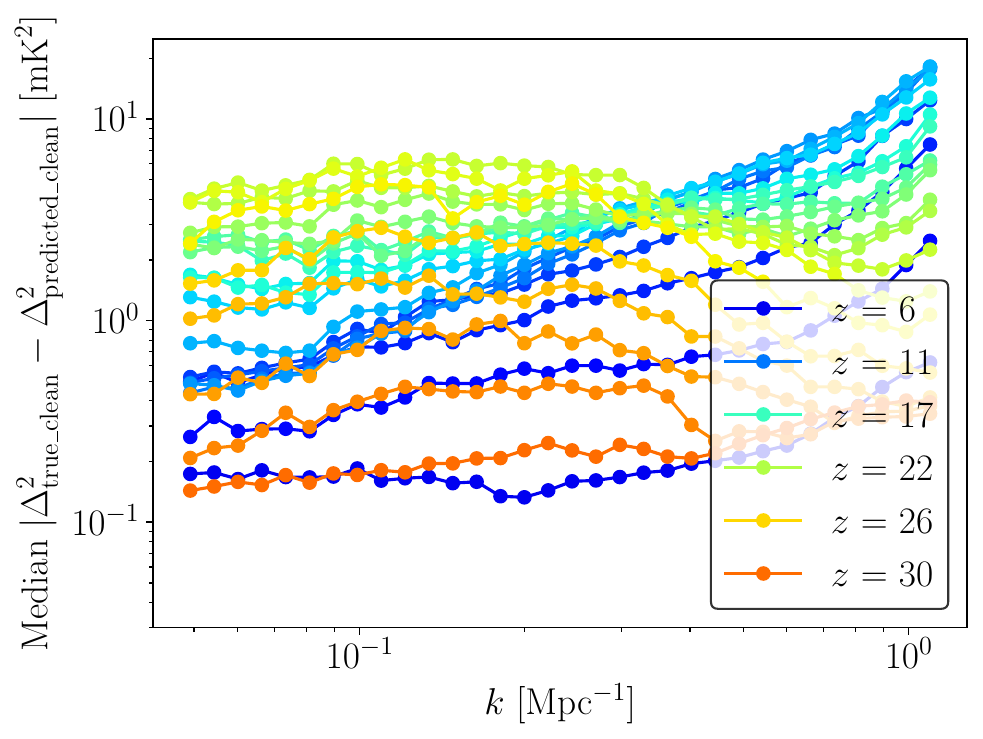}
    \end{minipage}\hfill
    
    \centering
    \begin{minipage}{0.38\textwidth}
    \includegraphics[scale=0.315]{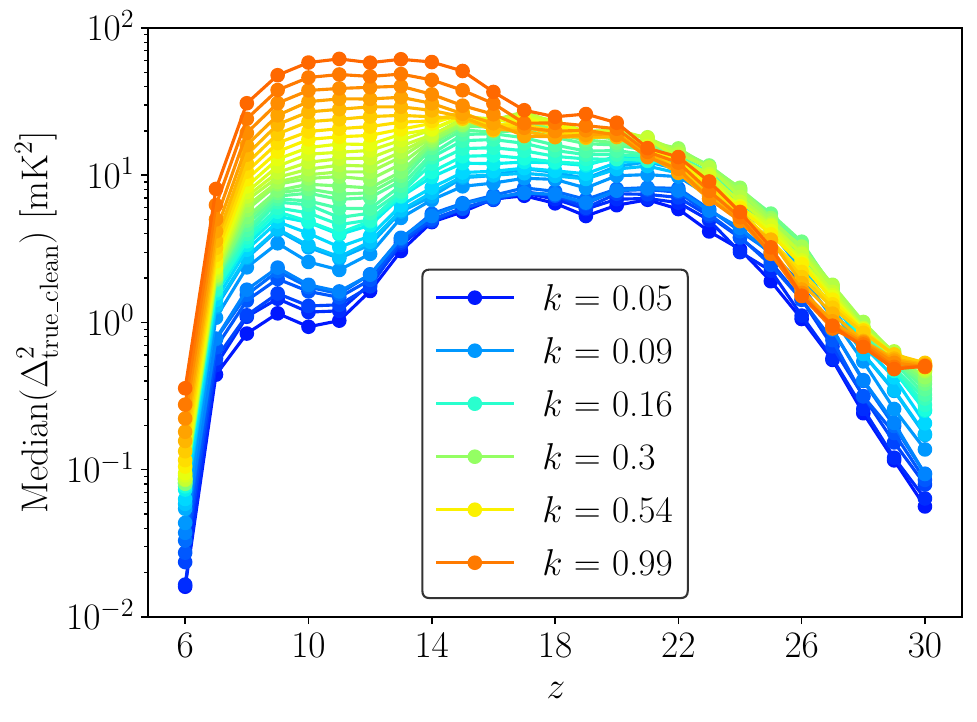}
    \hspace{0.25in}
    \rulesep
    \end{minipage}\hfill
    \begin{minipage}{0.30\textwidth}
    \includegraphics[scale=0.315]{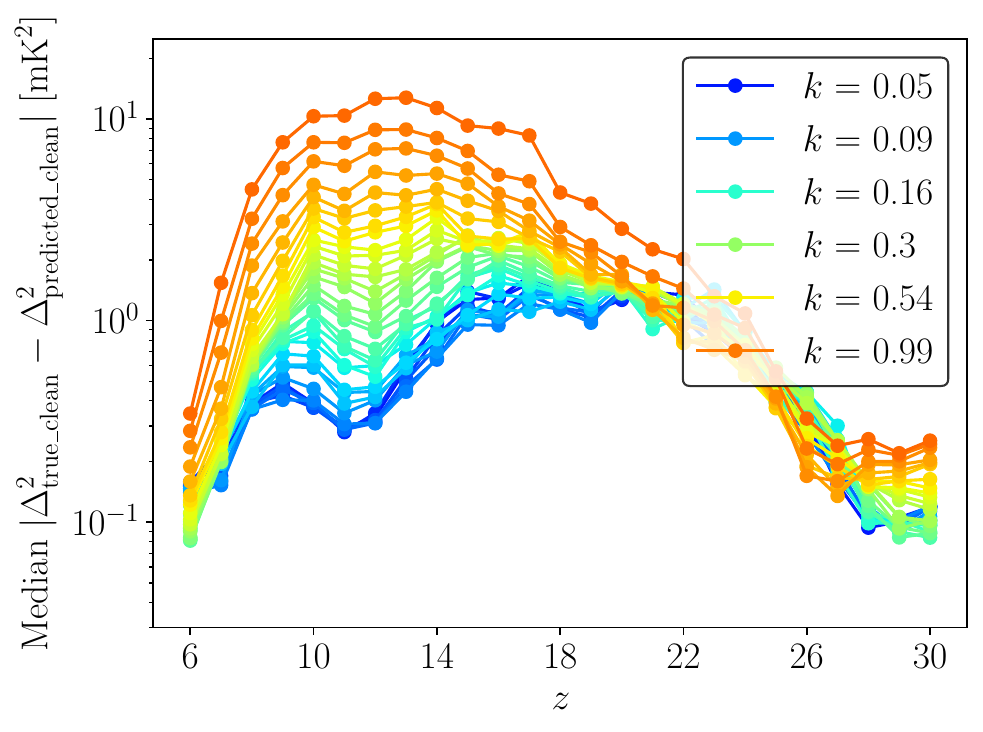}
    \end{minipage}\hfill
    \begin{minipage}{0.30\textwidth}
    \includegraphics[scale=0.315]{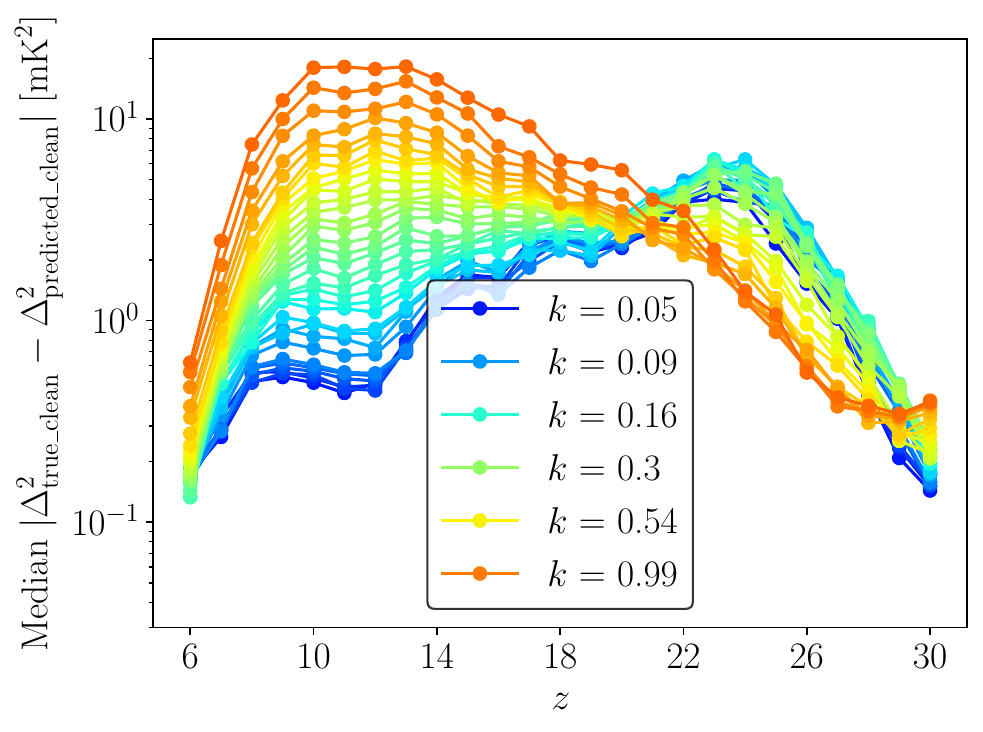}
    \end{minipage}\hfill

    \caption{\textbf{Left column:} Median of the true (clean) power spectrum (ideal dataset), $\Delta^2_{\rm{true\_clean}}$, as a function of wavenumber (upper panel) and redshift (lower panel). \textbf{Other columns :} The median of the absolute value of the difference between the true and predicted clean power spectrum. For the panels in the middle column, we emulate the power spectrum using the best-fit parameters derived from the network trained using ideal dataset. For the panels in the right column, the best-fit parameters are derived from the network trained using mock SKA dataset, but the error is measured by emulating the clean power spectrum. As in Fig.~\ref{fig: error_k_z}, the plots in this figure show all 25 $z$ values and 32 $k$ values.}
    \label{fig:median_ps_clean}
\end{figure*}

In order to get a better understanding of the span of the models over $k$ and $z$, we show in Fig.~\ref{fig:median_ps_clean} characteristic quantities that enter into the above calculation of the relative errors. In the left column, we show the median of the clean power spectrum (without any noise) as a function of the wavenumber (upper panel) and redshift (lower panel). In the other columns, the median of the absolute difference between the true and predicted clean power spectra is shown as a function of wavenumber (upper panels) and redshift (lower panels). For the panels in the middle column, the best-fit parameters are derived from the network without any noise (i.e., ideal dataset), whereas for the panels in the right column we use the best-fit parameters derived from the network trained using mock SKA dataset to emulate the clean power spectrum (without noise). This figure shows that the 21-cm power spectrum varies greatly as a function of $k$ and $z$, even when we take out the model-to-model variation by showing the median of the 639 random test cases. The variation is by three and a half orders of magnitude; even if we ignore the parameter space in which the fluctuation level is
lower than 0.5~mK (see eq.~\ref{eq:sn}), we are left with a range of more than two orders of magnitude. For the considered ranges, the overall variation with redshift at a given wavelength is much greater than the variation with wavelength at a given redshift. Over this large range, the relative error in each case (with or without SKA noise) remains relatively constant; this is seen by the panels in  Fig.~\ref{fig:median_ps_clean} that show the relative error, which overall follows a similar pattern (with $z$ and $k$) as the power spectrum except with a compressed range of values.

\subsection{Errors in the fitted astrophysical parameters}\label{sec: bayesian_analysis}

\begin{figure*}
    \centering
    \includegraphics[scale=0.4]{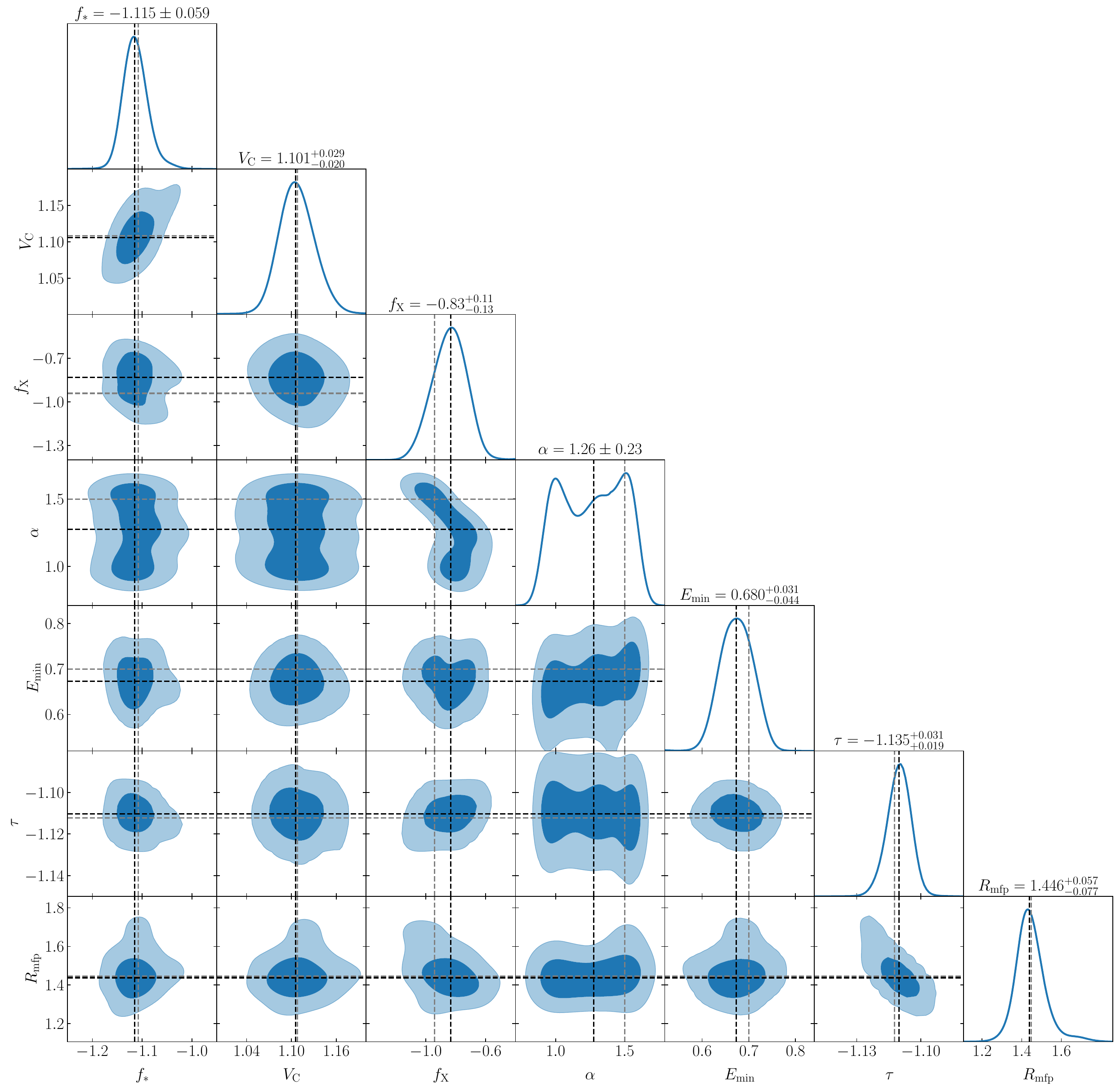}\llap{\raisebox{10cm}{\includegraphics[scale=0.5]{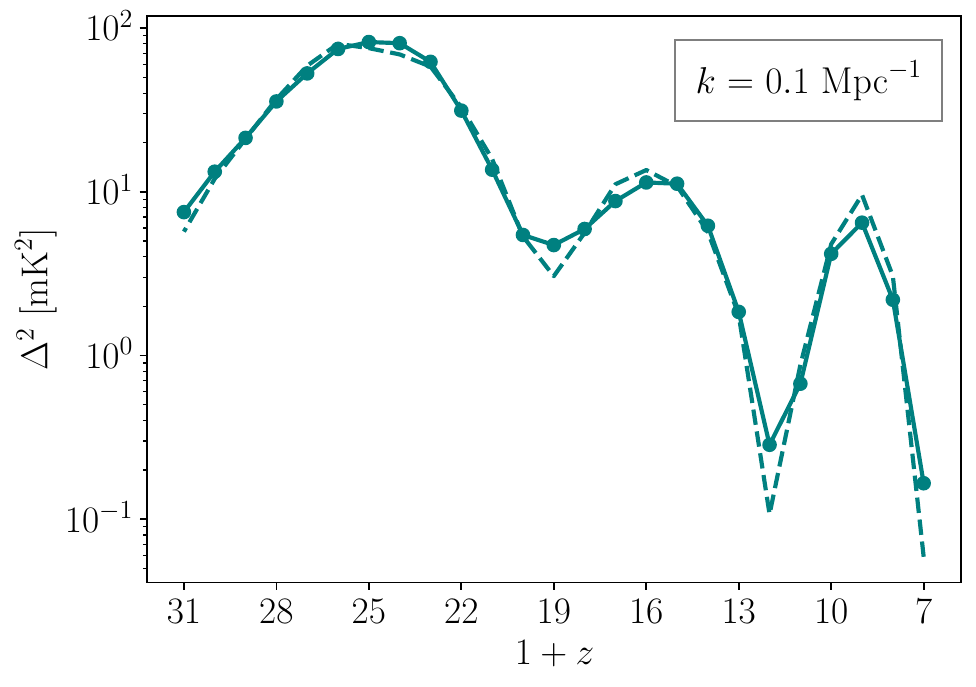}}}
    \caption{The posterior distribution of the seven parameter astrophysical model, showing the pairwise covariances (off-diagonal entries) and their marginalized distribution (diagonal entries) across each model parameter.  The black dashed lines denote the median of the distribution. The gray dashed lines denote the true parameters from the simulation. In the 2-D contours, there are only two levels of probability. The fraction of probability contained in each contour: 0.68 and 0.95 (dark through light blue). All the parameters are in $\log_{10}$ except $\alpha$ and $E_{\rm{min}}$. The upper right panel shows the true power spectrum (dashed) vs.\ the reconstructed best-fit one (solid), at one wavenumber. This figure shows a case that uses the power spectrum from the ideal dataset.}
    \label{fig: pos_distribution_clean1}
\end{figure*}

Up to now, we have examined the errors in emulating or reconstructing the 21-cm power spectrum. Of greater interest is, of course, the ability to extract astrophysical information from a given power spectrum. In addition to the unavoidable effect on the fitting of the emulation uncertainty, there are
also the SKA observational effects. 



 




In order to account for the emulation uncertainty, we employed a method in the spirit of $k$-fold cross validation, where the training dataset is randomly divided into $k$ portions, and each model is trained on various $k-1$ portions (we emphasize that this is {\it completely separate}\/ from the actual $k$-fold cross validation that we perform in the next subsection). We added the $k$ MCMC chains from all the runs and used the resulting combined chain to get parameter uncertainties and error contours. We used $k=40$ for the algorithm described in this subsection. The posterior probability for one astrophysical model is shown in figure~\ref{fig: pos_distribution_clean1}, for the ideal dataset (\ref{fig: pos_distribution_SKA1} shows the same example for the case of the mock SKA data). The figure shows the posterior distribution of the seven parameter astrophysical model for an example power spectrum from the ideal dataset. In this figure the gray dashed lines denote the true parameters from the simulation. As the posteriors are not perfectly Gaussian, we report the median of the distribution (black dashed lines) as the predicted parameter value from the MCMC sampler. The results are rather insensitive to $\alpha$, and so its value is not well determined, and thus extends to the limits of the prior assumed range, especially in the noisy cases (where the errors are larger). In the case of $R_{\rm{mfp}}$, the upper limit runs into the edge of the assumed range, in the the mock SKA case. As noted in sec.~2.1.1, we assumed a reasonable upper limit to $R_{\rm{mfp}}$ based on observational and theoretical constraints; in the future, we plan to explore more realistic modeling of the end stages of reionization.

More generally, while this type of result for individual models is interesting, we prefer to look at properties that more generally characterize the broad range of possible astrophysical models. In order to understand the general trends, we consider below the overall statistics of the fitting as calculated for a large number of models.

\subsection{$k$-fold cross validation and statistical analysis of the astrophysical parameter errors}

In order to test the overall performance in predicting each of the parameters, we use our test dataset of 639 models, which is $20\%$ of our full dataset. We calculate in each case the 1-$\sigma$ MCMC uncertainty.
In order to test whether this uncertainty is a realistic error estimate, we calculate the normalized error in predicting a parameter ($\rm{P_{predicted}}$) compared to the true value ($\rm{P_{true}}$), which we term the $z\rm{-score}$: 
\begin{equation}\label{eq: error}
z\rm{-score}\  = \frac{P_{\rm{true}} - P_{\rm{predicted}}}{\sigma}\ .
\end{equation}
If $\sigma$ is an accurate estimate then the actual values of this normalized error (i.e., $z\rm{-score}$) for the test dataset should have a standard deviation of unity. All these quantities, namely $\rm{P_{true}}$, $\rm{P_{predicted}}$ and $\sigma$, are measured in log space ($\log_{10}$) for all the parameters except for $\alpha$ and $E_{\rm{min}}$.

To test and improve the error estimation using our procedure, we perform $k$-fold cross validation. In $k$-fold cross validation, the training sample is partitioned into subsets, reserving one subset for testing while training the emulator on the remaining data. We then repeat this process $k$ times, each time using a different subset as the test data. Comparing the different cases gives validation (if the results do not vary too much), while taking the mean of various statistics among the $k$-folds gives better estimates. Here we choose to work with $k = 5$ so that each of our validation sets consists of a test dataset equal to $20\%$ of the whole dataset. We note that the PCA reduction (as described in section 2) is performed on the training data for each of the $k$-folds.

\begin{figure*}
    \centering
    \begin{minipage}{0.33\textwidth}
       \includegraphics[scale=0.37]{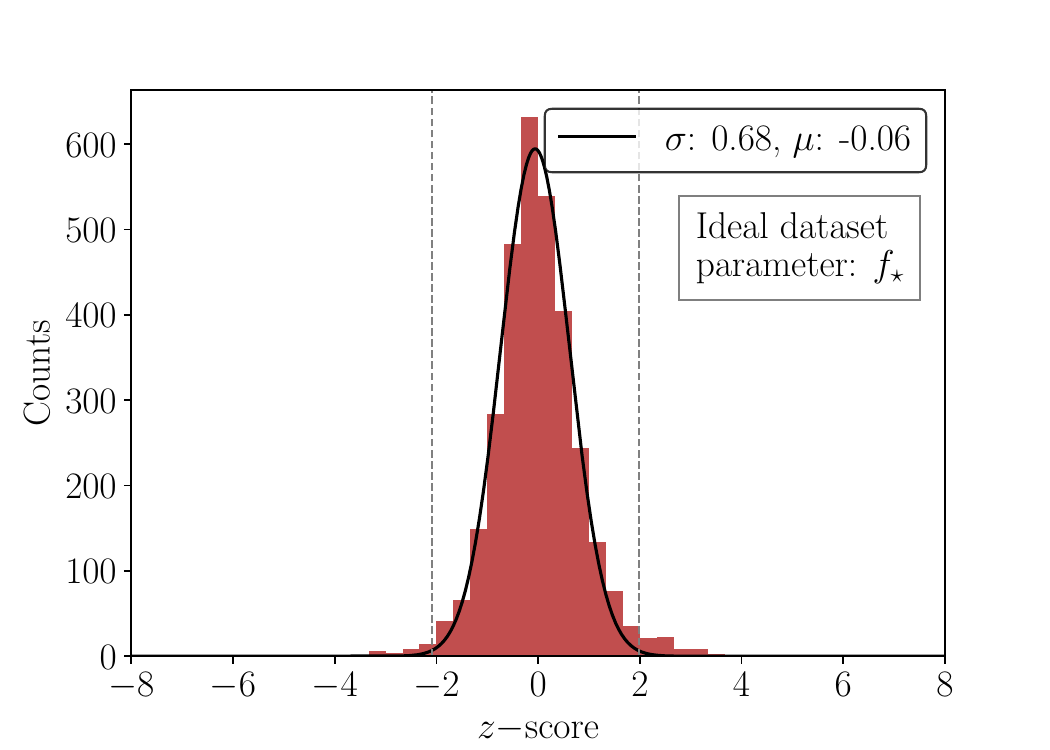}
       
    \end{minipage}
    \begin{minipage}{0.33\textwidth}
      \includegraphics[scale=0.37]{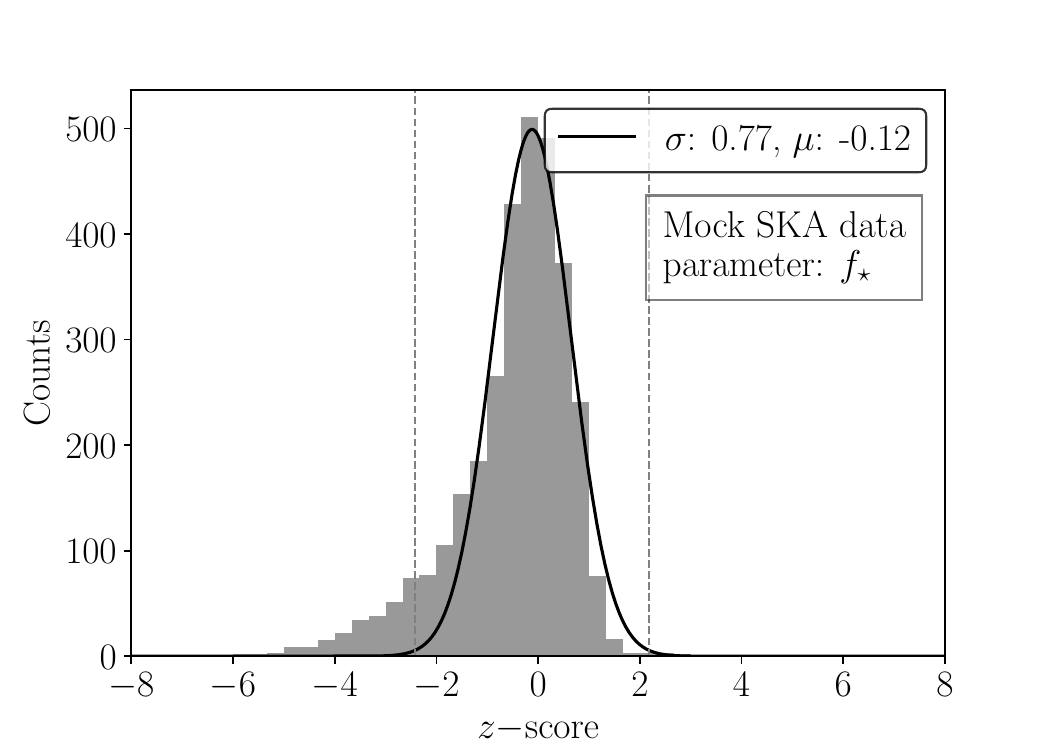}
      
    \end{minipage}
    \begin{minipage}{0.3\textwidth}
      \includegraphics[scale=0.37]
      {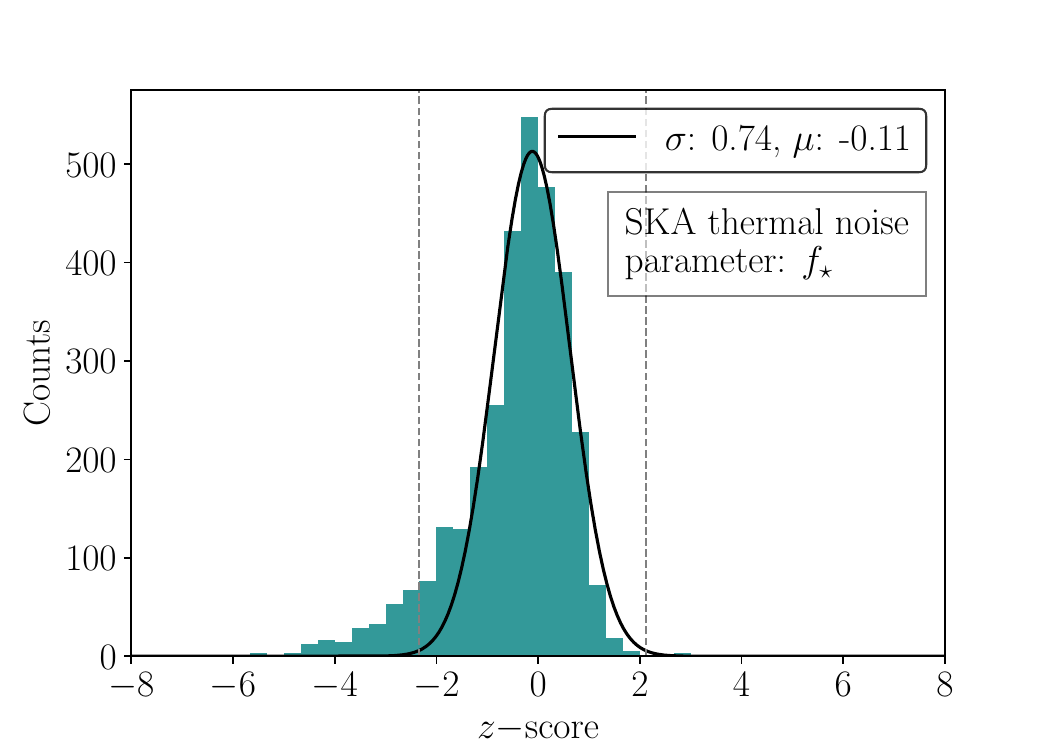}
      
    \end{minipage}
    \centering
    \begin{minipage}{0.33\textwidth}
       \includegraphics[scale=0.37]
    {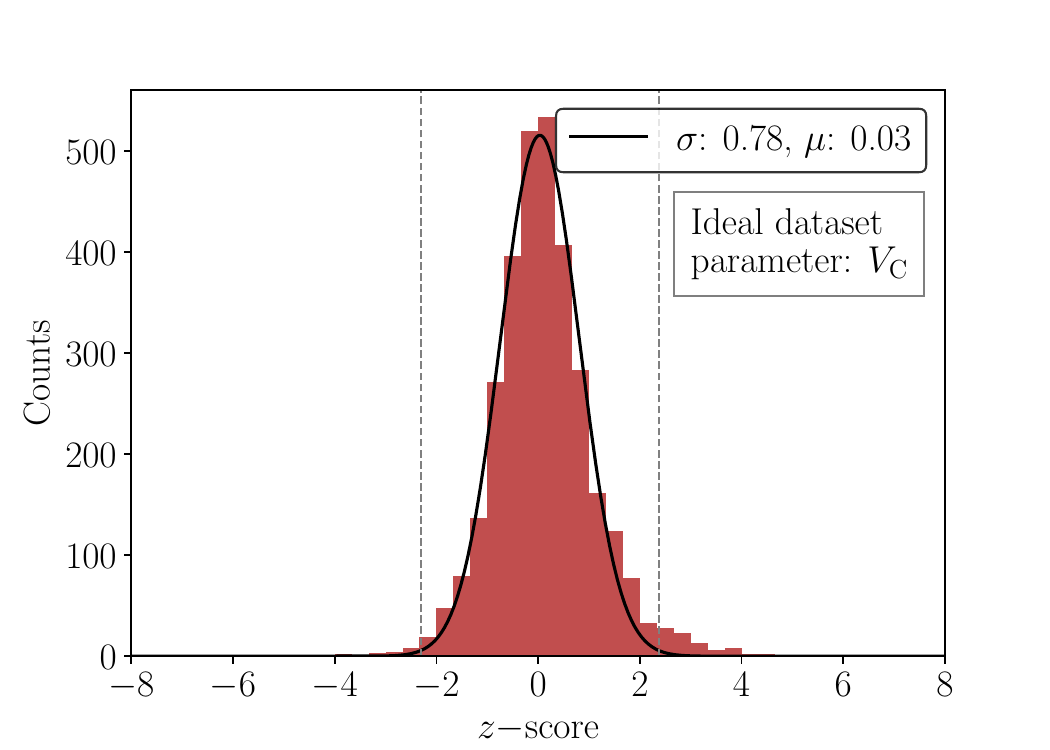}
       
    \end{minipage}
    \begin{minipage}{0.33\textwidth}
      \includegraphics[scale=0.37]
      {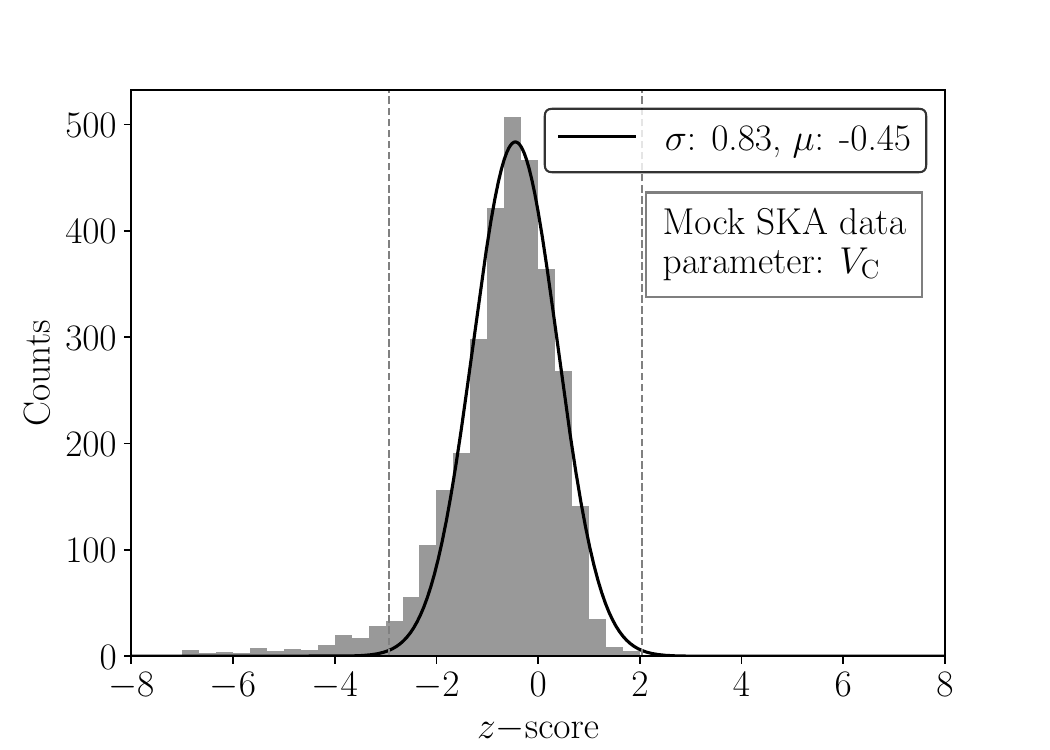}
      
    \end{minipage}
    \begin{minipage}{0.3\textwidth}
      \includegraphics[scale=0.37]
      {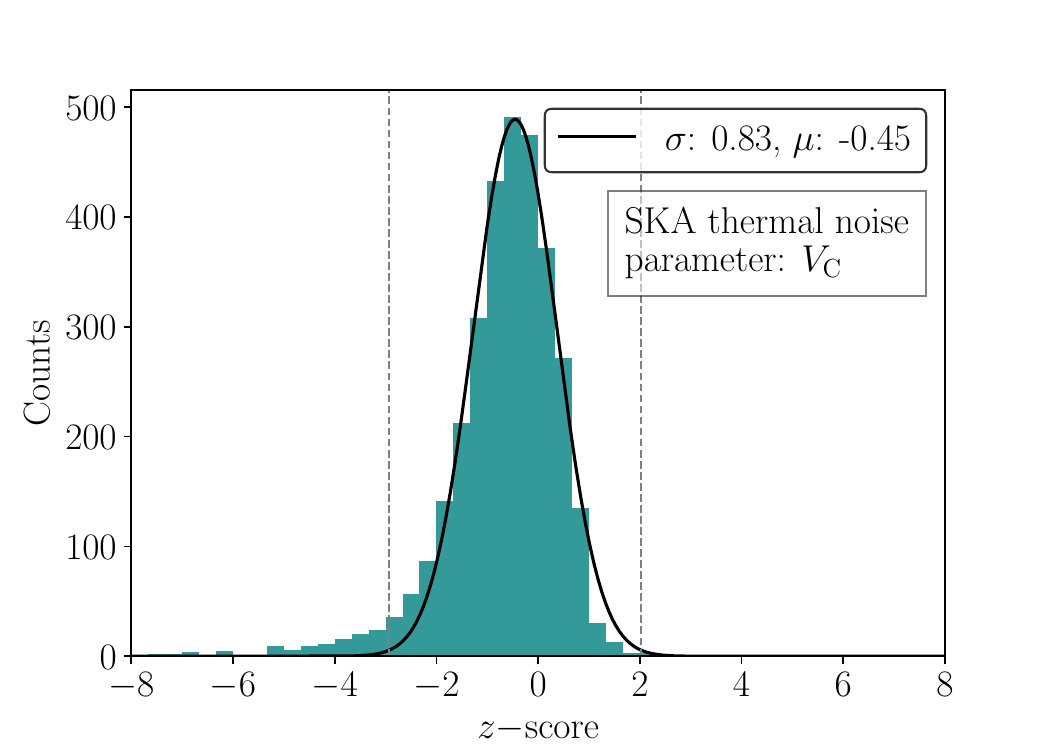}

    \end{minipage}
    \centering
    \begin{minipage}{0.33\textwidth}
       \includegraphics[scale=0.37]{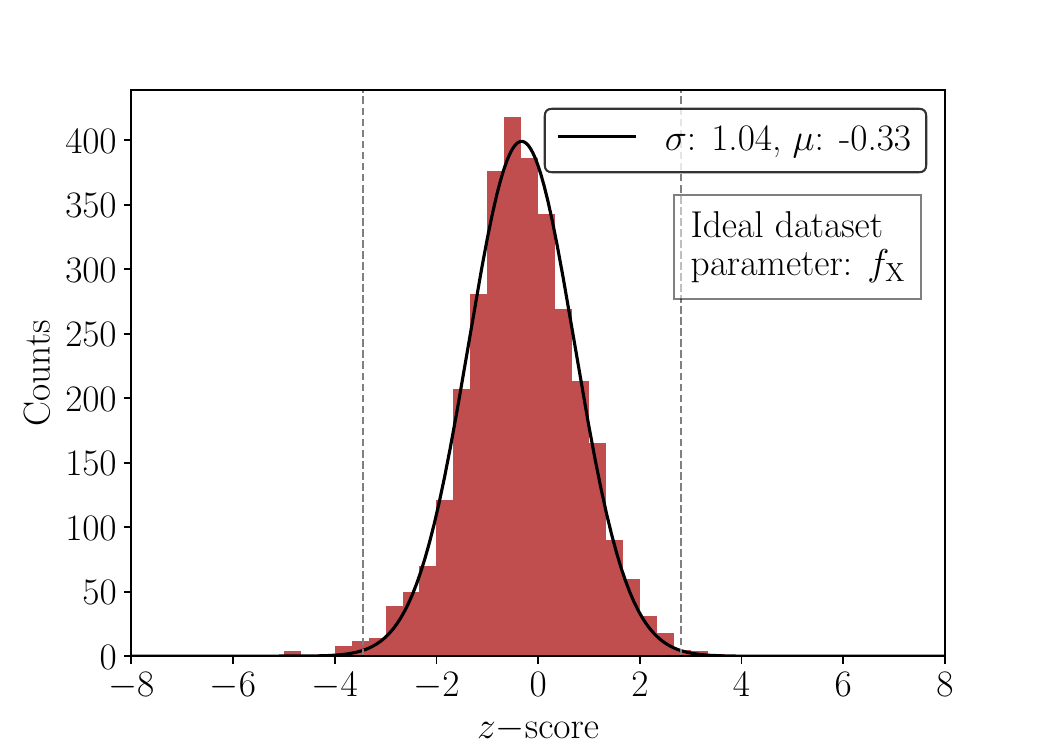}
    
    \end{minipage}
    \begin{minipage}{0.33\textwidth}
      \includegraphics[scale=0.37]{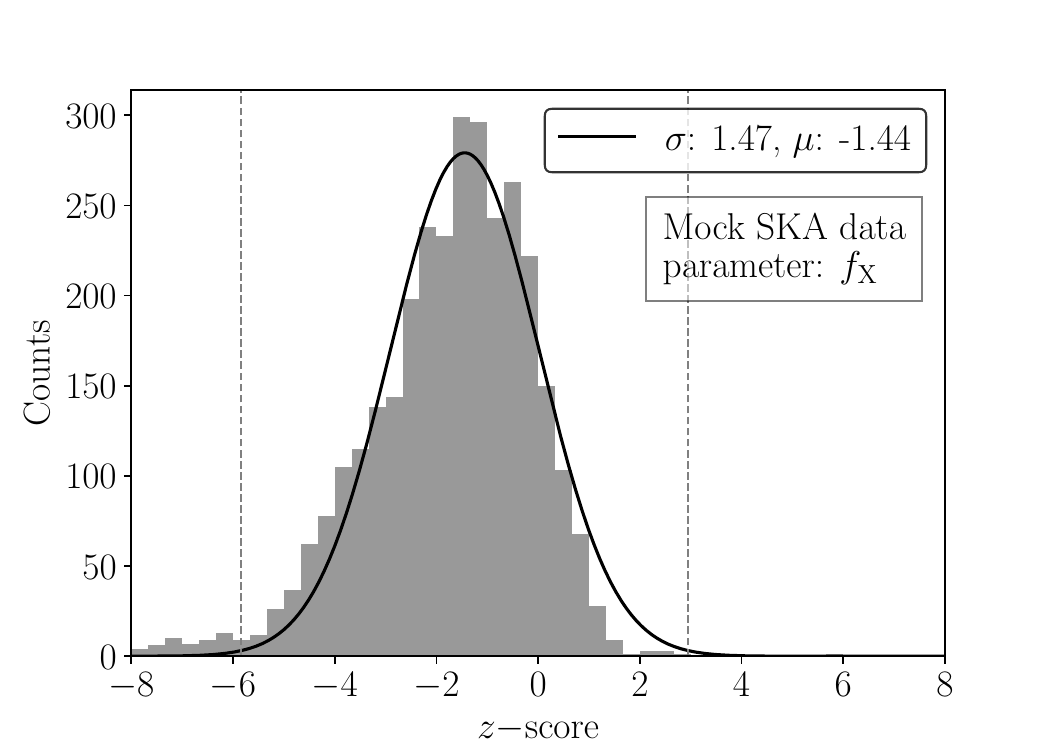}
      
    \end{minipage}
    \begin{minipage}{0.3\textwidth}
      \includegraphics[scale=0.37]{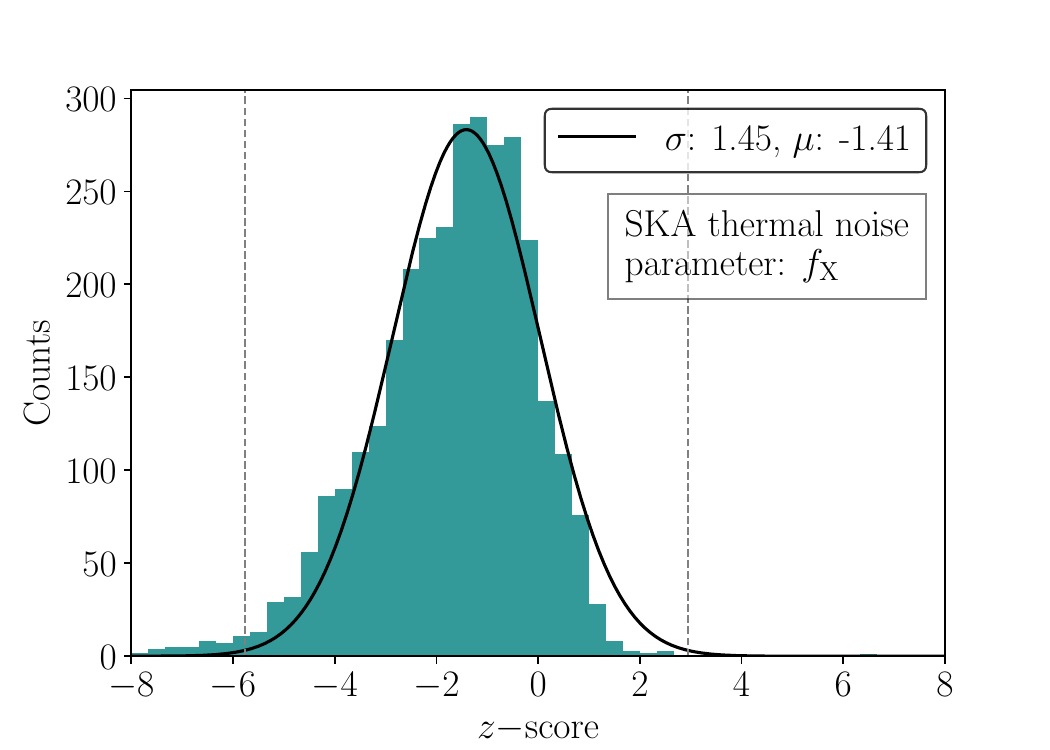}
    \end{minipage}
    
\caption{Histogram of the normalized errors / $z\rm{-scores}$ for the parameters: $f_{\star}$, $V_C$ and $f_X$ as defined in Eq. \ref{eq: error}. The distribution is shown over all the 5-folds. Also shown in each panel is the best-fit Gaussian, with its parameters listed within the panel. The parameter values here are in $\log_{10}$.}\label{fig: histogram1}
\end{figure*}

We show the combined histogram of the z\rm{-score} from all 5-folds in predicting each of the parameters in Figs.~\ref{fig: histogram1} (for the three most important parameters of high-redshift galaxies) and \ref{fig: histogram2} (for the four other parameters, shown in the Appendix).
In these figures, the left panels are for the ideal data, the middle panels are for the case with mock SKA data, and the right panels are for the SKA thermal noise case. The black solid line in each panel shows the best-fit Gaussian of the histogram, also listing its mean ($\mu$) and standard deviation ($\sigma$) within the panel. The two grey dashed lines in each panel represent the $3\sigma$ boundary of the respective Gaussian. Table~\ref{tab: table_clean_CV} (along with Table~\ref{tab: table_clean_CV_}) lists the corresponding parameters of best-fit Gaussians for each of the 5-folds. The $\sigma$ values are fairly consistent among the different folds, while there is more variation in some cases in the bias $\mu$. The best-fit values to the combined distributions (shown in the Figures) agree rather closely with the mean values of the Gaussian parameters (shown in the Tables).

\begin{table}
\begin{tabular}{ccccccc}
\hline
\multicolumn{7}{c}{Ideal dataset}   \\

\hline
Parameters   & \multicolumn{2}{c}{$f_{\star}$} & \multicolumn{2}{c}{$V_{\rm{C}}$} & \multicolumn{2}{c}{$f_{\rm{X}}$} \\
\hline
Gaussian fit & $\sigma$         & $\mu$        & $\sigma$         & $\mu$         & $\sigma$                  & $\mu$                  \\
\hline\hline
\centering
 Fold 1 & $0.78$ & $-0.00$ & $0.81$ & $+0.07$ & $1.01$ & $-0.25$\\ 
\centering
 Fold 2 & $0.85$ & $+0.15$ & $0.89$ & $+0.42$ & $1.05$ & $-0.41$     \\ 
\centering
 Fold 3 & $0.69$ & $+0.08$ & $0.75$ & $+0.28$ & $1.09$ & $-0.32$   \\ 
 \centering
 Fold 4 & $0.60$ & $-0.24$ & $0.66$ & $-0.25$ & $1.06$ & $-0.24$\\ 
\centering
 Fold 5 & $0.55$ & $-0.14$ & $0.67$ & $-0.14$ & $1.07$ & $-0.38$   \\ 
Mean   & 0.69 & $-0.03$ & 0.76 &  $+0.08$  & 1.06  &  $-0.32$     \\
\hline
\end{tabular}

\begin{tabular}{ccccccc}
\multicolumn{7}{c}{Mock SKA dataset}   \\

\hline
Parameters   & \multicolumn{2}{c}{$f_{\star}$} & \multicolumn{2}{c}{$V_{\rm{C}}$} & \multicolumn{2}{c}{$f_{\rm{X}}$} \\
\hline
Gaussian fit & $\sigma$         & $\mu$        & $\sigma$         & $\mu$         & $\sigma$                  & $\mu$                  \\
\hline\hline
\centering
 Fold 1 & $0.80$ & $-0.09$ & $0.83$ & $-0.44$ & $1.37$ & $-1.27$\\ 
\centering
 Fold 2 & $0.74$ & $-0.11$ & $0.72$ & $-0.19$ & $1.44$ & $-1.47$     \\ 
\centering
 Fold 3 & $0.73$ & $-0.23$ & $0.72$ & $-0.12$ & $1.48$ & $-1.23$   \\ 
 \centering
 Fold 4 & $0.85$ & $-0.10$ & $0.87$ & $-0.88$ & $1.48$ & $-1.61$\\ 
\centering
 Fold 5 & $0.73$ & $-0.04$ & $0.86$ & $-0.85$ & $1.48$ & $-1.65$   \\ 
Mean   & 0.77 & $-0.11$ & 0.80 &  $-0.50$  & 1.45  &  $-1.45$  \\
\hline
\end{tabular}

\begin{tabular}{ccccccc}
\multicolumn{7}{c}{SKA thermal noise case}   \\

\hline
Parameters   & \multicolumn{2}{c}{$f_{\star}$} & \multicolumn{2}{c}{$V_{\rm{C}}$} & \multicolumn{2}{c}{$f_{\rm{X}}$} \\
\hline
Gaussian fit & $\sigma$         & $\mu$        & $\sigma$         & $\mu$         & $\sigma$                  & $\mu$                  \\
\hline\hline
\centering
 Fold 1 & $0.79$ & $-0.11$ & $0.85$ & $-0.45$ & $1.44$ & $-1.25$\\ 
\centering
 Fold 2 & $0.76$ & $-0.16$ & $0.71$ & $-0.20$ & $1.44$ & $-1.45$     \\ 
\centering
 Fold 3 & $0.70$ & $-0.22$ & $0.73$ & $-0.12$ & $1.45$ & $-1.24$   \\ 
 \centering
 Fold 4 & $0.87$ & $-0.11$ & $0.82$ & $-0.83$ & $1.47$ & $-1.56$\\ 
\centering
 Fold 5 & $0.75$ & $-0.06$ & $0.85$ & $-0.84$ & $1.47$ & $-1.60$   \\ 
Mean   & 0.77 & $-0.13$ & 0.79 &  $-0.49$  & 1.45  &  $-1.42$     \\
\hline
\end{tabular}

\caption{Standard deviation ($\sigma$) and mean ($\mu$) of the best-fit Gaussian of the respective histogram of $f_{\star}$, $V_{\rm{C}}$ and $f_{\rm{X}}$ for 5-fold cross validation. We also calculate the mean value of each of the best-fit parameter for the Gaussian over all the 5-folds. The three parameters here are in $\log_{10}$. Here we show the results for all three cases: ideal dataset (top), mock SKA data (middle) and SKA thermal noise (bottom).}\label{tab: table_clean_CV}
\end{table}


The standard deviations ($\sigma$) for most of the seven parameters are close to unity (within $\sim 20\%$), which implies that our procedure generates a reasonable estimate of the uncertainties. The errors are significantly smaller than expected for $\tau$, and also (to a lesser extent) for $f_{\star}$ and $V_{\rm{C}}$. In the noisy cases, the error is significantly underestimated for $f_{\rm{X}}$ (which is the parameter that has the largest log uncertainty). The mean (which measures the bias in the prediction) is in every parameter at most $0.3 \sigma$ in size, for the ideal dataset. The mock SKA dataset and SKA thermal noise case give similar results to each other, consistent with the similar comparison in Fig.~\ref{fig: error_k_z}. With the noisy data, the mean values are biased by as much as $\sim 1.4 \sigma$ for some of the parameters ($f_X$ and $E_{\rm{min}}$), with significant skewness that favors low values (particularly for $f_{\star}$). Even though the thermal noise is assumed to be Gaussian, it is quite large (especially at high redshifts), and when this is combined with the highly non-linear dependence of the power spectrum on the astrophysical parameters, the resulting distributions are significantly non-Gaussian. When fitting real data, these results can be used directly as estimates of the expected error distributions. It may also be possible to improve the procedure in order to reduce the errors and the bias. In the case of noisy data, various regularization techniques such as dropout or weight decay, or exploring alternative network architectures, might be effective in improving the predictions. We leave for future work further exploration of these possibilities. We also note that most of the distributions are fairly Gaussian, in that only a small fraction of the samples yield best-fit parameter values that fall outside the 3$\sigma$ boundary of the respective Gaussian fit.

In Figs.~\ref{fig: sigma_prime1} and \ref{fig: sigma_prime2} (the latter in the Appendix), we show the combined histogram of the size of the actual error ($|P_{\rm{true}} - P_{\rm{predicted}}|$) from all 5-folds for each of the parameters. In the left panels we compare the histogram of the actual error for the cases: ideal data and mock SKA data, whereas in the right panels we compare the actual error for the cases: ideal data and SKA thermal noise. Again the actual errors are measured in $\log_{10}$ for all the parameters except for $\alpha$ and $E_{\rm{min}}$. 

Table~\ref{tab:table_median} shows the corresponding median of the actual error for each fold from the 5-fold cross validation for the cases: ideal dataset, mock SKA dataset and SKA thermal noise. We also calculate the mean (of the median) over all the 5-folds for all cases. In the theoretical case of no observational limitations ("ideal dataset"), the emulation errors still allow the parameters $V_C$ and $\tau$ to be reconstructed with a typical accuracy of 2.3\% and 0.9\%, respectively, and $f_{\star}$ to within 4.5\%. The ionizing mean free path ($R_{\rm{mfp}}$) is typically uncertain by a factor of 1.16, and $f_X$ by a factor of 1.78. For the linear parameters, the actual error is typically $\pm 0.19$ in $\alpha$ and $\pm 0.20$~keV in $E_{\rm min}$. 

One might worry that the dimensionality reduction using PCA on the dataset likely introduces some correlations in the resulting power spectrum predictions (see Fig. \ref{fig: error_k_z}) over the various $z$ and $k$-bins. In testing the sensitivity to the number of PCA components used to train the emulator, we focus on our main results. Specifically, for the case of the ideal dataset we tried doubling the number of PCA components used throughout the analysis. We found that the standard deviation ($\sigma$) and mean ($\mu$) of the best-fit Gaussian of the astrophysical parameters were as follows: $f_{\star}: (\sigma=0.67, \mu=0.00$), $V_{\rm{C}}: (\sigma=0.60, \mu=0.03)$, $f_{\rm{X}}: (\sigma=0.94, \mu=-0.24)$,  $\alpha: (\sigma=1.05, \mu=-0.03)$, $E_{\rm{min}}: (\sigma=0.79, \mu=-0.10)$, $\tau: (\sigma=0.59, \mu=0.05)$, and $R_{\rm{mfp}}: (\sigma=0.89, \mu=0.02)$. These values are similar (within $\sim 20\%$ or better) to the mean values shown for the ideal dataset in Tables \ref{tab: table_clean_CV} and \ref{tab: table_clean_CV_}. Thus, the main goal of this work, which is to constrain seven-parameter astrophysical models and obtain their uncertainties from the mock 21-cm power spectrum, is not very sensitive to the details of the PCA reduction; we leave for the future further analysis of correlations in the reconstructed power spectrum.

\begin{table}
\centering
\begin{tabular}{cllllll}

\hline
\multicolumn{7}{c}{Ideal dataset}         \\
\hline
parameters           & \multicolumn{1}{c}{Fold 1} & \multicolumn{1}{c}{Fold 2} & \multicolumn{1}{c}{Fold 3} & \multicolumn{1}{c}{Fold 4} & \multicolumn{1}{c}{Fold 5} & Mean \\
\hline\hline

$f_{\star}$         & 0.0199 & 0.0273 & 0.0226 & 0.0126 & 0.0114 &  0.0188\\
$V_{\rm{C}}$ [km/s] & 0.0080 & 0.0067 & 0.0061 & 0.0145 & 0.0138 & 0.0098\\
$f_{\rm{X}}$        & 0.1924 & 0.3458 & 0.3106 & 0.1912 & 0.2149 & 0.2510\\
$\alpha$            & 0.1928 & 0.1936 & 0.1902 & 0.1847 & 0.1921 & 0.1907\\
$E_{\rm{min}}$ [keV]& 0.1257 & 0.3631 & 0.2955 & 0.1048 & 0.1312  & 0.2041\\
$\tau$              & 0.0036 & 0.0045 & 0.0043 & 0.0034 & 0.0034  & 0.0038\\
$R_{\rm{mfp}}$ [Mpc]& 0.0575 & 0.0613 & 0.0643 & 0.0731 & 0.0741  & 0.0661\\
\hline
\end{tabular}

\begin{tabular}{cllllll}

\multicolumn{7}{c}{Mock SKA dataset}         \\
\hline
parameters           & \multicolumn{1}{c}{Fold 1} & \multicolumn{1}{c}{Fold 2} & \multicolumn{1}{c}{Fold 3} & \multicolumn{1}{c}{Fold 4} & \multicolumn{1}{c}{Fold 5} & Mean \\
\hline\hline

$f_{\star}$         & 0.1355 & 0.0691 & 0.0764 & 0.1894 & 0.1661 & 0.1273 \\
$V_{\rm{C}}$ [km/s] & 0.0605 & 0.0192 & 0.0209 & 0.1752 & 0.1921 & 0.0936 \\
$f_{\rm{X}}$        & 0.8439 & 0.9416 & 0.7955 & 1.1552 & 1.1771 & 0.9827 \\
$\alpha$            & 0.2210 & 0.2167 & 0.2222 & 0.2242 & 0.2292 & 0.2227 \\
$E_{\rm{min}}$ [keV]& 0.7906 & 0.9044 & 0.8302 & 1.0050 & 1.0200 & 0.9100 \\
$\tau$              & 0.0108 & 0.0076 & 0.0081 & 0.0158 & 0.0159 & 0.0117 \\
$R_{\rm{mfp}}$ [Mpc]& 0.1030 & 0.0750 & 0.0729 & 0.1151 & 0.1121 & 0.0956 \\
\hline
\end{tabular}

\begin{tabular}{cllllll}

\multicolumn{7}{c}{SKA thermal noise case}         \\
\hline
parameters           & \multicolumn{1}{c}{Fold 1} & \multicolumn{1}{c}{Fold 2} & \multicolumn{1}{c}{Fold 3} & \multicolumn{1}{c}{Fold 4} & \multicolumn{1}{c}{Fold 5} & Mean \\
\hline\hline

$f_{\star}$         & 0.1266 & 0.0687 & 0.0768 & 0.1931 & 0.1528 & 0.1236 \\
$V_{\rm{C}}$ [km/s] & 0.0572 & 0.0187 & 0.0211 & 0.1725 & 0.1828 & 0.0905 \\
$f_{\rm{X}}$        & 0.7845 & 0.9113 & 0.7805 & 1.0742 & 1.1113 & 0.9323 \\
$\alpha$            & 0.2220 & 0.2179 & 0.2181 & 0.2235 & 0.2276 & 0.2218 \\
$E_{\rm{min}}$ [keV]& 0.7572 & 0.8882 & 0.8236 & 0.9760 & 1.0110 & 0.8912 \\
$\tau$              & 0.0107 & 0.0078 & 0.0080 & 0.0150 & 0.0151 & 0.0113 \\
$R_{\rm{mfp}}$ [Mpc]& 0.1011 & 0.0751 & 0.0736 & 0.1118 & 0.1104 & 0.0944 \\
\hline
\end{tabular}
\caption{The median (over 639 test models) of the actual error ($|P_{\rm{true}} - P_{\rm{predicted}}|$) for each parameter for 5-fold cross validation. As before, all the parameter values are in $\log_{10}$ except $\alpha$ and $E_{\rm{min}}$. Here we show the results for all three cases: ideal dataset (top), mock SKA data (middle) and SKA thermal noise (bottom). }\label{tab:table_median}
\end{table}

The actual error with mock SKA data and with SKA thermal noise are nearly identical, with the mock SKA case increasing the error by up to $10\%$ (but much less in most cases). The errors are substantially larger compared to the ideal dataset, with the mock SKA case giving a median error of 24\% in $V_C$, 2.8\% in $\tau$, 34\% in $f_{\star}$, factors of 1.25 in $R_{\rm{mfp}}$ and 9.6 in $f_X$, and errors in the linear parameters of $\pm 0.22$ in $\alpha$ and $\pm 0.91$~keV in $E_{\rm min}$.
Of course, currently our knowledge of most of these parameters is uncertain by large factors (orders of magnitude in some cases), so these types of constraints would represent a remarkable advance.

\begin{figure*}
    \centering
    \begin{minipage}{0.4\textwidth}
       \includegraphics[scale=0.42]{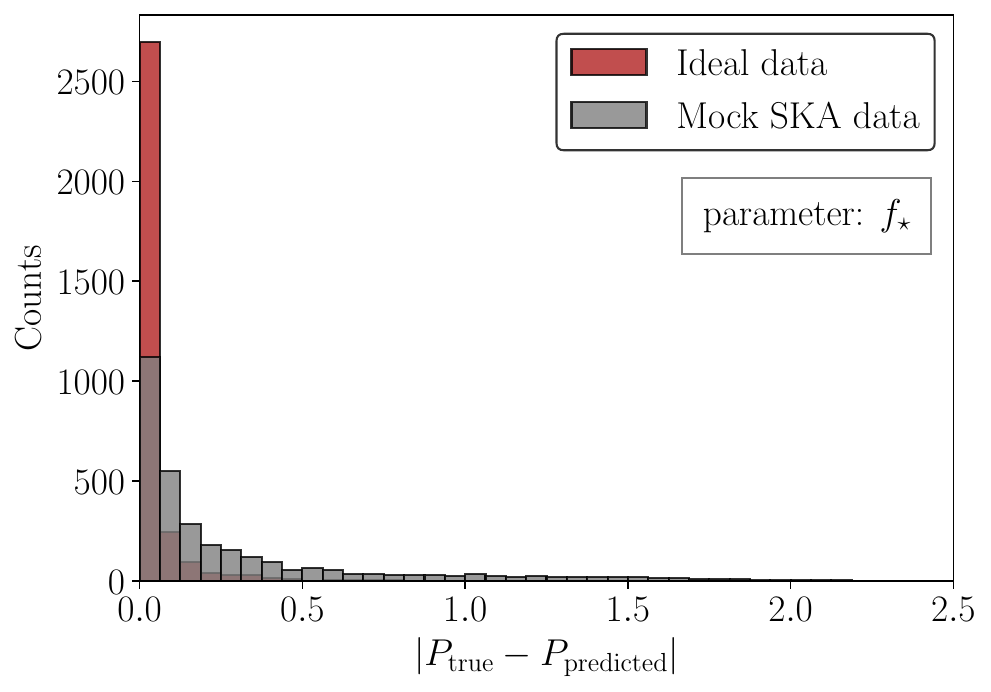}

    \end{minipage}
    \begin{minipage}{0.4\textwidth}
      \includegraphics[scale=0.42]{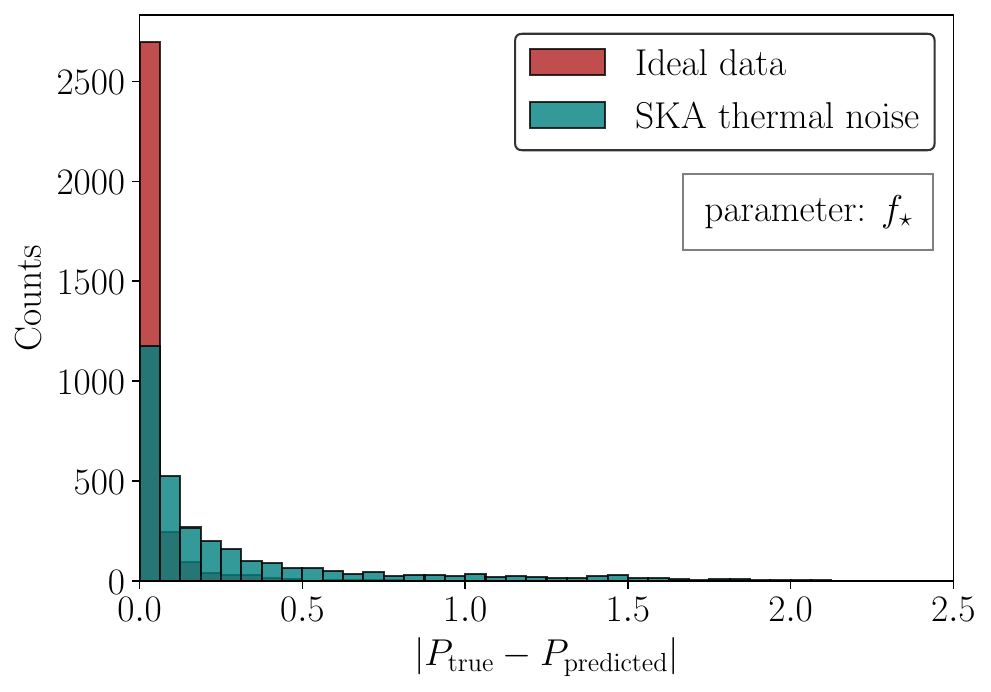}
    
    \end{minipage}
    \centering
    \begin{minipage}{0.4\textwidth}
       \includegraphics[scale=0.42]{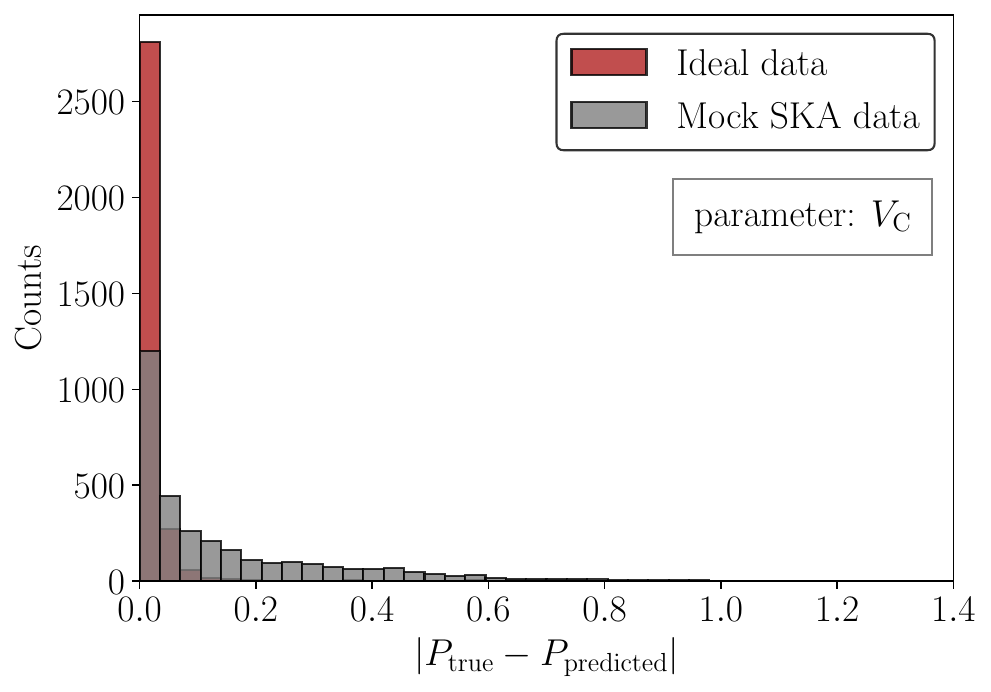}
       
    \end{minipage}
    \begin{minipage}{0.4\textwidth}
      \includegraphics[scale=0.42]{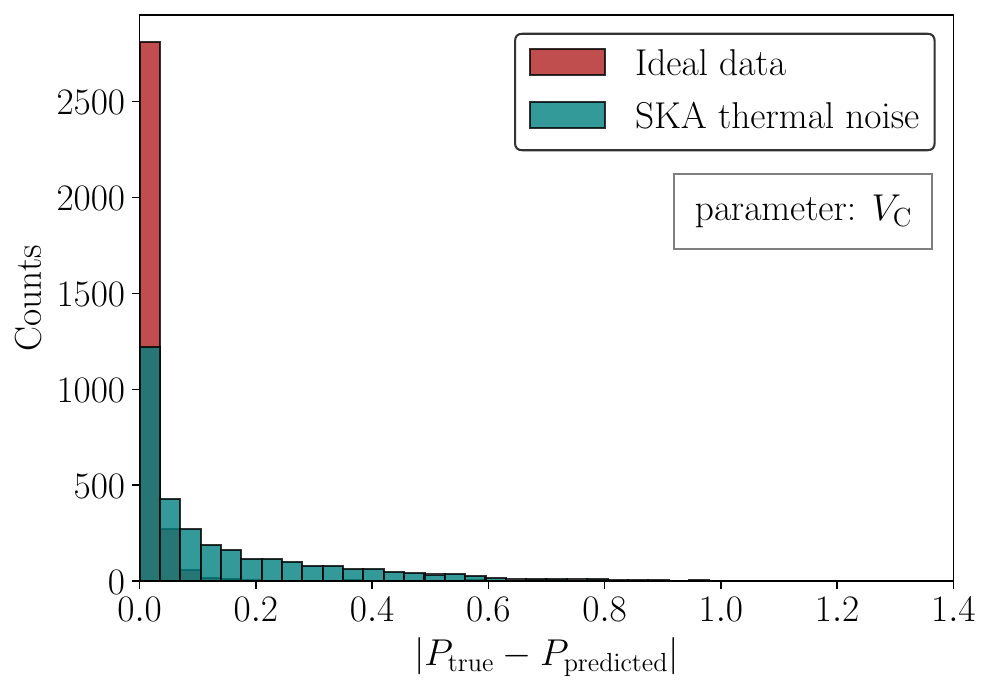}
      
    \end{minipage}
    \centering
    \begin{minipage}{0.4\textwidth}
       \includegraphics[scale=0.42]{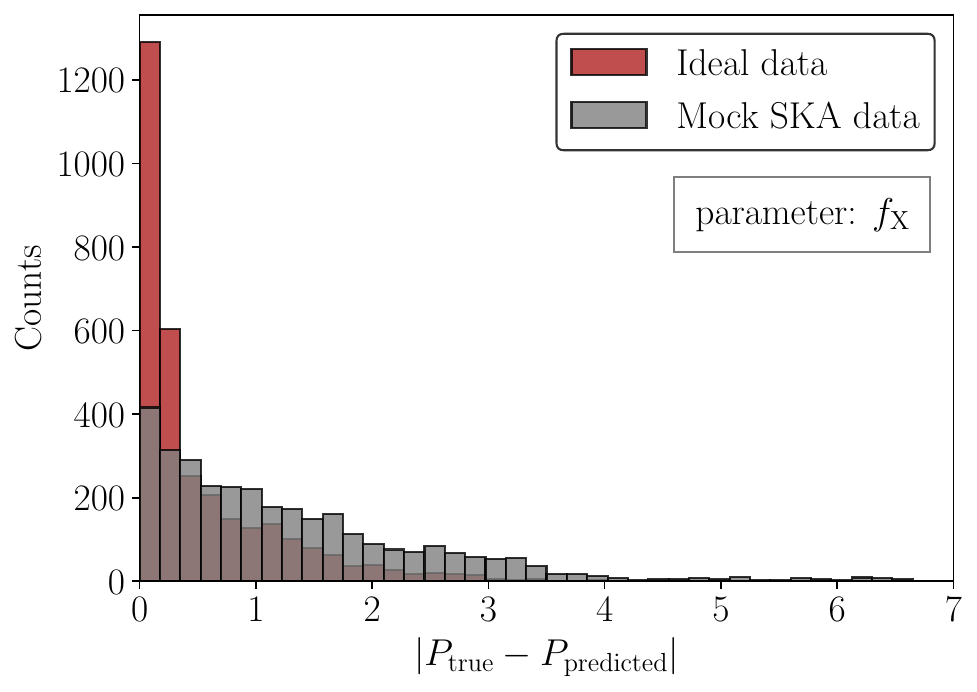}
    \end{minipage}
    \begin{minipage}{0.4\textwidth}
      \includegraphics[scale=0.42]{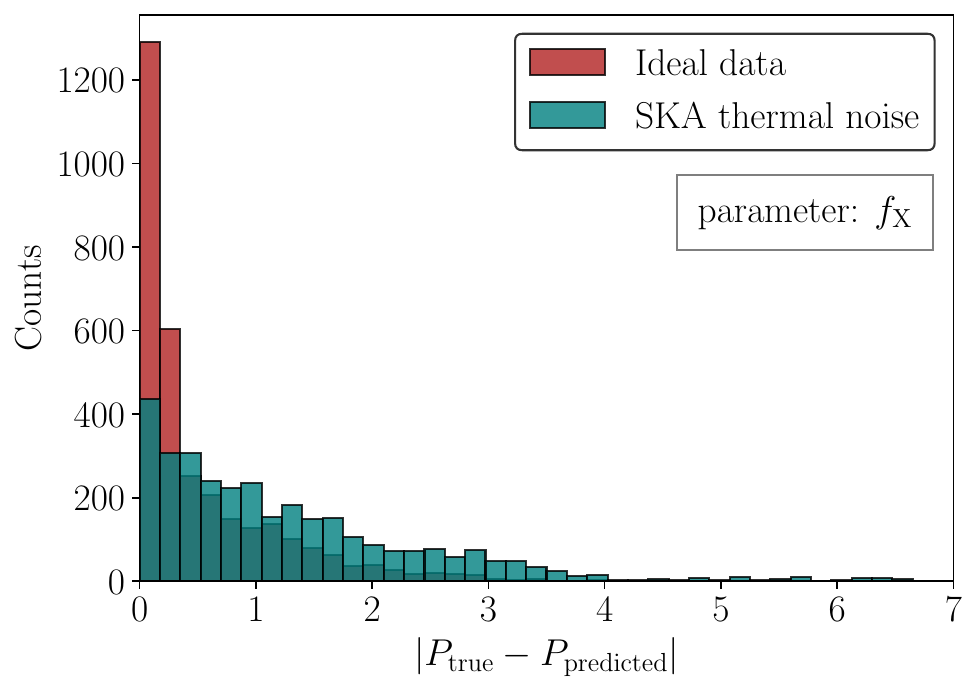}
    \end{minipage}
\caption{Histogram of the actual errors ($|P_{\rm{true}} - P_{\rm{predicted}}|$) in predicting the parameters: $f_{\star}$, $V_C$ and $f_X$. The distribution is shown over all the 5-folds. The parameter values here are in $\log_{10}$.}\label{fig: sigma_prime1}
\end{figure*}

\subsection{Classification of the radio backgrounds}

As noted in the introduction, the possible observation of the absorption profile of the 21-cm line centered at 78 MHz with an amplitude of $-500$ K by the EDGES collaboration is incompatible with the standard astrophysical prediction. One of the possible explanations for this unexpected signal is that the excess radio background above the CMB enhances the contrast between the spin temperature and the background radiation temperature. \citet{fialkov19} considered a uniform external radio background (not related to the astrophysical sources directly), with a synchrotron spectrum of spectral index $\beta = -2.6$ and amplitude parameter $A_{\rm{r}}$ measured relative to the CMB at the reference frequency of 78 MHz. Another potential model for the excess radio background is that it comes from the high redshift radio galaxies. The effect of the inhomogeneous galactic radio background on the 21-cm signal has been explored by \citet{Reis2020}. They used the galactic radio background model to explain the unexpected EDGES low band signal. In our work, we use both the external and galactic radio models and train a neural network to try to infer the type of the radio background given the 21-cm power spectrum. For this purpose, we create a training dataset of 9500  models (where there are $\sim 5000$ models with a galactic radio background and $\sim 4500$ models with an external radio background), with the astrophysical parameters varying over the following ranges: $f_{\star} = 0.01 - 0.5$, $V_C = 4.2 - 60$ km s$^{-1}$, $f_X = 0.0001 - 1000$, $\alpha = 1.0 - 1.5$, $E_{\rm{min}} = 0.1 - 3.0$, $\tau = 0.033 - 0.089$, $R_{\rm{mfp}} = 10.0 - 70.0$. For the models with a galactic radio background, the normalization of the radio emissivity (measured relative to low-redshift galaxies), $f_{\rm{R}}$, varies over the range $f_R = 0.01 - 10^7$, and the range for the amplitude of the radio background, $A_{\rm{r}}$, for the external radio models is is $0.0001 - 0.5$.

\begin{figure*}
    \centering
    \begin{minipage}{0.4\textwidth}
       \includegraphics[scale=0.5]{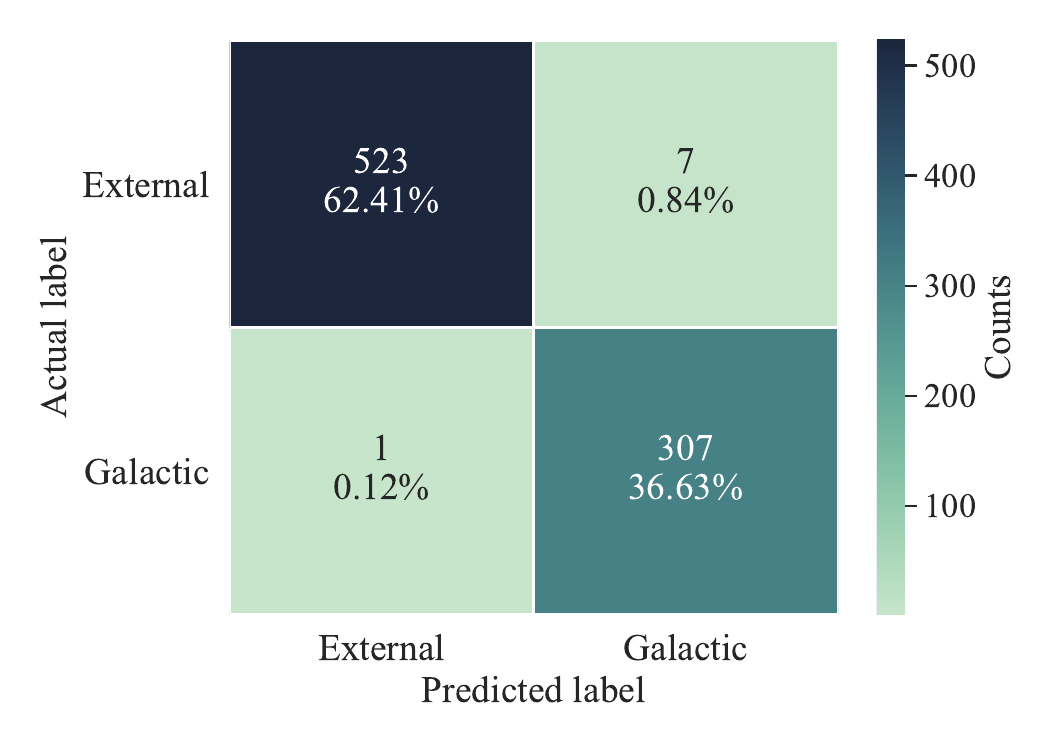}
    \end{minipage}\hfill
    \begin{minipage}{0.48\textwidth}
       \includegraphics[scale=0.5]{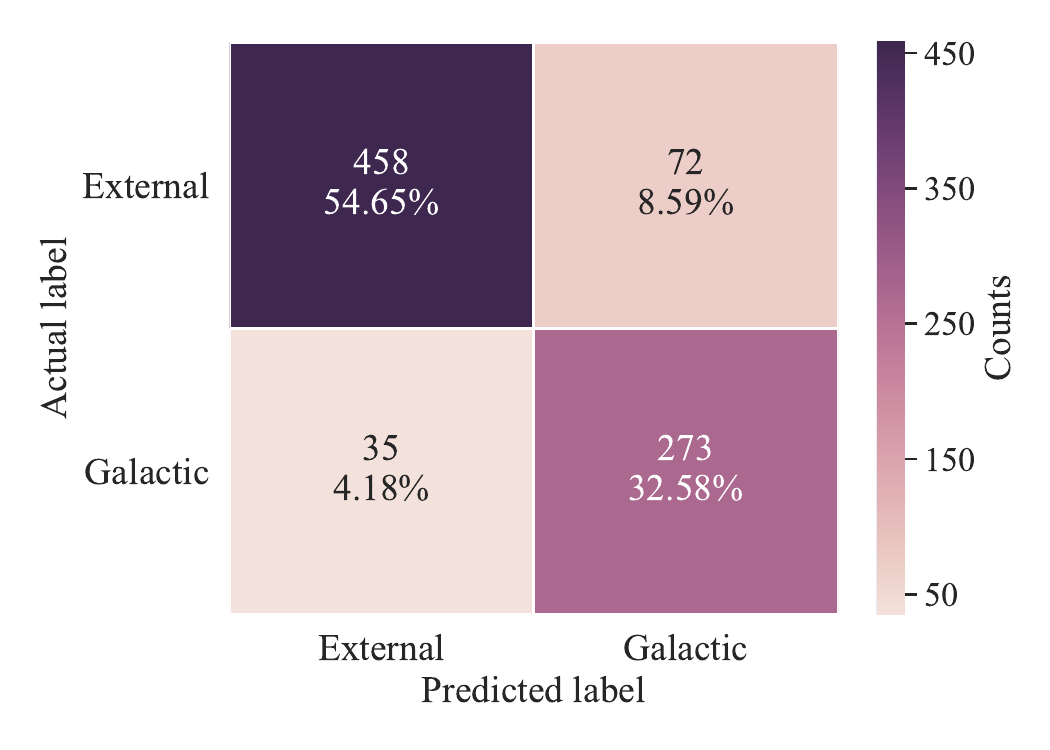}
    \end{minipage}
    \caption{Confusion matrix depicts the performance of our classification method to distinguish the type of the radio background given the 21-cm power spectrum. Left panel: ideal dataset. $62.41\%$ of the test models, i.e., 523 models out of 838 models are labeled as having the external radio background and $36.63\%$ of the test models, i.e., 307 models out of 838 models are labeled as having the galactic radio background. The number of misclassified cases is 8, i.e., eight models have been classified as having the other radio background rather than the true radio background. In this case, the overall accuracy is $99\%$. Right panel: When we train the ANN using the 21-cm power spectrum from the mock SKA dataset, the overall accuracy is $87.2\%$, as seen in the confusion matrix.}\label{fig: confusion_matrix}
\end{figure*}

We apply an EDGES-compatible test dataset to the two trained networks. The models that we refer to as EDGES-compatible satisfy the criteria adopted by \citet{fialkov19} as representing a rough compatibility with the 99\% limits of the detected signal in the EDGES low band experiment, in terms of the overall decline and rise without regard to the precise shape of the absorption (which is much more uncertain). 
The enhanced radio emission must strictly be a high redshift phenomena, in order to not over-produce the observed radio background \citep{fialkov19}, so we assume a cut-off redshift, $z_{\rm{cutoff}} = 15$ \citep{Reis2020} below which $f_R = 1$ as for present-day radio sources. So we only consider here redshifts from 15 to 30 (or the highest SKA redshift in the case with SKA noise). In our training dataset, we treat the radio background parameters $A_r$ or $f_R$ on an equal footing and add an extra column of a binary parameter that specify the type of radio background: 0 for the external radio background and 1 for the galactic radio background. In our EDGES compatible test dataset, we have 530 models and 308 models with an external and a galactic radio background, respectively. We apply this test dataset to the trained NN. In the predicted parameters, we round off the binary parameter either to zero (when it is $\leq 0.5$)  which is the label for the external radio background, or to unity (when it is $> 0.5$) which is the label of the galactic radio background. The confusion matrix shown in Fig.~\ref{fig: confusion_matrix} indicates the performance of our classification method for identifying the type of radio background. In the case without noise, the accuracy is 99\%. The information available in the whole $k$ range, i.e., 0.05 Mpc$^{-1}$ $< k < $1.0 \ Mpc$^{-1}$,  helps yield this high classification accuracy, despite the level of emulation errors as seen in Fig.~\ref{fig: error_k_z}. The classification accuracy drops to $87.2\%$ if we use the power spectra from the mock SKA dataset with excess radio background; the accuracy remains fairly high as these EDGES-inspired models have high 21-cm power spectra that are not so strongly affected by the SKA thermal noise. 

\subsection{Accuracy of fitting the excess radio background models}

\begin{figure*}
    \centering
    \begin{minipage}{0.4\textwidth}
       \includegraphics[scale=0.35]{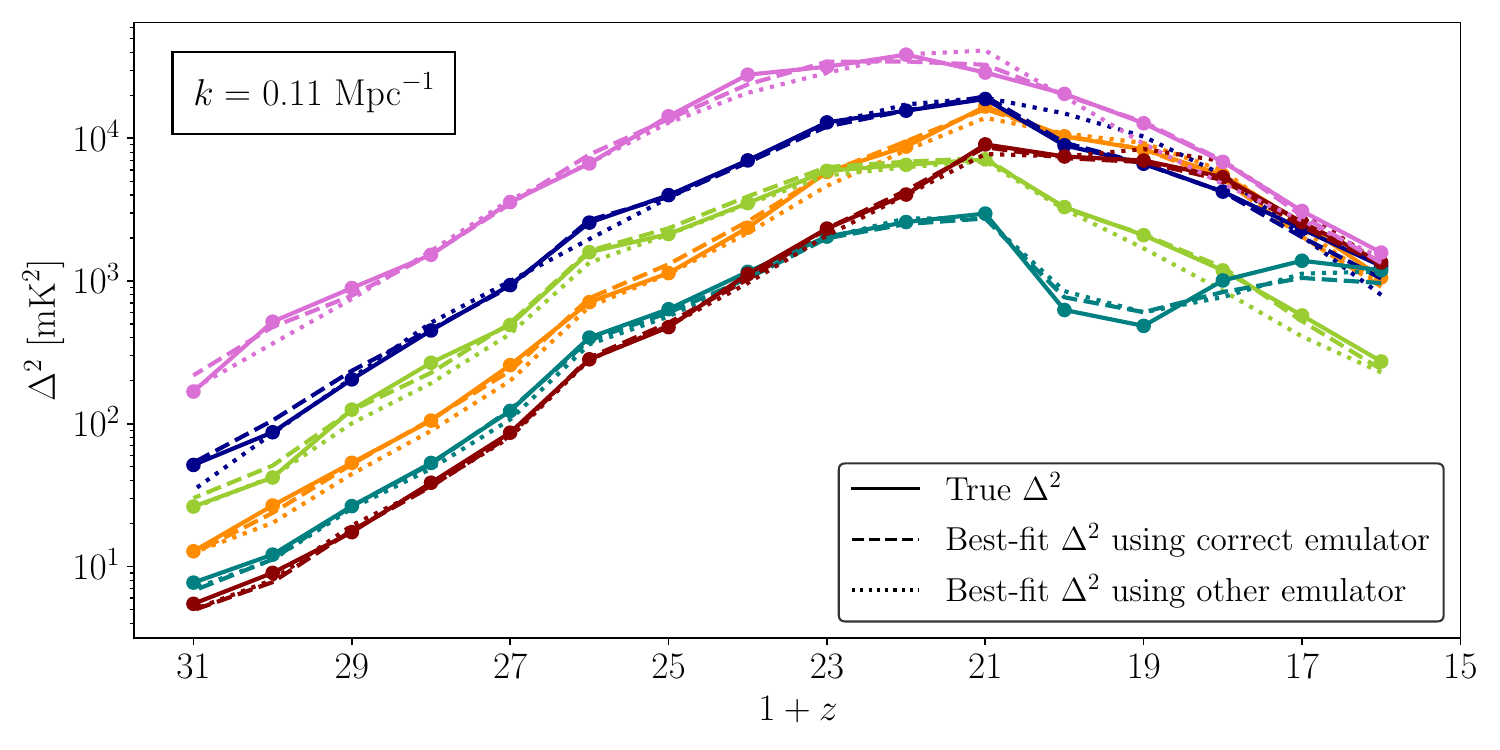}
    \end{minipage}\hfill
    \begin{minipage}{0.5\textwidth}
       \includegraphics[scale=0.35]{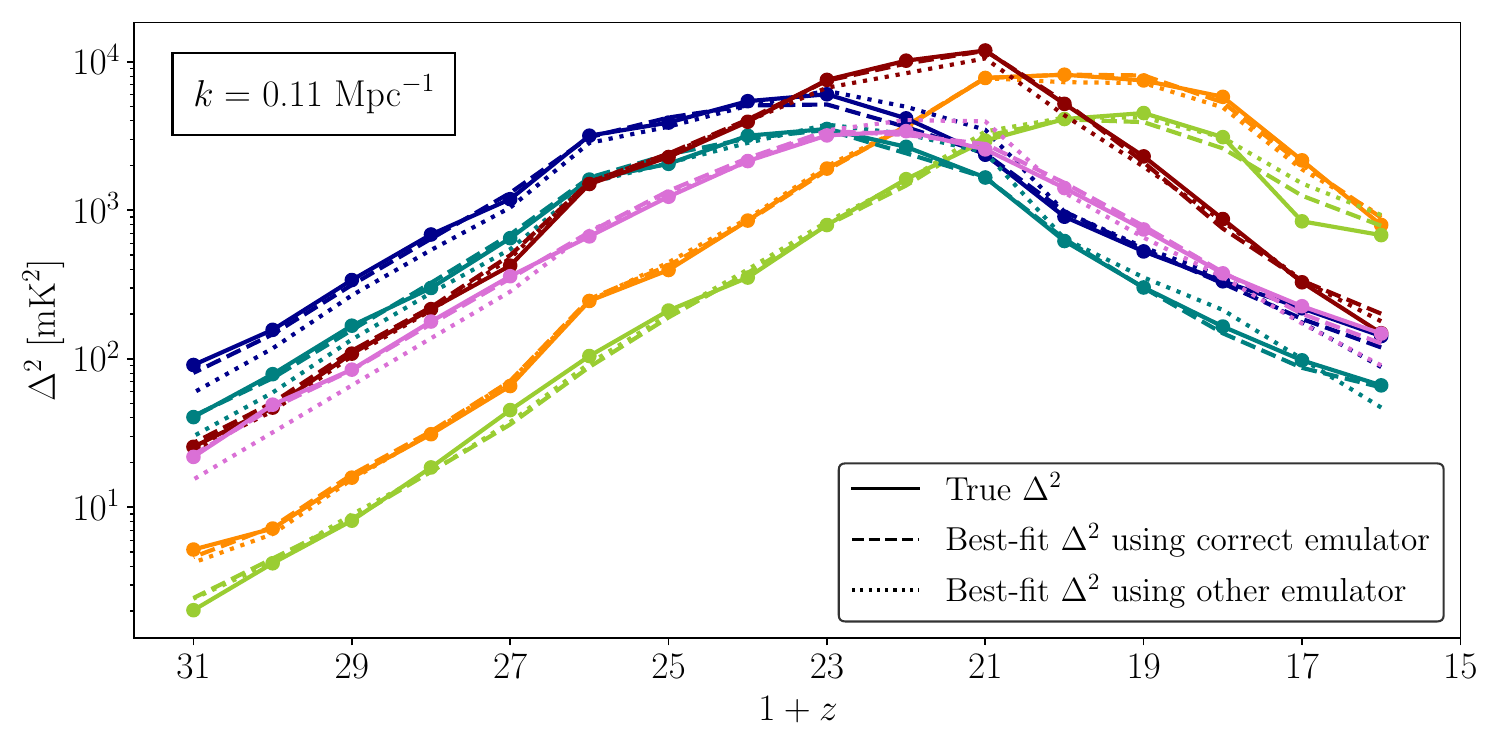}
    \end{minipage}
    \caption{A few examples of the comparison of the emulated power spectra and the true power spectra with the two different models for an excess radio background. Left panel: The true power spectrum comes from the galactic radio background model, and the other emulator uses the external radio background model. Right panel: The true power spectrum comes from the external radio background, and the other emulator uses the galactic radio background model. Shown here is $k = 0.11$ Mpc$^{-1}$. Different colors refer to different models (with no relation between the models in the two panels).}\label{fig: fitting_example1}
\end{figure*}

In the previous subsection we found that our NN works well to infer the type of  radio background present in the 21-cm power spectrum. In order to understand cases of misclassification, we now ask whether we can fit a model with the galactic radio background with parameters ($f_{\star}$, $V_C$, $f_X$, $\alpha$, $E_{\rm{min}}$, $\tau$, $R_{\rm{mfp}}$, $A_r$) of the external radio background model. To address this question, we train an NN that can predict the parameters in the parameter space of an external radio background given the 21-cm power spectra with an excess radio background. We also construct an emulator and train it using the dataset with the external radio background. We then employ our trained emulator in the MCMC sampler. If we apply a dataset generated with a galactic background, the NN will find the approximate best-fit parameters in the parameter space of the external radio background, and we use this predicted set of parameters from the NN as the initial guess in the MCMC sampler. The output of the MCMC sampler is the posterior distribution of the best-fit parameters. We take the median of the distribution for each of the parameters and report it as the predicted best-fit parameter value. We use these best-fit parameter values in the emulator trained on the external radio background dataset to emulate the 21-cm power spectrum (which, again, is actually based on the galactic radio background model). The left panel of the Fig.~\ref{fig: fitting_example1} shows a few random examples of the quality of fitting the galactic models with an external radio background. In the plot, the solid line is the true power spectrum with the galactic radio background, the dashed line is the best-fit emulated power spectrum using the correct emulator (galactic) and the dotted line is the best-fit emulated power spectrum using the other emulator (external). The right panel of Fig.~\ref{fig: fitting_example1} shows a few examples of a similar setup except that the roles of the two types of radio models have been reversed. 

\begin{figure*}
    \centering
    \begin{minipage}{0.4\textwidth}
       \includegraphics[scale=0.42]{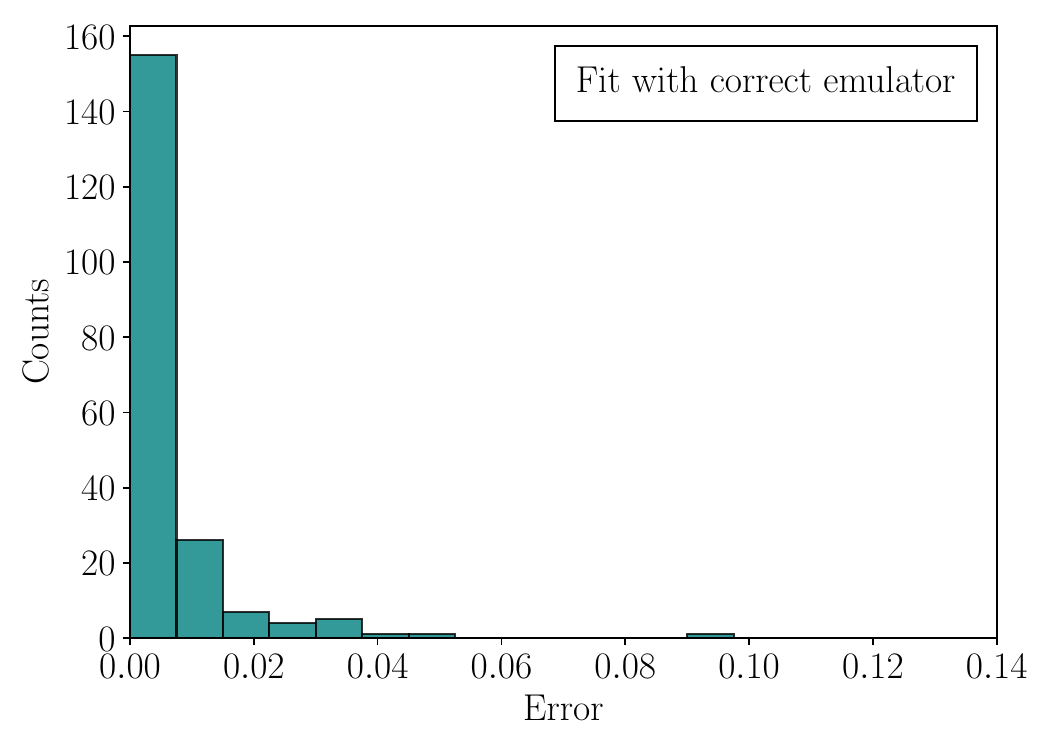}
    \end{minipage}
    \begin{minipage}{0.4\textwidth}
      \includegraphics[scale=0.42]{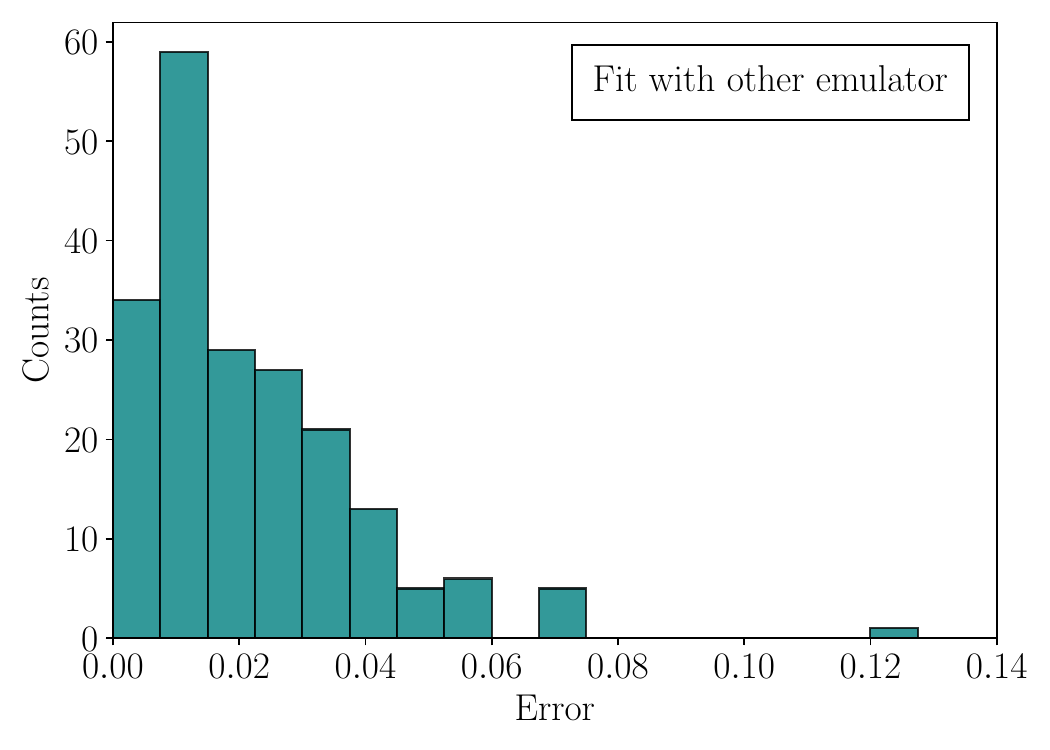}
    \end{minipage}
    \centering
    \begin{minipage}{0.4\textwidth}
       \includegraphics[scale=0.42]{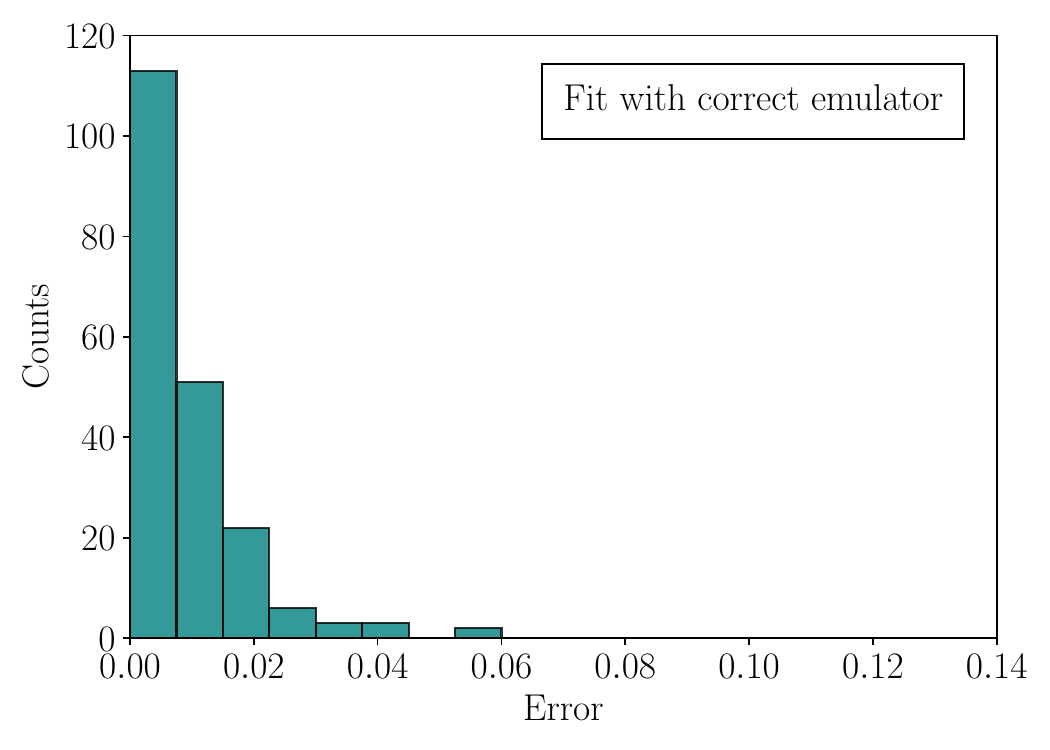}
    \end{minipage}
    \begin{minipage}{0.4\textwidth}
      \includegraphics[scale=0.42]{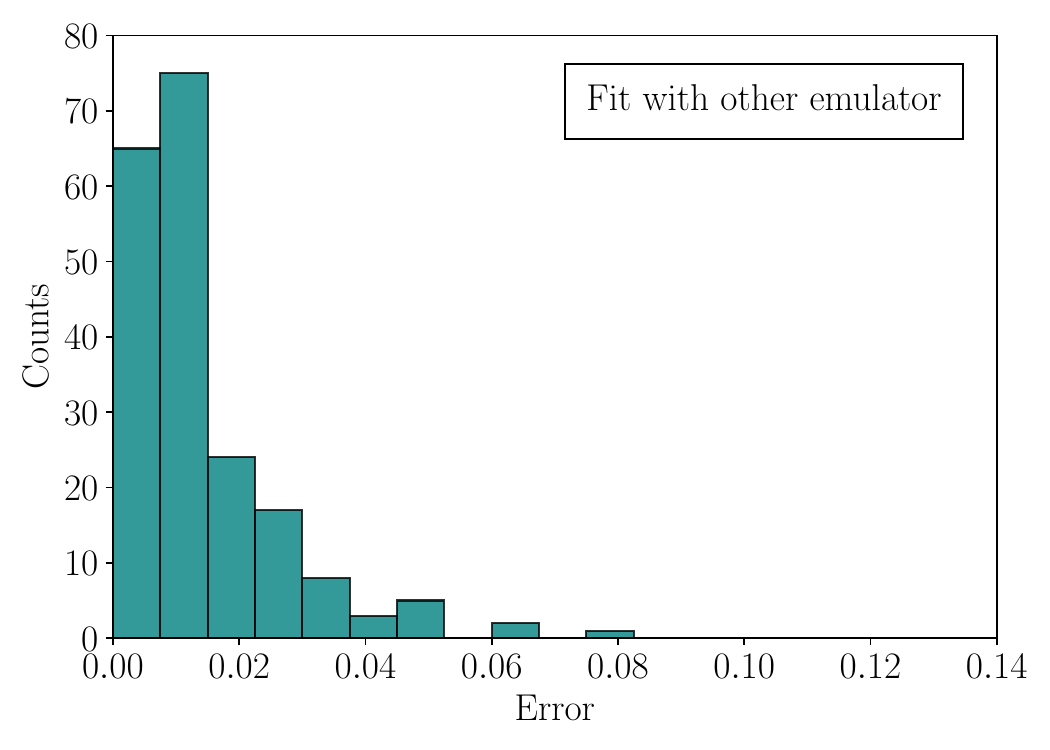}
    \end{minipage}
    \caption{Histogram of errors showing the overall performance of the fitting procedure. Top panels: We use a test dataset of 200 models with an external radio background. In the top left panel, the histogram shows the error when we employ the emulator trained using the 21-cm power spectra with the external radio background. In the top right panel, the histogram shows the error when we use the emulator trained using 21-cm power spectra with the wrong, galactic radio background model. Bottom panels: We use a test dataset of 200 models with a galactic radio background. In the bottom left panel, the histogram represents the error when we apply the emulator trained using the 21-cm power spectra with a galactic radio background. The histogram in the bottom right panel shows the error when we use the emulator trained using the 21-cm power spectra with the wrong, external radio background model. Here we consider the 21-cm power spectra from the ideal dataset.}\label{fig: emulator_performance_radio}
\end{figure*}

To test the overall, statistical performance of the emulators in fitting power spectra with different radio backgrounds, we estimate the relative error using the equation 
\begin{equation}\label{eq: emulator__error__}
\rm{Error} = \frac{\sqrt{\rm{Mean\left[\left(\Delta^2_{\rm{predicted}} - \Delta^2_{\rm{true}}\right)^2\right]}}}{Max\left[\Delta^2_{\rm{true}}\right]}\ .
\end{equation}
As explained before, this yields a simple, optimistic estimate of the overall error in a simple number. It suffices here for our interest in comparing various cases, as follows:\\
Case A: We have a test dataset of 200 models with an external radio background. We fit the power spectra of the test dataset using the emulator trained with power spectra with the external radio background (i.e., the correct model in this case). The top left panel of Fig.~\ref{fig: emulator_performance_radio} shows the histogram of errors in the fitting procedure. If we compare the true and predicted power spectra, $98.5\%$ of the cases in the test dataset give a relative error less than 0.04. \\
Case B: We fit the power spectra of the same test dataset as in case A but using the wrong emulator, trained with power spectra with a galactic radio background. The top right panel of Fig.~\ref{fig: emulator_performance_radio} shows the histogram of errors in the fitting procedure. In this case, we find that the relative error is still lower than 0.04 for $87\%$ of cases.\\
Case C: We use a test dataset of 200 models with a galactic radio background. We fit the power spectra of the test dataset using the correct emulator, i.e., trained using the power spectra with a galactic radio background. The histogram of errors is shown in the bottom left panel of Fig.~\ref{fig: emulator_performance_radio}.  Here the relative error is lower than 0.04 for $98.5\%$ cases.\\
Case D: We fit the power spectra of the same test dataset as in case C but using the wrong emulator, trained with the external radio background model. The bottom right panel of Fig.~\ref{fig: emulator_performance_radio} shows the histogram of error for this case. We find that the relative error is still lower than 0.04 in $95\%$ of cases.

\begin{figure*}
    \centering
    \begin{minipage}{0.4\textwidth}
       \includegraphics[scale=0.42]{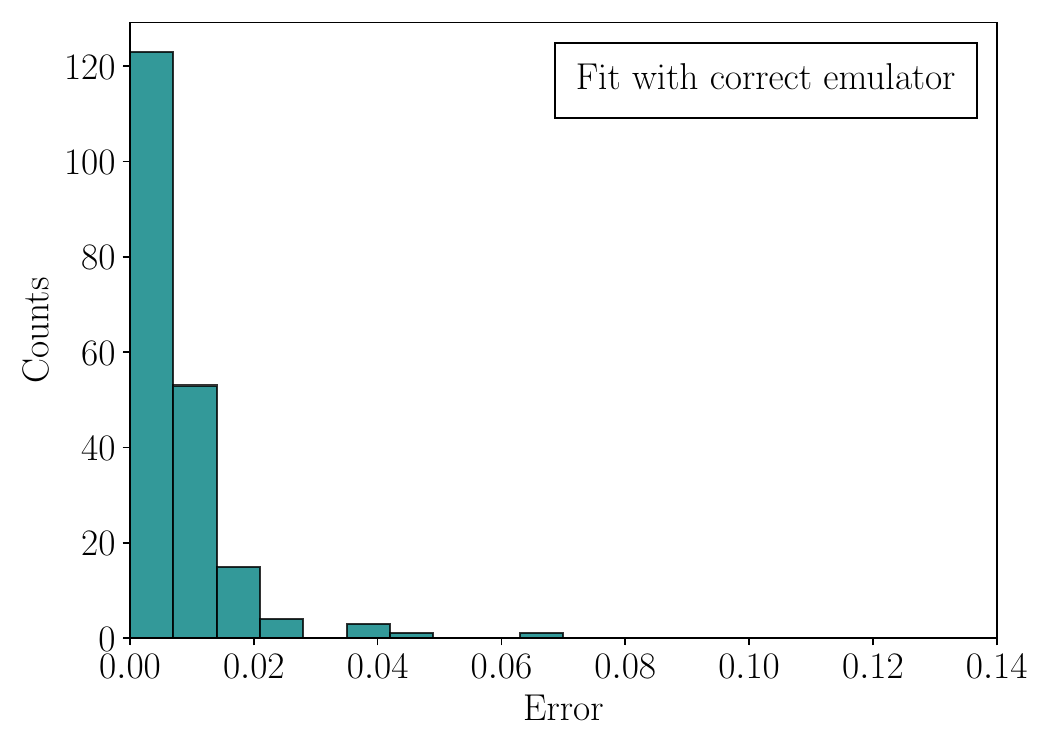}
    \end{minipage}
    \begin{minipage}{0.4\textwidth}
      \includegraphics[scale=0.42]{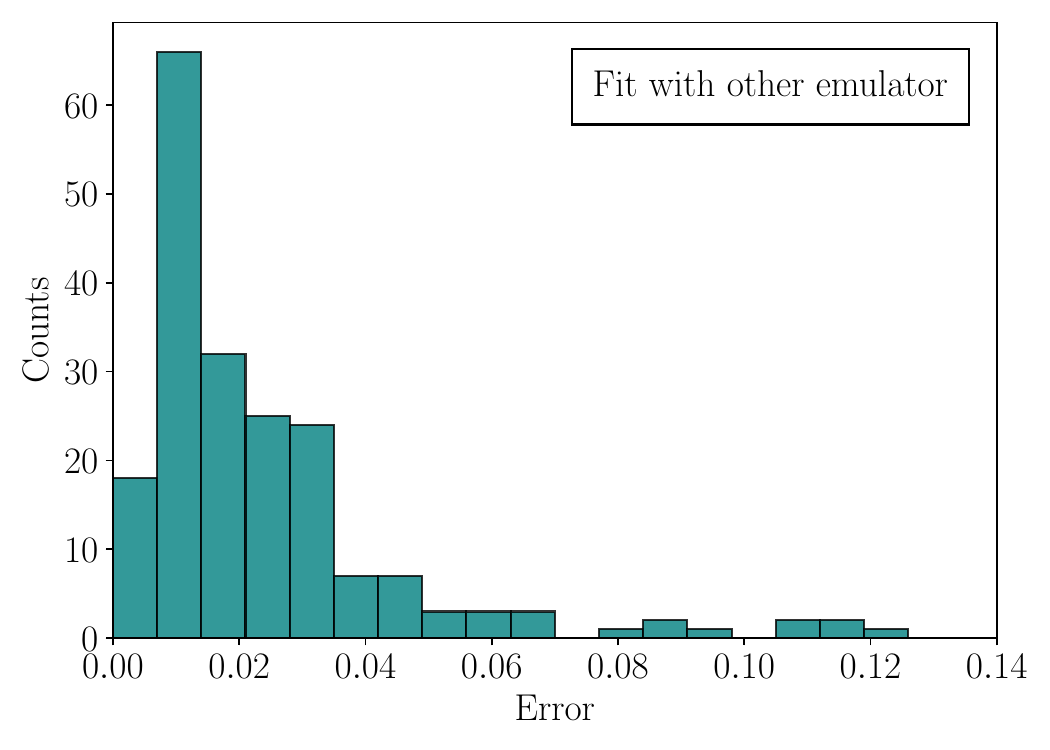}
    \end{minipage}
    \centering
    \begin{minipage}{0.4\textwidth}
       \includegraphics[scale=0.42]{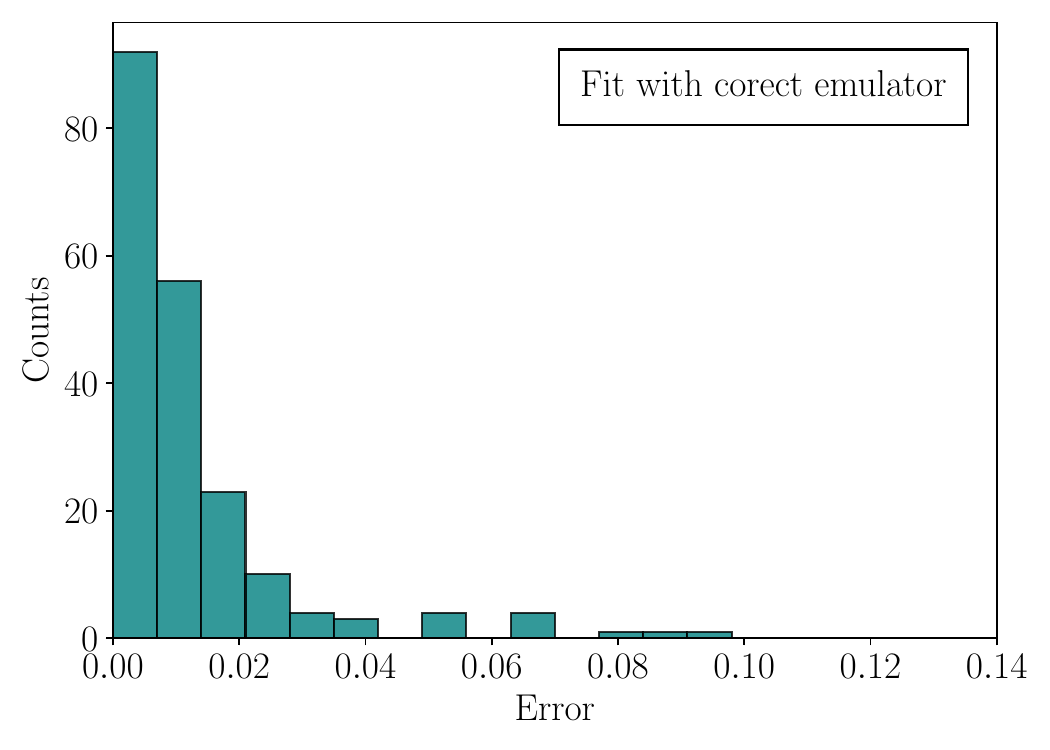}
    \end{minipage}
    \begin{minipage}{0.4\textwidth}
      \includegraphics[scale=0.42]{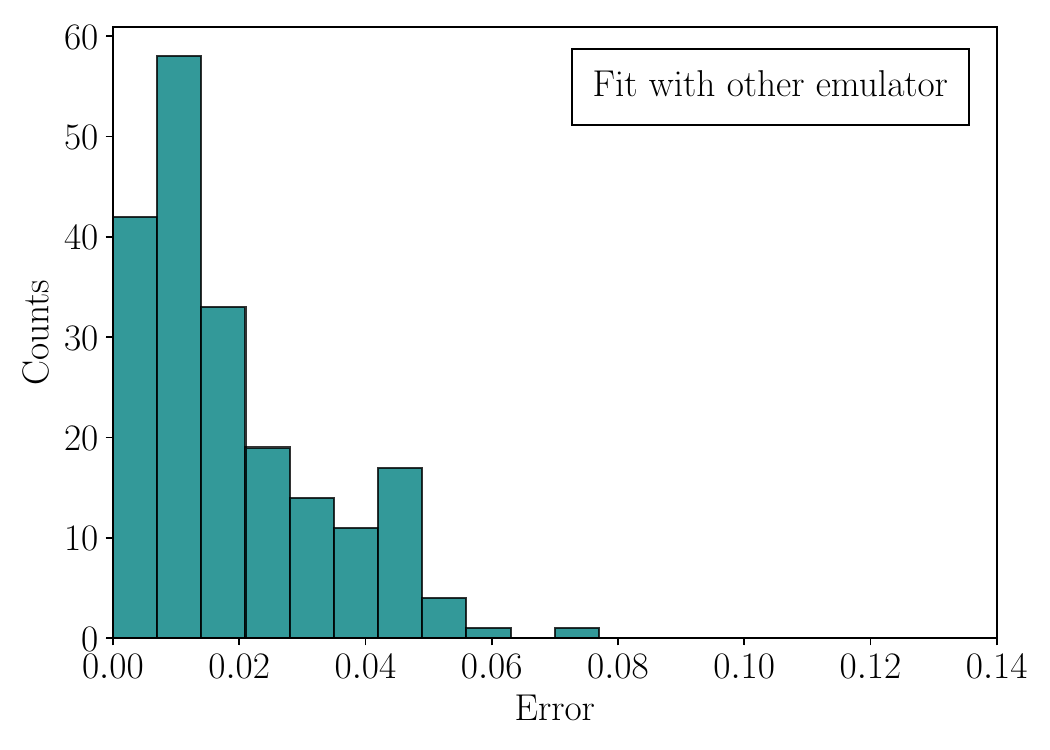}
    \end{minipage}
    \caption{Same as Fig.~\ref{fig: emulator_performance_radio}, except using mock SKA dataset. In each case we use the best-fit parameters derived from NN's trained using mock SKA data, but for the error we measure the prediction of the clean power spectrum, i.e., we apply the emulators (both external and galactic) trained using ideal dataset to emulate the best-fit 21-cm power spectrum, and compare this to the true clean power spectrum.}\label{fig: emulator_performance_radio_SKA}
\end{figure*}

Table~\ref{tab:table_radio} lists the mean and the median of the relative errors in the various cases mentioned above. The errors increase significantly when the incorrect emulator is used, by factors between 1.5 and 4 in the various cases. Fig.~\ref{fig: emulator_performance_radio_SKA} and Table~\ref{tab:table_radio_SKA} show similar results to Fig.~\ref{fig: emulator_performance_radio} and Table~\ref{tab:table_radio}, but using the mock SKA power spectrum. SKA noise increases the errors, but typically by only tens of percent, up to a factor of 1.5. 
This is generally reminiscent of the magnitude of the effect of SKA noise that we saw in a different context in Fig.~\ref{fig: error_k_z}.

\begin{table}
\centering
\begin{tabular}{ccccc} 
\hline
Case  & Excess background type   & \ \ Emulator \ \ \           & \ Mean\ \  & Median  \\ 
\hline\hline
A  & External  & External   & 0.0071  & 0.0043     \\
B  & External  & Galactic   & 0.0217  & 0.0159     \\
C  & Galactic  & Galactic   & 0.0092  & 0.0060    \\
D  & Galactic  & External   & 0.0142  & 0.0109     \\

\hline
\end{tabular}
\caption{The mean and median of the error (using equation \ref{eq: emulator__error__}) in the fitting procedure. External emulator: The emulator trained using the 21-cm power spectra with an external radio background. Galactic emulator: The emulator trained using the 21-cm power spectra with a galactic radio background. Here we use the 21-cm power spectrum from the ideal dataset.}\label{tab:table_radio}
\end{table}

\begin{table}
\centering
\begin{tabular}{ccccc} 
\hline
Case  & Excess background type   & \ \ Emulator \ \ \           & \ Mean\ \  & Median  \\ 
\hline\hline
A  & External  & External   & 0.0079  & 0.0056     \\
B  & External  & Galactic   & 0.0265  & 0.0170     \\
C  & Galactic  & Galactic   & 0.0136  & 0.0075    \\
D  & Galactic  & External   & 0.0188  & 0.0140    \\

\hline
\end{tabular}
\caption{Same as table \ref{tab:table_radio}, but here we use the 21-cm power spectra from the mock SKA dataset. As in Fig.~\ref{fig: emulator_performance_radio_SKA}, the errors are measured on the ability to predict the clean power spectrum.}\label{tab:table_radio_SKA}
\end{table}

\section{Conclusions}\label{sec: conclusions}

In this work, we applied  machine learning techniques to analyze a dataset of power spectra from mock 21-cm observations. We developed a numerical emulation that is a computationally efficient method that speeds up the data analysis by bypassing the need to run very large numbers of semi-numerical simulations. We trained our neural network over a wide range of possible values of the seven astrophysical parameters that include the star formation efficiency, the minimum circular velocity of star-forming halos, the X-ray radiation efficiency, the power law slope and low energy cutoff of the X-ray SED, the CMB optical depth from reionization, and the mean free path of the ionizing photons. We constructed our algorithm in a way that approximately accounts for emulation error (i.e., the uncertainty due to the finite size of the training set), and also tested the accuracy (and improved the error estimates) using 5-fold cross validation.

As a result we obtained an emulator that can predict the 21-cm power spectrum given the set of astrophysical parameters. We analyzed the overall performance of the emulator by comparing, for 639 test models, the emulated power spectrum to the true power spectrum generated in the semi-numerical simulation. We showed how the error varies with wavenumber and redshift, using the definition in eq.~\ref{eq: error_k_z}. We found that the typical emulation error of the power spectrum in each bin is $10-20\%$ over a broad range of $k$ and $z$, but it rises above $20\%$ at the lowest and highest $k$ values (for most redshifts), and at the lowest redshift for all $k$ values (i.e., at $z = 6$, near the end of reionization). SKA noise reduces the accuracy of the reconstruction of the astrophysical parameters but not by too much, increasing the typical errors by a fairly uniform factor of $\sim 1.5$.

In order to find constraints on the astrophysical parameters, we employed the emulator to an MCMC sampler to fit parameters to a given 21-cm power spectrum. We quantified the error in predicting the parameters using equation~\ref{eq: error}. We found that the standard deviations ($\sigma$) for most of the seven parameters (Fig.~\ref{fig: histogram1} and \ref{fig: histogram2}) were within 20\% of our error estimate. Also, the mean (which measures the bias) was small in every parameter for the ideal dataset, and the distribution was fairly well described by a Gaussian. The mock SKA dataset or SKA thermal noise case gave nearly identical results to each other; the noisy data made the bias as large as $\sim 1.4\sigma$ for some of the parameters, and added significant skewness in some cases. Table~\ref{tab: table_clean_CV} (along with Table~\ref{tab: table_clean_CV_}) listed the corresponding parameters of best-fit Gaussians for each of the 5-folds, along with means of these parameters over the 5-folds. In Fig.~\ref{fig: sigma_prime1} and \ref{fig: sigma_prime2}, we showed the histograms of the size of the actual error in predicting each of the parameters, with the medians listed in Table~\ref{tab:table_median}.
We measured the parameters and uncertainties in log scale ($\log_{10}$) for all the parameters except for $\alpha$ and $E_{\rm{min}}$. In the case of the ideal dataset, we found that the emulation errors still allow the parameters $V_c$ and $\tau$ to be reconstructed with a typical accuracy of 2.3\% and 0.9\% respectively, and $f_{\star}$ to within 4.5\%. The ionizing mean free path ($R_{\rm{mfp}}$) is typically uncertain by a factor of 1.16, and $f_X$ by a factor of 1.78. For the linear parameters, the uncertainty is typically $\pm 0.19$ in $\alpha$ and $\pm 0.20$ in $E_{\rm min}$. 

Noisy SKA data only marginally affected the uncertainty in $\alpha$, indicating the importance of the emulation error and also the overall lack of sensitivity of the power spectrum to this parameter. However, SKA noise increased the errors to 24\% in $V_{C}$, 2.8\% in $\tau$, and 34\% in $f_{\star}$. The uncertainty factor increased mildly (to 1.25) in $R_{\rm{mfp}}$ but greatly (to 9.6) in $f_X$, and also in $E_{\rm min}$ (to $\pm 0.91$~keV). Currently we have almost no observational information about the values of most of the astrophysical parameters we used in this work, except for the optical depth ($\tau$) of the CMB photons. The detection of the polarized CMB signal by WMAP \citep{Spergel} and Planck \citep{planckcollaboration18} has provided constraints on the optical depth, but their precision is limited by the cosmic variance of the large-angle polarization effect on the CMB. We have shown that 21-cm power spectrum observations can potentially produce a precise measurement of this and other astrophysical parameters. Of course we have used a relatively simple astrophysical model here. If such precise measurements are indeed achieved, they will motivate comparisons with more complicated models where, for example, the various efficiencies (for star formation, X-rays, and ionizing photons) depend on redshift, halo mass, and perhaps halo environment or merger history.

In another part of this work, we applied a neural network to classify the nature of the excess radio background, if it is present in the 21-cm signal. We compared models with an external radio background (assumed to be primordial and homogeneous) to a galactic radio background (produced by the same galaxies as other radiation backgrounds, thus generally reflecting the galaxy distribution). The accuracy of the classification was $99\%$ for the 21-cm power spectrum without SKA noise (ideal dataset), going down to a still high accuracy of $87\%$ if we use the 21-cm power spectrum from the mock SKA dataset. When fitting data with either the correct emulator or the one from the other type of radio background, we found that the fits were in all cases rather accurate  (Table~\ref{tab:table_radio}). However, the errors increased significantly when the incorrect emulator was used, by factors between 1.5 and 4 in the various cases in the case without SKA noise. Adding SKA noise increased the errors, but typically by only tens of percent, up to a factor of 1.5 (Table~\ref{tab:table_radio_SKA}).

In summary, emulating and fitting the 21-cm power spectrum using artificial neural networks is a rapid and accurate method. One of the potential extensions of this work will be to improve the accuracy of the emulator, e.g., by trying to change various parameters such as the number of layers of the NN. Another possible improvement is to use the current procedure as the first step of a fit, and then zoom in on a smaller region of the parameter space in order to achieve higher accuracy (noting that we have covered a far wider range of astrophysical parameters than most similar work in the literature). A further direction is to make the astrophysical model more realistic by adding a significant number of parameters, and seeing whether the computational speed and fitting accuracy are maintained. NN's will clearly remain valuable in this field, given the highly non-linear dependence of the 21-cm power spectrum on astrophysical parameters, the wide range of possible values for these parameters, and the relative slowness of realistic simulations, even semi-numerical ones.

\section*{Acknowledgements}

This project was made possible for S.S., R.B., and I.R.\ through the support of the Israel Science Foundation (grant No. 2359/20). R.B.\ also acknowledges the support of The Ambrose Monell Foundation and the Institute for Advanced Study. AF was supported by the Royal Society University Research Fellowship. 
This research made use of: \texttt{Numpy} \citep{harris2020array}, \texttt{Scipy} \citep{2020SciPy-NMeth}, \texttt{matplotlib} \citep{Hunter:2007}, \texttt{seaborn} \citep{Waskom2021}, \texttt{getdist} \citep{getdist},
and the NASA Astrophysics Data System Bibliographic Services.
\section*{Data Availability}

 The data underlying this article will be shared on reasonable request to the corresponding author.



\bibliographystyle{mnras}
\bibliography{example} 




\appendix

\section{Posterior distribution of the seven parameter astrophysical model}

\begin{figure*}
    \centering
    \vspace*{3cm}
    \includegraphics[scale=0.4]{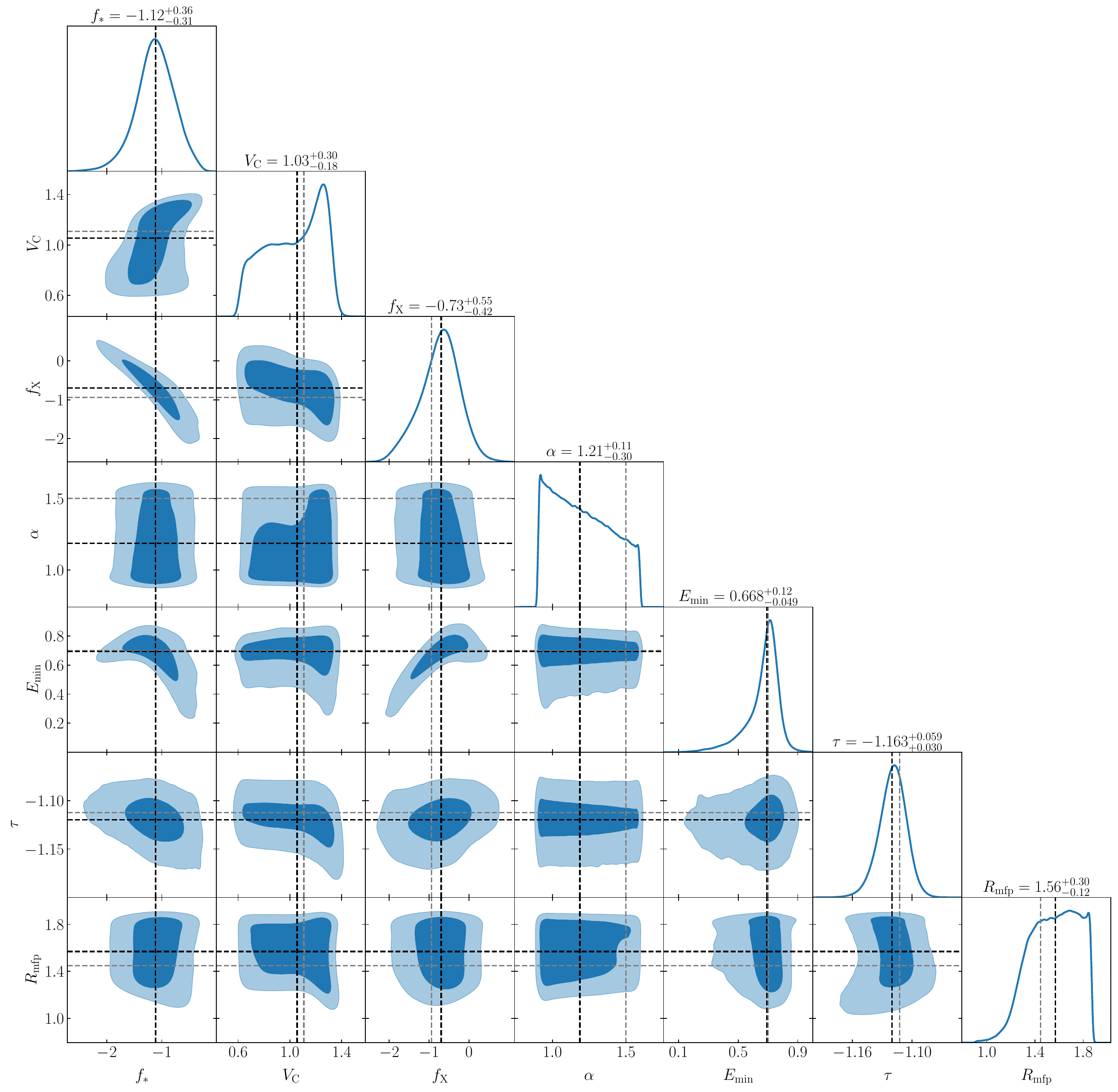}\llap{\raisebox{10cm}{\includegraphics[scale=0.5]{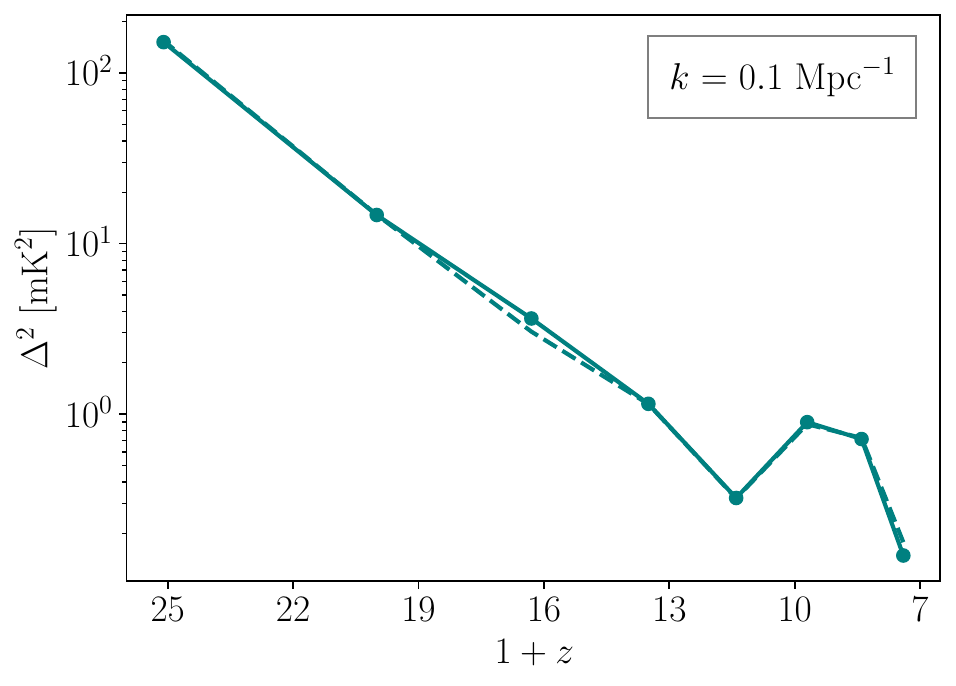}}}
    \caption{Similar to Fig.~\ref{fig: pos_distribution_clean1}. Here we show the same seven parameter astrophysical model as in Fig.~\ref{fig: pos_distribution_clean1} but using the power spectrum from the mock SKA dataset. All the parameter values are in $\log_{10}$ except $\alpha$ and $E_{\rm{min}}$. The upper right panel compares the true and reconstructed mock SKA power spectrum.}
    \label{fig: pos_distribution_SKA1}
 \end{figure*}

Fig.~\ref{fig: pos_distribution_SKA1} shows the posterior distribution of the same seven parameter astrophysical models as in Fig.~\ref{fig: pos_distribution_clean1}, but for the power spectrum from the mock SKA dataset. With SKA noise there are more distorted, non-Gaussian contours. However, all the true values of the parameters are lying well inside the MCMC error contours.

\section{Histograms of the normalized error in predicting the astrophysical parameters}

\begin{figure*}
    \centering
    \begin{minipage}{0.33\textwidth}
       \includegraphics[scale=0.37]{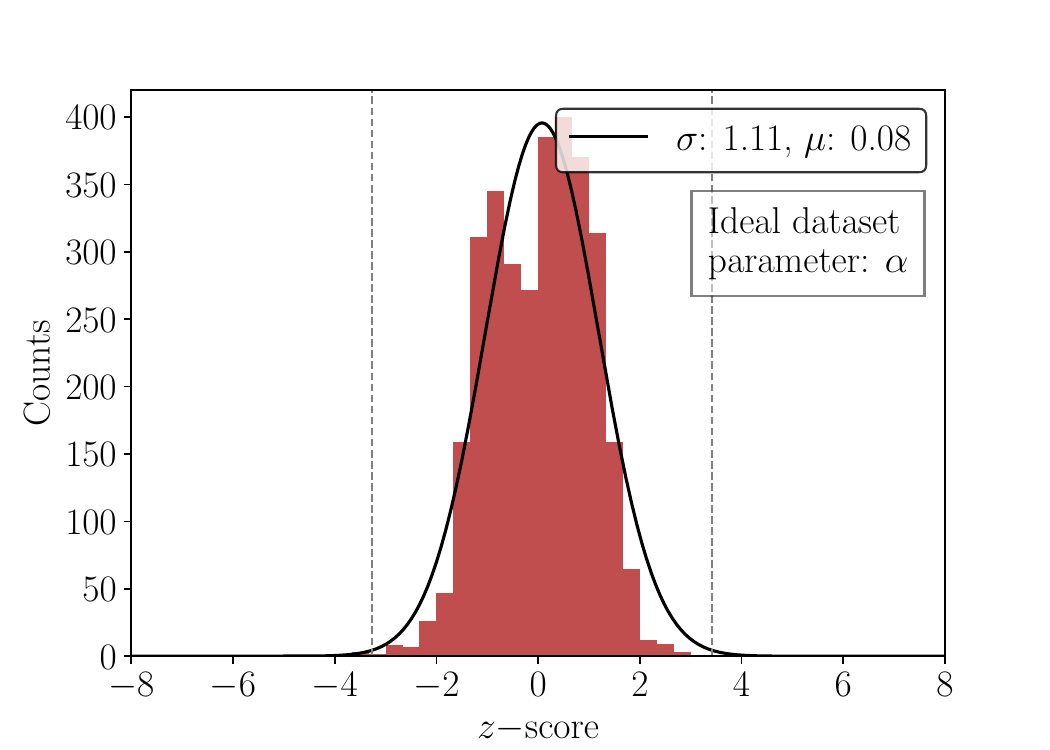}
       
    \end{minipage}\hfill
    \begin{minipage}{0.33\textwidth}
      \includegraphics[scale=0.37]{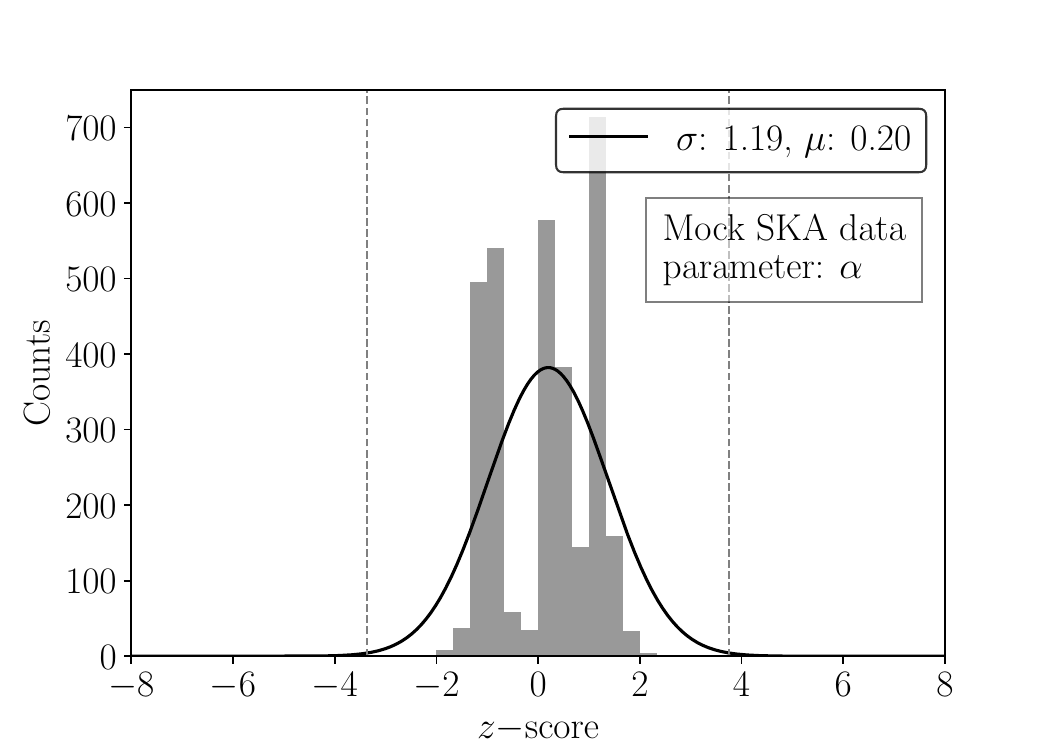}
    \end{minipage}\hfill
    \begin{minipage}{0.3\textwidth}
      \includegraphics[scale=0.37]{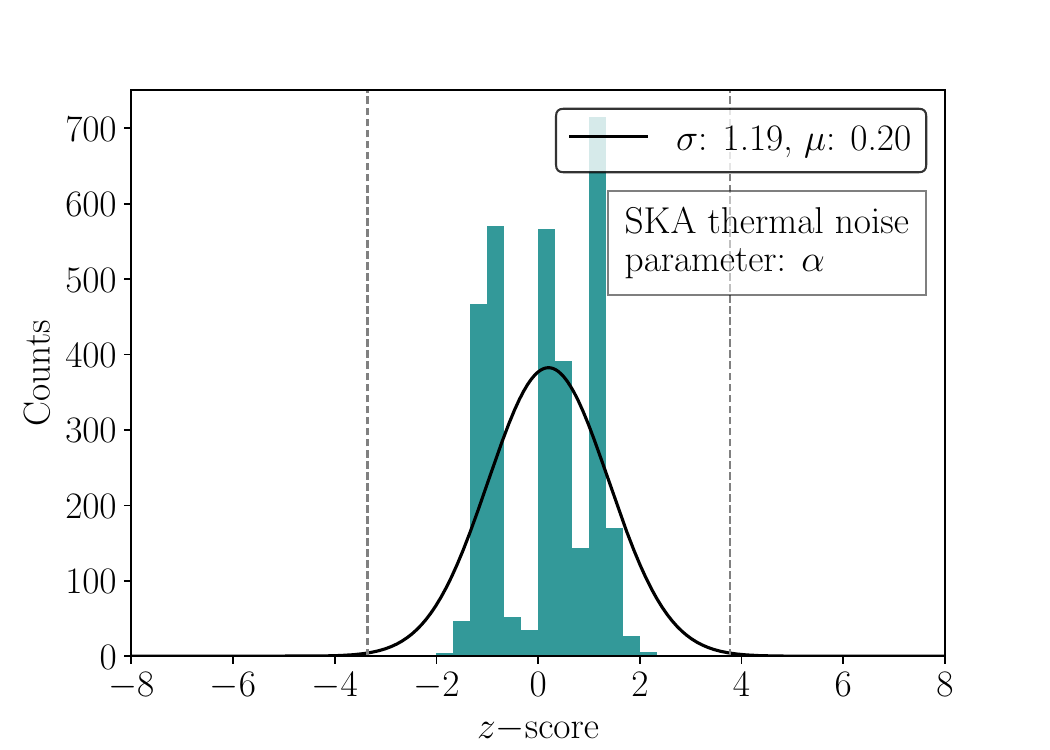}
      
    \end{minipage}
    \begin{minipage}{0.33\textwidth}
       \includegraphics[scale=0.37]{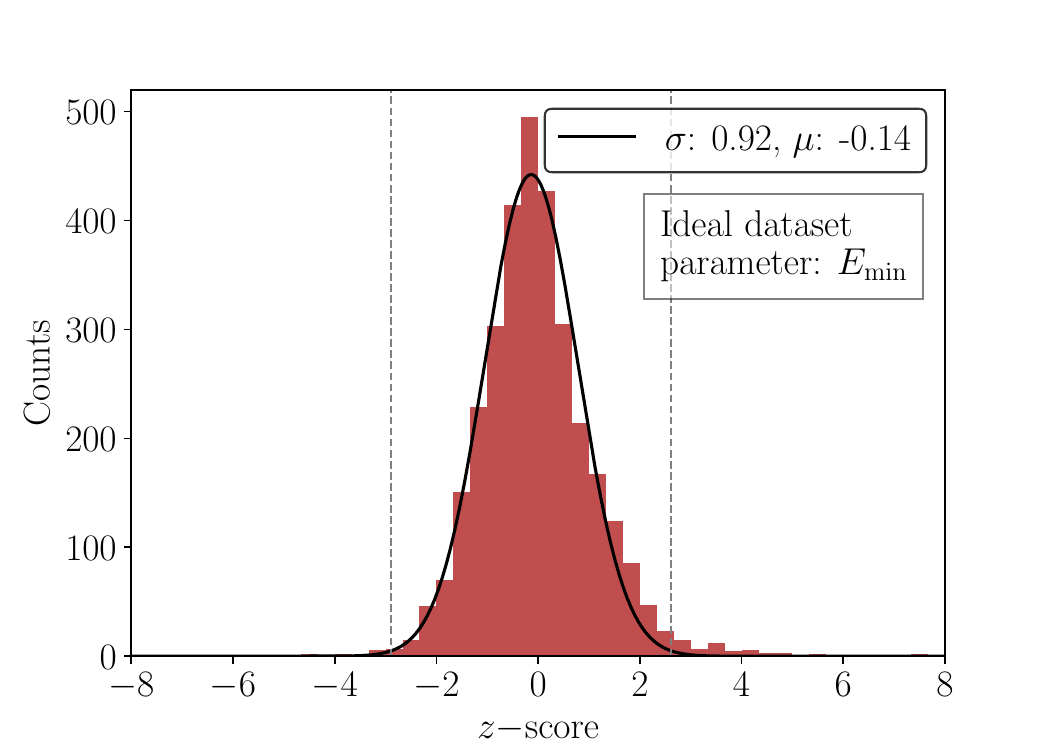}
    \end{minipage}\hfill
    \begin{minipage}{0.33\textwidth}
      \includegraphics[scale=0.37]{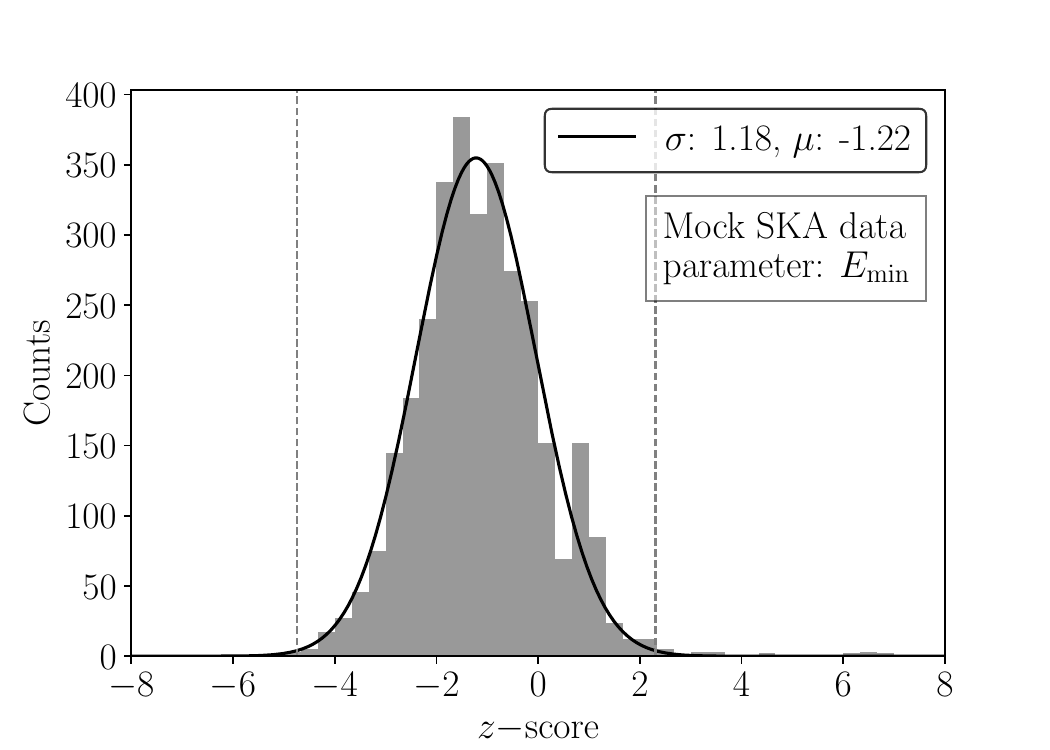}
    \end{minipage}\hfill
    \begin{minipage}{0.3\textwidth}
      \includegraphics[scale=0.37]{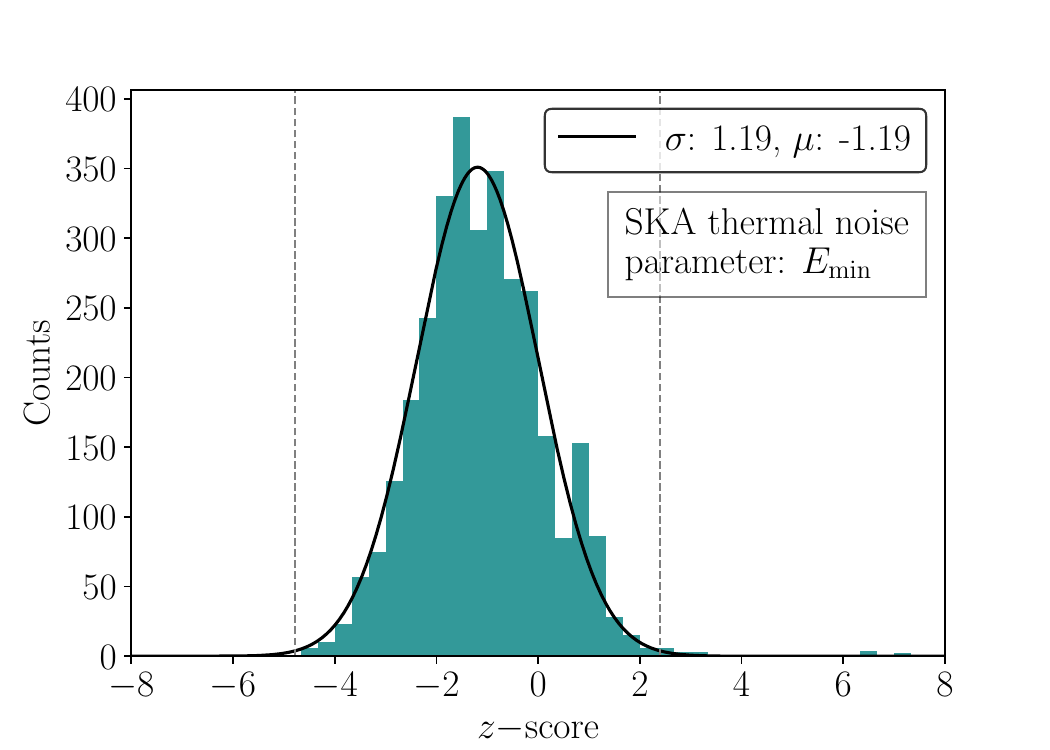}

    \end{minipage}
    \centering
    \begin{minipage}{0.33\textwidth}
       \includegraphics[scale=0.37]{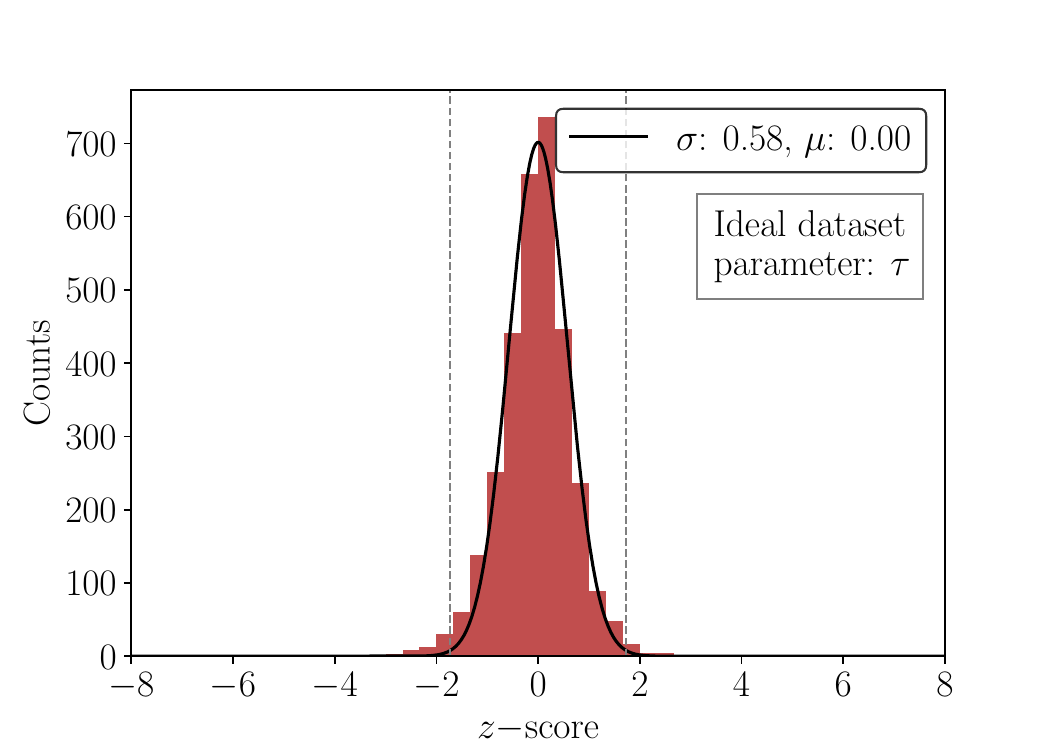}
       
    \end{minipage}\hfill
    \begin{minipage}{0.33\textwidth}
      \includegraphics[scale=0.37]{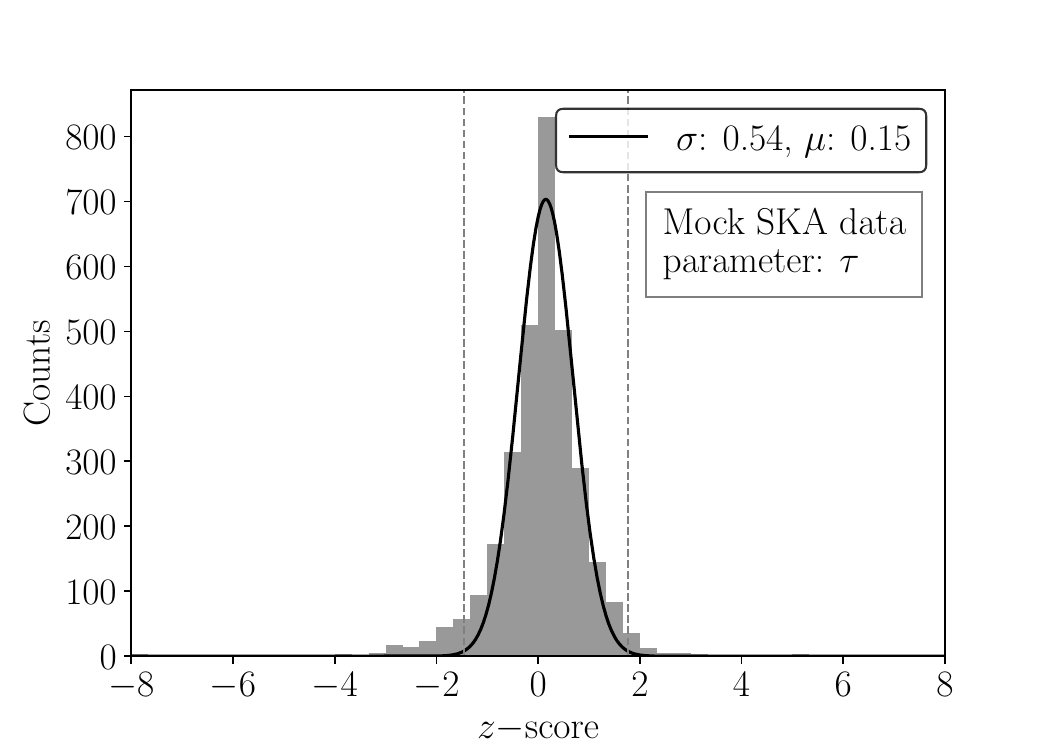}

    \end{minipage}\hfill
    \begin{minipage}{0.3\textwidth}
      \includegraphics[scale=0.37]{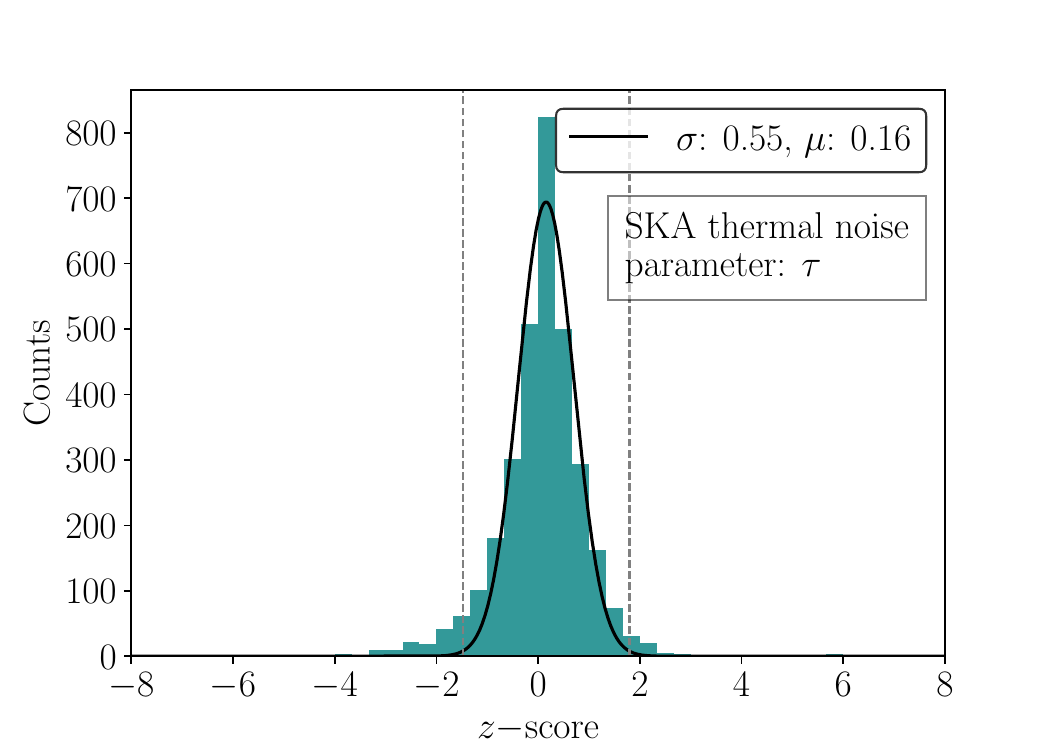}

    \end{minipage}
    \centering
    \begin{minipage}{0.33\textwidth}
       \includegraphics[scale=0.37]{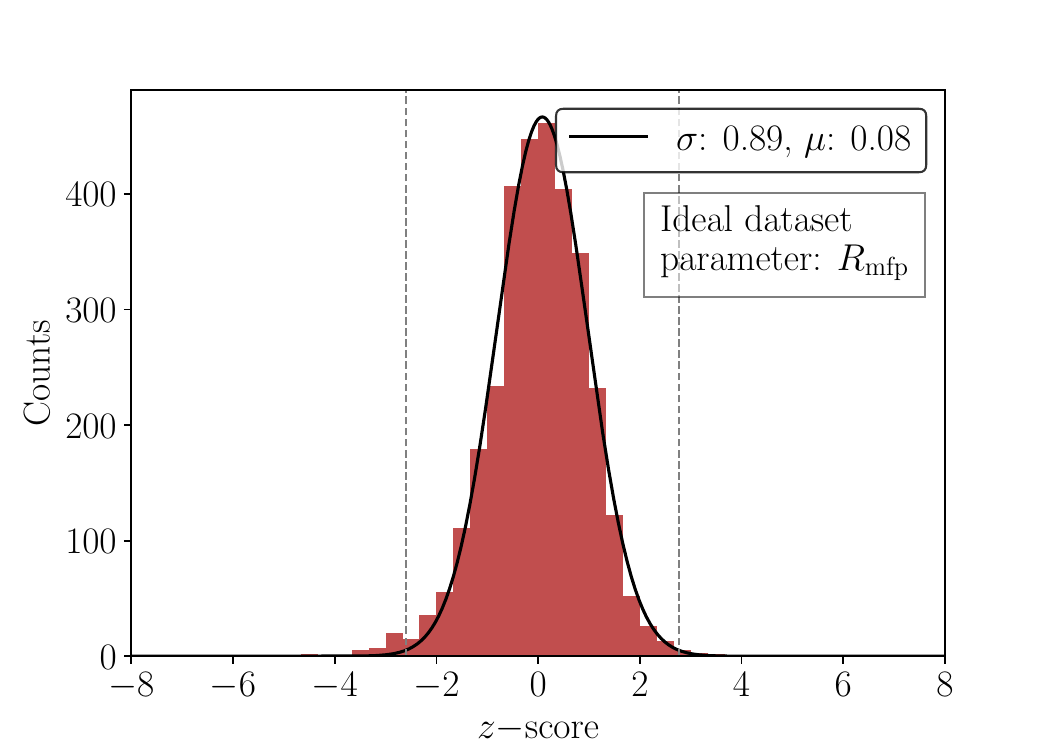}
    
    \end{minipage}\hfill
    \begin{minipage}{0.33\textwidth}
      \includegraphics[scale=0.37]{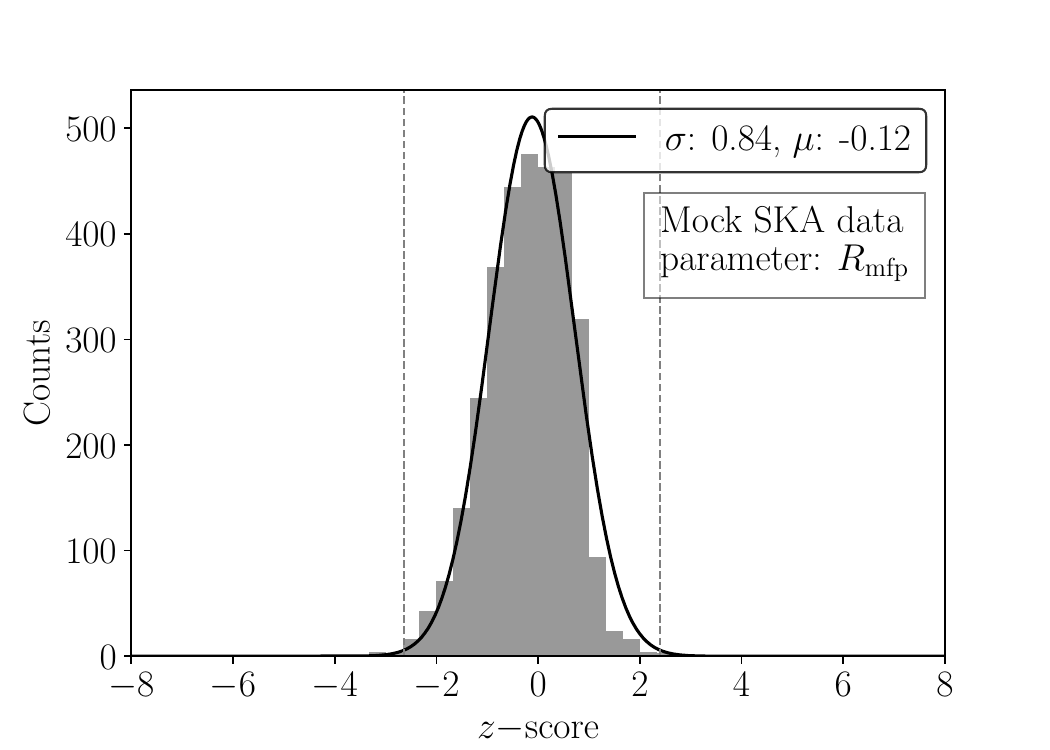}

    \end{minipage}\hfill
    \begin{minipage}{0.3\textwidth}
      \includegraphics[scale=0.37]{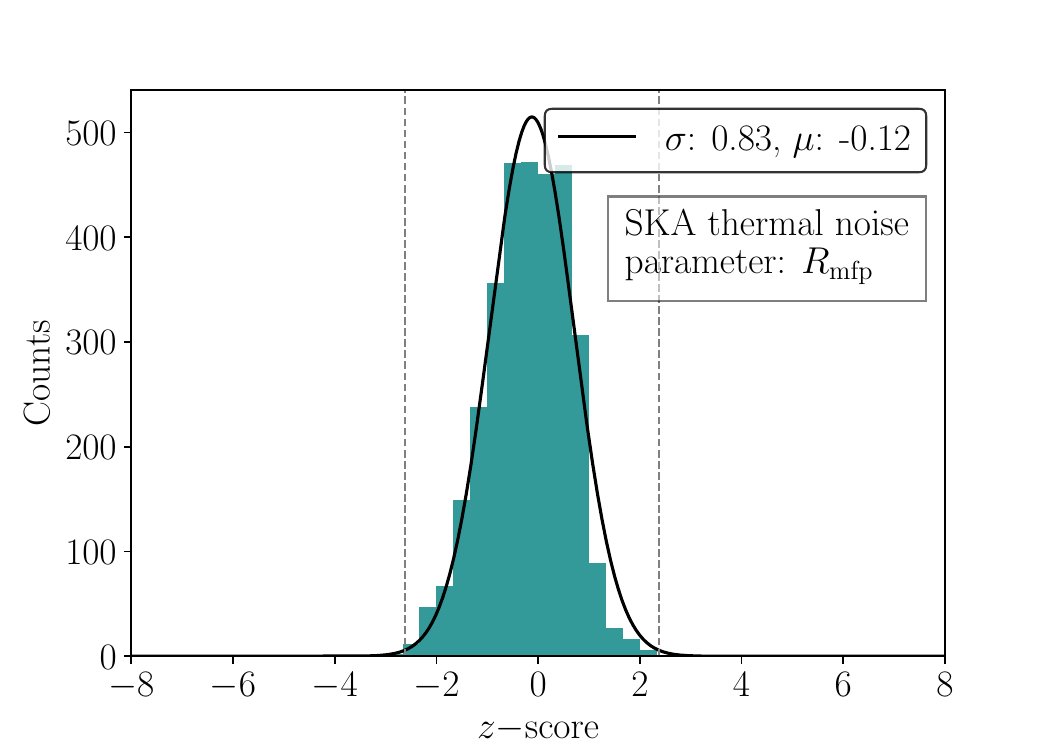}
      
    \end{minipage}
    \caption{Histogram of the normalized errors / $z\rm{-scores}$ for the parameters: $\alpha$, $E_{\rm{min}}$, $\tau$ and $R_{\rm{mfp}}$ as defined in Eq. \ref{eq: error}, similarly to Fig.~\ref{fig: histogram1}. The distribution is shown over all the 5-folds. Also shown in each panel is the best-fit Gaussian, with its parameters listed within the panel. All the parameter values are in $\log_{10}$ except $\alpha$ and $E_{\rm{min}}$.}\label{fig: histogram2}
\end{figure*}

\begin{figure*}
    \centering
    \begin{minipage}{0.4\textwidth}
       \includegraphics[scale=0.42]{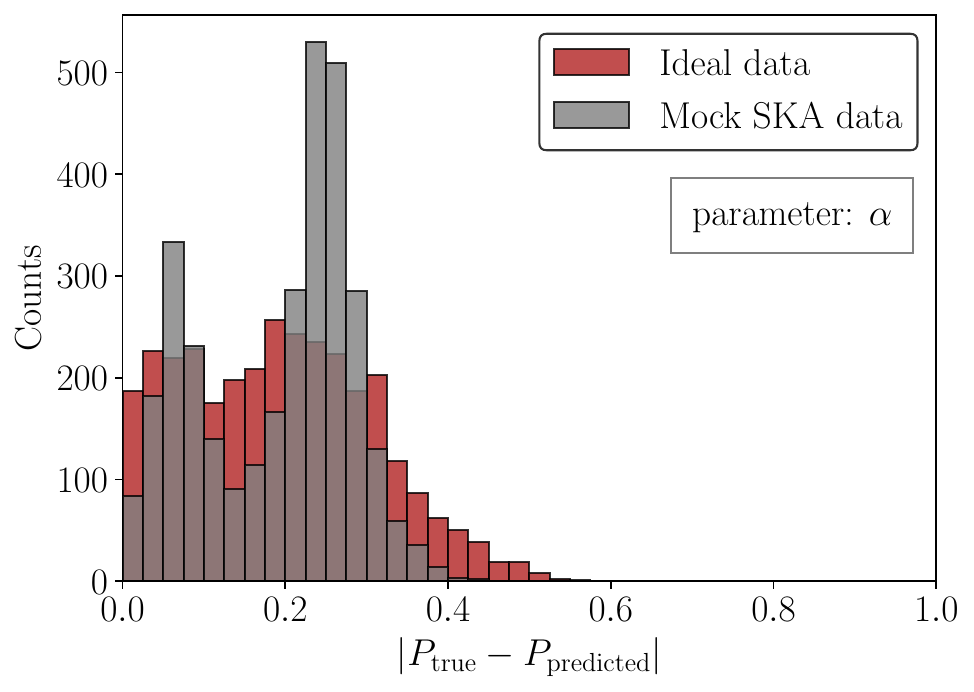}

    \end{minipage}
    \begin{minipage}{0.4\textwidth}
      \includegraphics[scale=0.42]{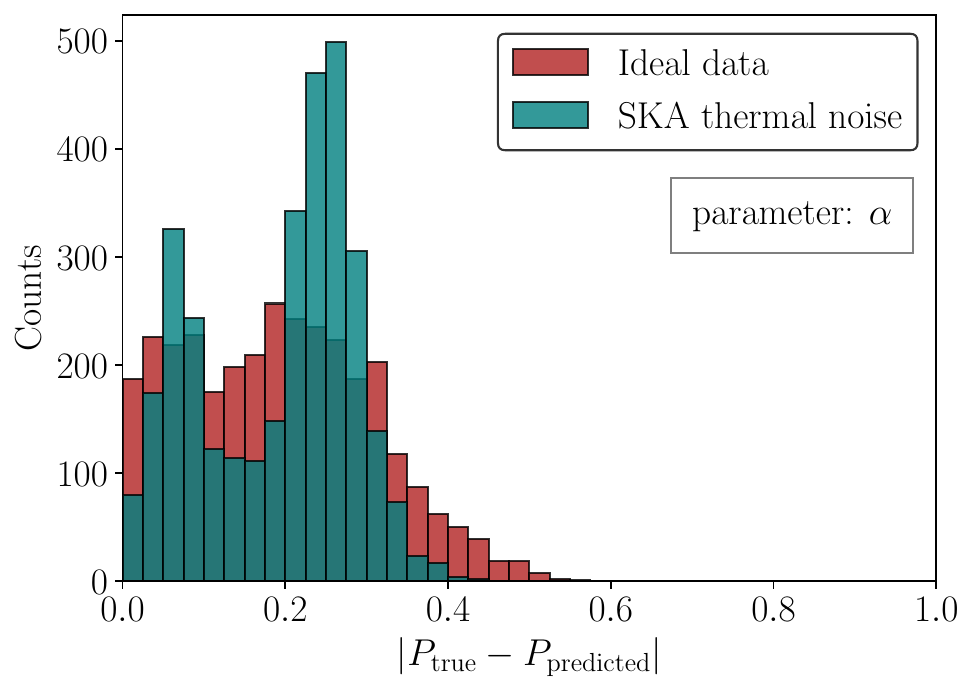}
      
    \end{minipage}
    \begin{minipage}{0.4\textwidth}
       \includegraphics[scale=0.42]{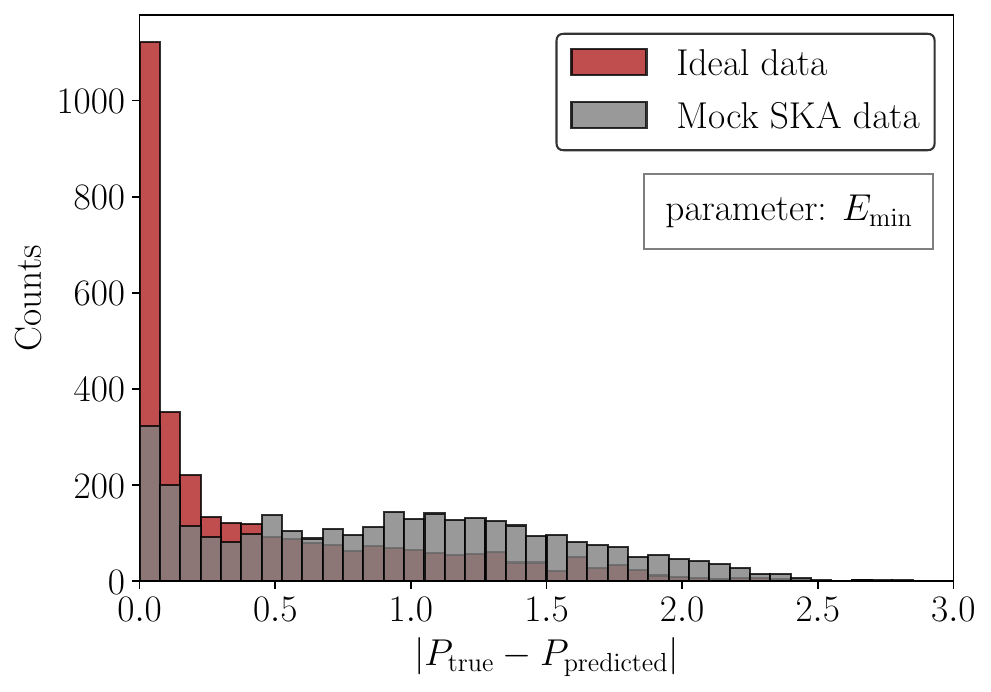}
    
    \end{minipage}
    \begin{minipage}{0.4\textwidth}
      \includegraphics[scale=0.42]{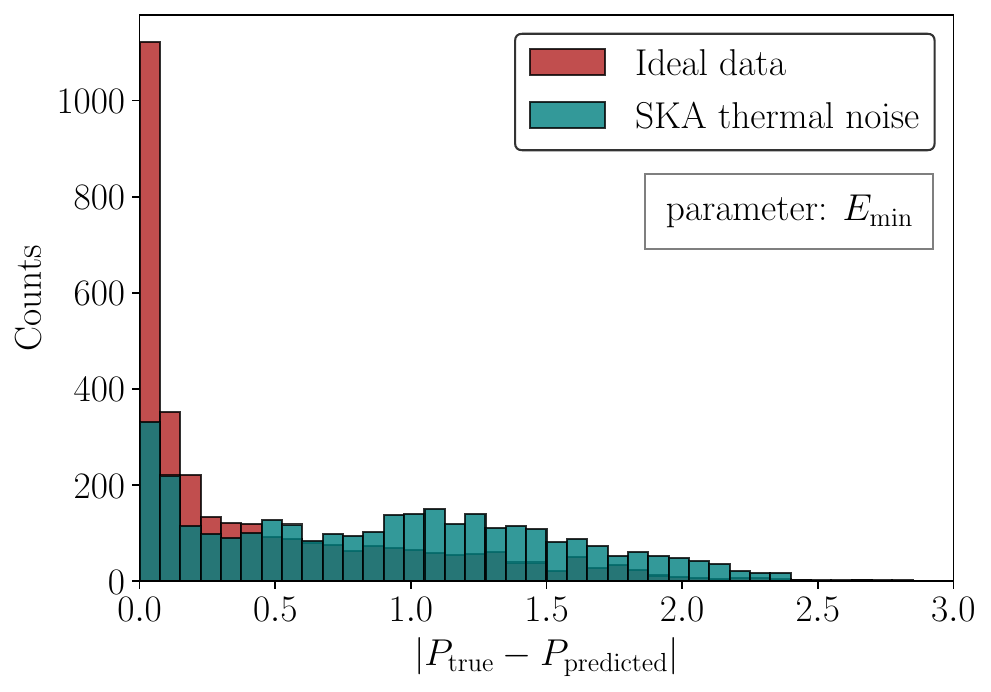}
      
    \end{minipage}
    
        \centering
    \begin{minipage}{0.4\textwidth}
       \includegraphics[scale=0.42]{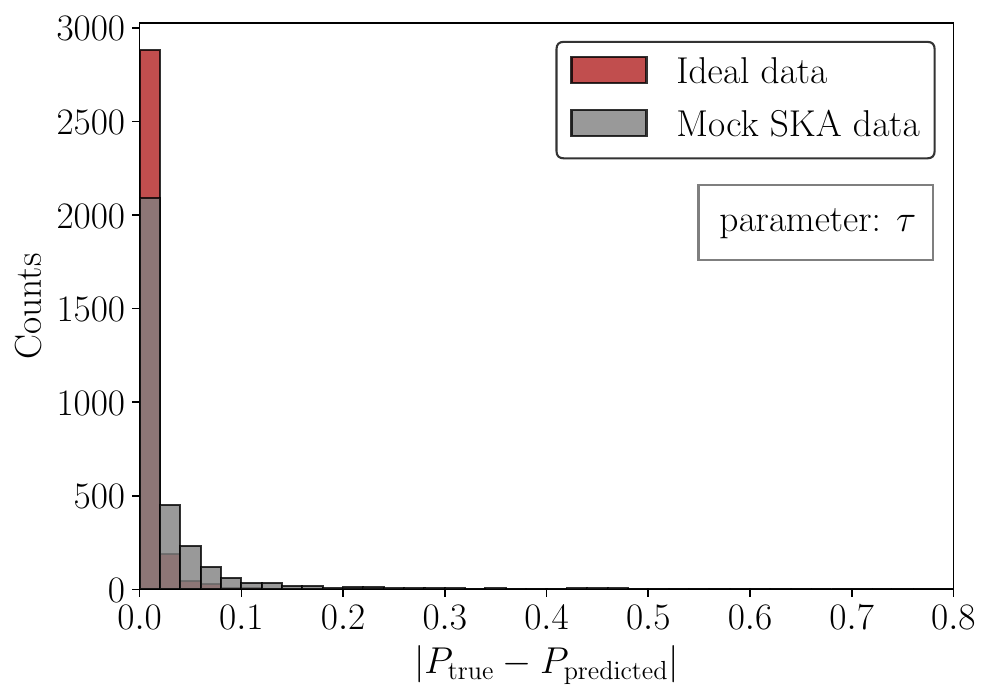}
       
    \end{minipage}
    \begin{minipage}{0.4\textwidth}
      \includegraphics[scale=0.42]{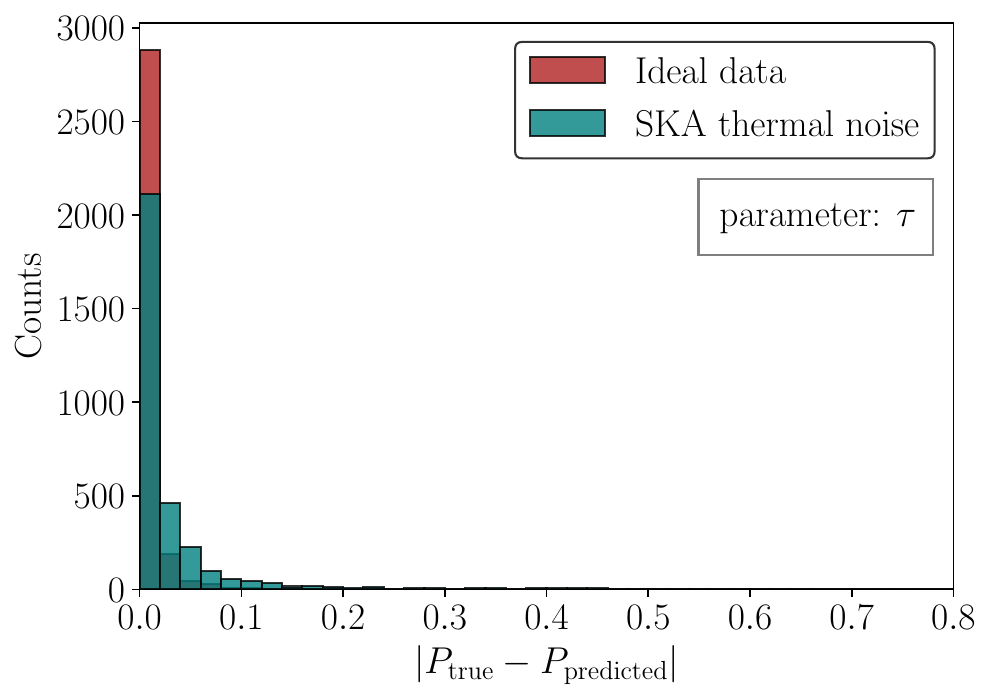}
      
    \end{minipage}
    
    \centering
    \begin{minipage}{0.4\textwidth}
       \includegraphics[scale=0.42]{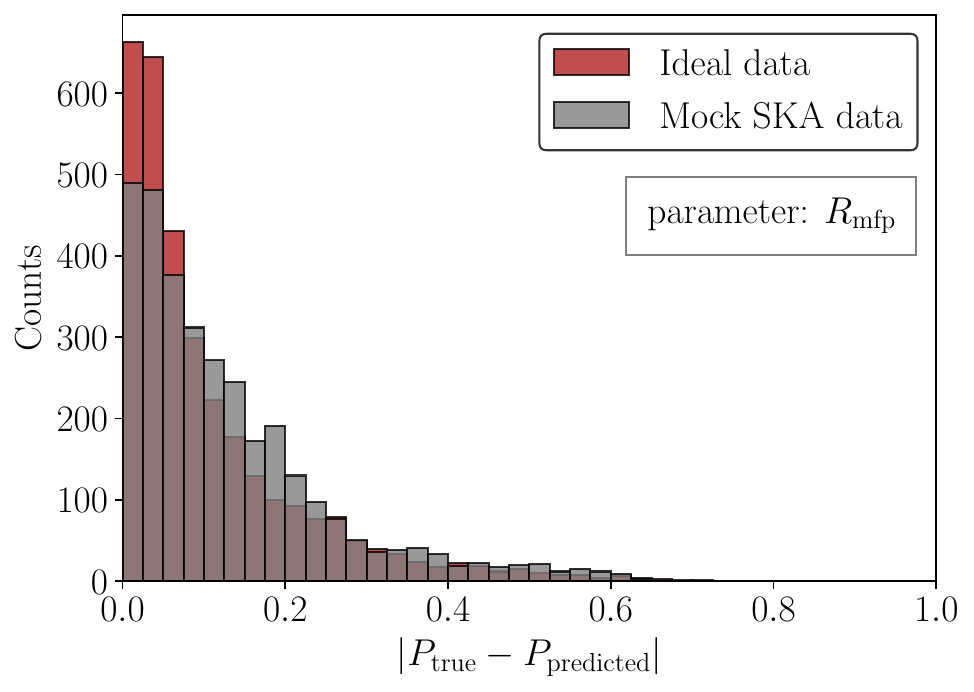}
       
    \end{minipage}
    \begin{minipage}{0.4\textwidth}
      \includegraphics[scale=0.42]{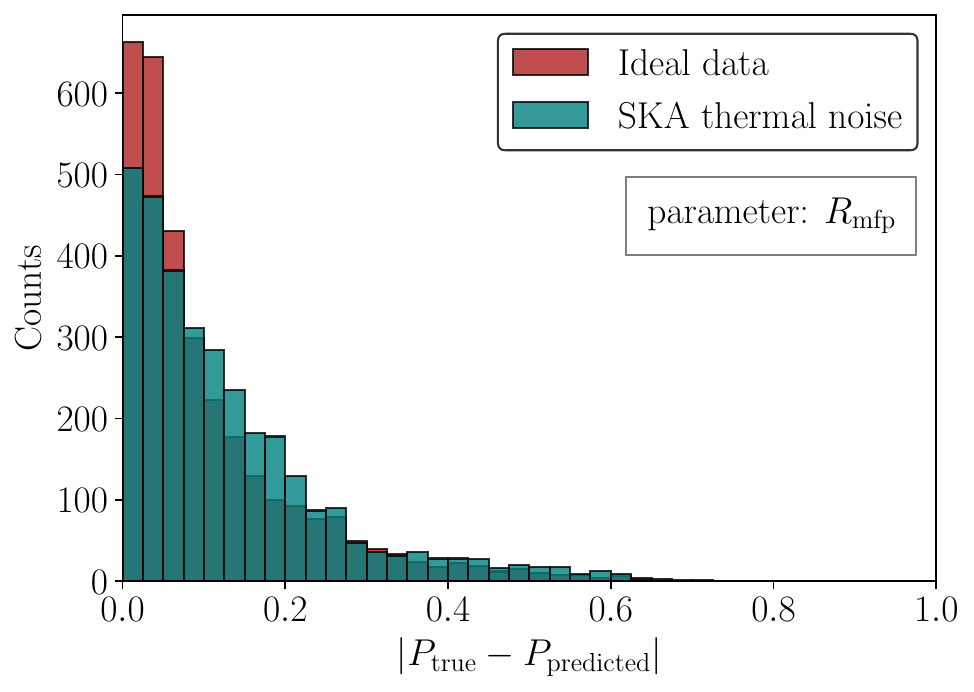}
      
    \end{minipage}
\caption{Histogram of the actual errors ($|P_{\rm{true}} - P_{\rm{predicted}}|$) in predicting the parameters: $\alpha$, $E_{\rm{min}}$, $\tau$ and $R_{\rm{min}}$, similarly to Fig.~\ref{fig: sigma_prime1}. The distribution is shown over all the 5-folds. All the parameter values are in $\log_{10}$ except $\alpha$ and $E_{\rm{min}}$.}\label{fig: sigma_prime2}

\end{figure*}

In order to avoid having too many figures in the main paper, we moved to this Appendix the histograms for four of the astrophysical parameters: $\alpha$, $E_{\rm{min}}$, $\tau$ and $R_{\rm{mfp}}$. Fig.~\ref{fig: histogram2} shows histograms of the normalized error or $z\rm{-scores}$  in predicting the parameters, similarly to Fig.~\ref{fig: histogram1}. Fig.~\ref{fig: sigma_prime2} shows histograms of the actual errors ($|P_{\rm{true}} - P_{\rm{predicted}}|$), similarly to 
Fig.~\ref{fig: sigma_prime1}.

\begin{table*}
\begin{tabular}{ccccccccc}
\hline
\multicolumn{9}{c}{Ideal dataset}   \\

\hline
Parameters   & \multicolumn{2}{c}{$\alpha$} & \multicolumn{2}{c}{$E_{\rm{min}}$} & \multicolumn{2}{c}{$\tau$} & \multicolumn{2}{c}{$R_{\rm{mfp}}$} \\
\hline
Gaussian fit & $\sigma$  & $\mu$  & $\sigma$  & $\mu$  & $\sigma$    & $\mu$  & $\sigma$    & $\mu$       \\
\hline\hline
\centering
 Fold 1 & $1.13$ & $+0.05$ & $0.86$ & $-0.07$ & $0.62$ & $+0.02$ & 0.83 & $+0.06$\\ 
\centering
 Fold 2 & $1.10$ & $+0.19$ & $0.94$ & $-0.30$ & $0.61$ & $+0.02$ & 0.90 & $+0.08$\\ 
\centering
 Fold 3 & $1.14$ & $+0.05$ & $0.94$ & $-0.22$ & $0.65$ & $-0.06$ & 0.94 & $+0.04$\\ 
 \centering
 Fold 4 & $1.10$ & $+0.09$ & $0.95$ & $+0.04$ & $0.61$ & $+0.00$ & 0.89 & $+0.11$\\ 
\centering
 Fold 5 & $1.10$ & $+0.00$ & $0.87$ & $-0.17$ & $0.54$ & $+0.06$ & 0.90 & $+0.11$\\ 
Mean    & 1.11   & $+0.08$ &  0.91 &  $-0.14$  & 0.61  & $+0.01$ & 0.89 & $+0.08$ \\
\hline
\end{tabular}

\begin{tabular}{ccccccccc}
\multicolumn{9}{c}{Mock SKA dataset}   \\

\hline
Parameters   & \multicolumn{2}{c}{$\alpha$} & \multicolumn{2}{c}{$E_{\rm{min}}$} & \multicolumn{2}{c}{$\tau$} & \multicolumn{2}{c}{$R_{\rm{mfp}}$} \\
\hline
Gaussian fit & $\sigma$  & $\mu$  & $\sigma$  & $\mu$  & $\sigma$    & $\mu$  & $\sigma$    & $\mu$       \\
\hline\hline
\centering
 Fold 1 & $1.18$ & $+0.19$ & $1.17$ & $-1.09$ & $0.45$ & $+0.20$ & 0.87 & $-0.15$\\ 
\centering
 Fold 2 & $1.18$ & $+0.28$ & $1.19$ & $-1.14$ & $0.54$ & $+0.17$ & 0.82 & $-0.07$\\ 
\centering
 Fold 3 & $1.23$ & $+0.17$ & $1.23$ & $-1.06$ & $0.58$ & $+0.11$ & 0.82 & $-0.14$\\ 
 \centering
 Fold 4 & $1.18$ & $+0.22$ & $1.11$ & $-1.37$ & $0.60$ & $+0.18$ & 0.87 & $-0.15$\\ 
\centering
 Fold 5 & $1.19$ & $+0.17$ & $1.14$ & $-1.42$ & $0.52$ & $+0.15$ & 0.83 & $-0.08$\\ 
Mean    &  1.19 &  $+0.21$ &  1.17 &  $-1.22$ &  0.54  & $+0.16$ & 0.84 & $-0.12$ \\
\hline
\end{tabular}

\begin{tabular}{ccccccccc}
\multicolumn{9}{c}{SKA thermal noise case}   \\

\hline
Parameters   & \multicolumn{2}{c}{$\alpha$} & \multicolumn{2}{c}{$E_{\rm{min}}$} & \multicolumn{2}{c}{$\tau$} & \multicolumn{2}{c}{$R_{\rm{mfp}}$} \\
\hline
Gaussian fit & $\sigma$  & $\mu$  & $\sigma$  & $\mu$  & $\sigma$    & $\mu$  & $\sigma$    & $\mu$       \\
\hline\hline
\centering
 Fold 1 & $1.18$ & $+0.20$ & $1.17$ & $-1.05$ & $0.52$ & $+0.20$ & 0.85 & $-0.16$\\ 
\centering
 Fold 2 & $1.16$ & $+0.29$ & $1.16$ & $-1.09$ & $0.52$ & $+0.18$ & 0.83 & $-0.09$ \\ 
\centering
 Fold 3 & $1.23$ & $+0.18$ & $1.24$ & $-1.05$ & $0.61$ & $+0.10$ & 0.83 & $-0.17$\\ 
 \centering
 Fold 4 & $1.18$ & $+0.22$ & $1.19$ & $-1.32$ & $0.61$ & $+0.20$ & 0.87 & $-0.17$\\ 
\centering
 Fold 5 & $1.20$ & $+0.19$ & $1.17$ & $-1.40$ & $0.52$ & $+0.16$ & 0.81 & $-0.05$\\ 
Mean    &  1.19  & $+0.22$ &  1.19 &  $-1.18$  & 0.56  & $+0.17$ & 0.84 & $-0.13$ \\
\hline
\end{tabular}

\caption{Same as Table \ref{tab: table_clean_CV}, but for the parameters: $\alpha$, $E_{\rm{min}}$, $\tau$ and $R_{\rm{mfp}}$. All the parameter values are in $\log_{10}$ except $\alpha$ and $E_{\rm{min}}$.}\label{tab: table_clean_CV_}
\end{table*}




\bsp	
\label{lastpage}
\end{document}